# On-surface synthesis and aromaticity of large cyclocarbons

**Authors:** Lisanne Sellies[1,#], Marco Vitek[2,#], Yueze Gao[3,#], Fabian Paschke[1,#], Florian Albrecht[1], Jakob Eckrich[1], Beren Dempsey[4], Stefano Barison[1,5], Leonard-Alexander Lieske[1], Samuele Piccinelli[1,6], Alberto Baiardi[1], Ivano Tavernelli[1], Harry L. Anderson[3,*], Igor Rončević[4,*], Leo Gross[1,*]

**Affiliations:**

[1]IBM Research Europe – Zurich, CH-8803 Rüschlikon, Switzerland

[2] Institute of Organic Chemistry and Biochemistry of the Czech Academy of Sciences, Flemingovo nám. 542/2, 160 00 Prague 6, Czechia

[3] Department of Chemistry, Oxford University, Chemistry Research Laboratory, Oxford, UK

[4] Department of Chemistry, University of Manchester, Manchester M13 9PL, UK

[5] Institute of Physics, Eidgenössische Technische Hochschule Zürich (ETHZ), CH-8092 Zurich, Switzerland

[6] Institute of Physics, École Polytechnique Fédérale de Lausanne (EPFL), CH-1015 Lausanne, Switzerland

[#] Equally contributing first authors

* Corresponding authors. Email: igor.roncevic@manchester.ac.uk; harry.anderson@chem.ox.ac.uk; lgr@zurich.ibm.com

**Abstract:** Molecular rings of $N$ carbon atoms, that is, cyclo[$N$]carbons, or $C_N$, can be formed by tip-induced chemistry [1-7]. Because of their monocyclic geometry, cyclocarbons are fundamentally important for testing theoretical models of aromaticity [8-11]. Here, we synthesized large cyclo[$N$]carbons, with $N$ up to 88, by tip-induced chemistry on a NaCl surface and studied their aromaticity by measuring their transport gaps by scanning tunnelling spectroscopy. We first generated $C_{20}$ and $C_{22}$, and then fused multiple cyclocarbons [5-7] by means of atom manipulation, obtaining $C_{42}$, $C_{44}$, $C_{46}$, $C_{66}$ and $C_{88}$. In agreement with predictions obtained using a finely tuned density functional [12-15] and large active space approximate configuration interaction calculations executed on quantum hardware [16, 17], we observe a substantially smaller transport gap for $C_{20}$ ($N = 4n$) compared to $C_{22}$ ($4n+2$), and for $C_{44}$ ($4n$) relative to $C_{42}$ ($4n+2$). In larger cyclocarbons, the oscillation of the transport gap between anti-aromatic $N = 4n$ and aromatic $N = 4n+2$ cyclocarbons becomes smaller, and is expected to eventually vanish with increasing $N$, indicating non-aromaticity. Our experimental results show that aromaticity persists at $N = 42$, and theory predicts ring currents comparable in magnitude to that of benzene in cyclocarbons of this size. In the future, such large cyclocarbons could be used to study conductance, quantum interference, and the effects of aromaticity in single atomic carbon wires and circuits.

## Introduction

Several cyclo[$N$]carbons, rings of $N$ carbon atoms, denoted as $C_N$, have been generated by on-surface synthesis on ultrathin NaCl films up to $N$ = 50 [1-7]. Cyclocarbons are excellent model systems for benchmarking theoretical predictions, as their high symmetry leads to multiple orbital degeneracies that can be broken by various distortions [8-11]. Moreover, their monocyclic structure renders them highly suitable for studying aromaticity and anti-aromaticity. Aromaticity is a manifestation of electron delocalisation in closed circuits, which affects several chemical and physical properties. For example, aromatic systems tend to have increased stability, reduced reactivity, diatropic ring currents, and a tendency towards bond-length equalization [18], whereas anti-aromatic systems are associated with the opposite effects. Hückel's rule states that rings with $4n+2$ delocalised electrons (with integer $n$) may be expected to be aromatic, whereas rings with $4n$ delocalised electrons tend to be anti-aromatic. Even-$N$ cyclocarbons, up to some size, follow this rule [9, 11]. They are often called doubly aromatic ($N = 4n+2$) or doubly anti-aromatic ($N = 4n$), because of their two π-systems, one of which is in-plane (of the atoms) and the other out-of-plane, with both being occupied by $N$ electrons in the ground state. However, the question at which ring size cyclocarbons become non-aromatic remains open [9, 19]. For example, $C_{18}$ was experimentally found to display bond-length alternation (BLA) [1], despite having a Hückel-aromatic electron count of 18 in both π-systems. This has led some theorists to question the aromaticity of $C_{18}$ [19], implying that cyclocarbons above $N \approx 20$ would be non-aromatic, essentially behaving like linear polyynes [20, 21]. Conversely, a recent study showed that the experimental solution-phase UV-vis absorption spectrum of a $C_{48}$ catenane is very different from that of its linear counterpart [13].

Even-$N$ cyclocarbons up to $N$ = 20 have been synthesized on surfaces by unmasking suitable precursors [1-4]. Increasing $N$ much further is challenging with this approach, because intact surface-deposition of precursors by thermal sublimation becomes difficult with increasing size of the precursors. Larger $C_N$ have been studied as cations in the gas phase (up to $N$ = 20–80) [22-26], but they are unstable with respect to fragmentation or rearrangement to polycyclic structures, eventually leading to fullerenes [24, 27, 28]. An alternative approach for the generation of large cyclocarbons is by unmasking and fusing cyclocarbon precursors, as some of us demonstrated by fusing two $C_{13}$ to form $C_{26}$ [5]. Recently, a similar approach was used by Xu and co-workers, who formed $C_{20}$ and $C_{30}$ by fusing $C_{10}$ [6], as well as $C_{25}$ and $C_{50}$ from precursors of $C_5$ and $C_6$, respectively [7].

To target large cyclocarbons, here we used custom-designed precursors to form $C_{20}$ and $C_{22}$ and then fused them using tip-induced chemistry to obtain larger cyclocarbons up to $C_{88}$. We measured the transport gaps of the created cyclocarbons, serving as observables of their aromaticity. In line with theory, using the finely tuned OX-BLYP30 density functional approximation [12, 13, 15] and the *GW* method [29], we observe oscillations on the order of a few hundreds of meV in the transport gap between $N = 4n+2$ and $N = 4n$ up to $N$ = 42, which result from aromaticity and indicate that the wavefunction is coherently delocalised over the whole ring. These calculations predict substantial aromatic stabilisation energies and ring currents at $N$ = 42, supporting this estimate of the ring size at which aromaticity persists. We confirm that the transport gap differences arise from the π-electron correlation via large active space approximate configuration interaction calculations executed on quantum hardware using the SqDRIFT algorithm [16, 17].

## Generation of large cyclocarbons

First, using solution-based chemistry, we synthesized $C_{22}(CO)_8$ (compound **1**, **Fig. 1a**), a precursor for the on-surface synthesis of $C_{22}$ (see **SI section 1 to 5** and Supplementary **Figs. S1** to **S7** for more details). Precursor **1** was sublimed onto an Au(111) single crystal surface partially covered with monolayer and bilayer NaCl at a sample temperature of about $T$ = 10 K, with the sample placed in a home-built scanning tunnelling microscopy (STM)/atomic force microscopy (AFM) system. On-surface synthesis and characterisation were performed at $T$ = 5 K using STM and AFM with CO-functionalized tips [30, 31]. In AFM images, the CO masking groups of **1** appear as dark features [1], whereas triple bonds appear as bright features, due to bond-order related AFM contrast obtained with CO tip-functionalisation [1, 31, 32], see **Fig. 1a**.

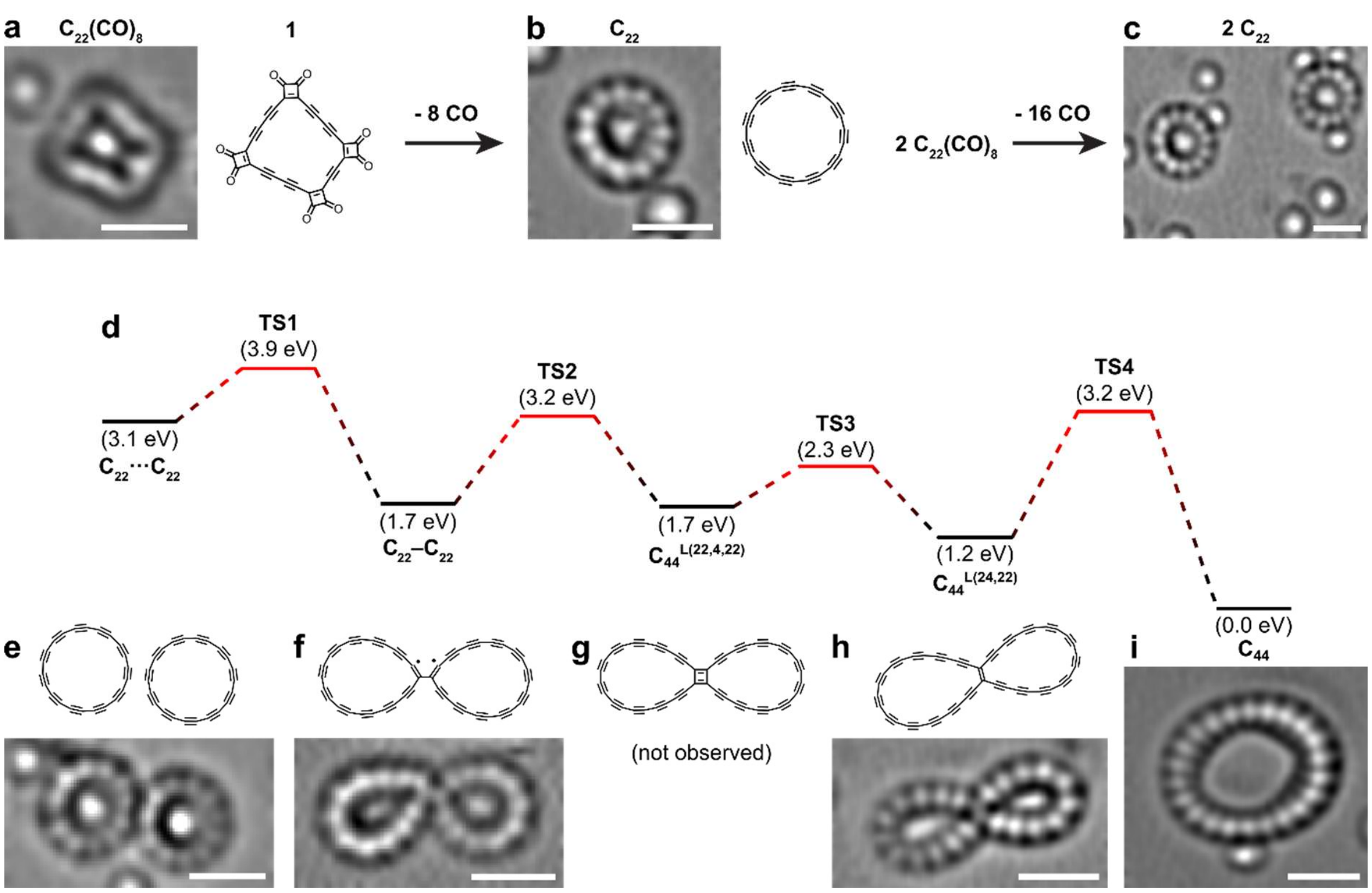

**Figure 1. On-surface synthesis of $C_{22}$ and $C_{44}$.** Laplace-filtered AFM data measured with CO-functionalized tips and corresponding chemical structures. **a**, Precursor **1**. **b**, Cyclo[22]carbon, $C_{22}$. **c**, Two $C_{22}$ molecules with a centre-to-centre-distance of 5 nm. **d–i**, Generation of $C_{44}$ from two $C_{22}$, along with an energy diagram (**d**) calculated using OX-BLYP30, including energies of transition structures (red plateaus, for details see Supplementary **Fig. S12**). **e**, Two adjacent $C_{22}$ molecules. **f**, Singly-linked $C_{22}$–$C_{22}$ intermediate. **g**, The $C_{44}^{L(22,4,22)}$ lemniscate intermediate (not observed). **h**, The $C_{44}^{L(24,22)}$ lemniscate intermediate. **i**, Cyclo[44]carbon, $C_{44}$. CO molecules appear as circular bright features in the AFM images. Reactions and lateral manipulations were induced by voltage pulses using the STM/AFM tip (see text for parameters). Scale bars: 1 nm. For AFM parameters and raw data see Supplementary **Figs. S9** and **S11**.

By on-surface synthesis, we generated $C_{22}$ on monolayer NaCl on Au(111). CO masking groups were dissociated [1] from **1** (**Fig. 1a** and Supplementary **Fig. S8**) by applying voltage pulses of typically $V_P$ = 4.5 V for a duration of $t_P$ = 100 ms, with the tip placed above the molecule and retracted by a tip-height offset $\Delta z$ = 12 to 13 Å from the STM-controlled setpoint of $V$ = 0.2 V, $I$ = 0.3 pA. $C_{22}$ was obtained with a yield of approximately 73% and successfully generated in 172 instances. In unsuccessful attempts, the ring opened to form linear polyynic chains, or a side group remained attached to the ring. The STM setpoint and pulse duration were kept constant throughout this work, but different pulse voltages $V_P$ and tip-height offsets $\Delta z$, resulting in different tunnelling currents $I_P$, were used for different steps of the on-surface synthesis. During the application of $V_P$ for unmasking, tunnelling currents $I_P$ were too small to be measured, but are extrapolated to be on the order of fA. **Fig. 1b** shows a generated $C_{22}$ with eleven bright lobes, indicating the location of the triple bonds.

We moved $C_{22}$ laterally on the surface with voltage pulses [33, 34], typically applied with the tip positioned about 1 nm next to the molecule in the direction of the intended lateral movement, using typically voltages of the pulse $V_P$ = 5.5 to 6.5 V at tip-height offsets $\Delta z$ = 13 to 19 Å, resulting in currents during the pulse $I_P$ in the range of fA to tens of pA. We moved $C_{22}$ (**Fig. 1c**) over distances of several nm, until two $C_{22}$ were adjacent to each other (**Figs. 1e** and Supplementary **Figs. S9** and **S10**). To covalently fuse two molecules [1, 5-7, 34], we laterally positioned the tip between two adjacent carbon rings and applied pulses with $V_P$ = 4.5 V, $\Delta z$ = 8 to 9 Å, resulting in $I_P$ of a few pA. Applying such pulses initially produced a $C_{22}$–$C_{22}$ intermediate that we assign as a diradical (**Figs. 1f** and Supplementary **Fig. S11**). Additional pulses with similar parameters – probably *via* the (not observed) intermediate with a central 4-membered ring [6] shown in **Fig. 1g** – produced two fused carbon rings that share a C=C bond, see **Fig. 1h**. We term such multicyclic carbon molecules, that share carbon bonds, “carbon lemniscates” and label this lemniscate as $C_{44}^{L(24,22)}$, with the numbers in the superscript denoting the numbers of carbon atoms in each of the connected rings. Often, $C_{44}^{L(24,22)}$ was formed by a single pulse from two adjacent $C_{22}$. Applying further voltage pulses with the tip placed above the shared C=C bond of the lemniscate, using similar parameters as for fusing, led to ring-opening and thus yielded cyclo[44]carbon, $C_{44}$, see **Fig. 1i**.

We modelled a reaction pathway from two $C_{22}$ to $C_{44}$, see **Fig. 1d** and Supplementary **Fig. S12**. For each reaction intermediate in the sequence, the energy is lower than that of the previous one, and all barriers are lower than the energy supplied to the molecule by a single electron with $eV_P$ in the experiment. The 4-membered ring intermediate $C_{44}^{L(22,4,22)}$ was not observed in our experiments, probably because the voltage pulses used to overcome the barrier to generate it from the $C_{22}$–$C_{22}$ diradical (1.5 eV, **TS2**), supply enough energy to overcome the next barrier (0.6 eV, **TS3**) as well, generating the $C_{44}^{L(24,22)}$ lemniscate. For smaller carbon lemniscates, e.g. $C_{20}^{L(10,4,10)}$, intermediates with 4-membered rings were observed [6], which may be related to strain. The monotonous decrease of the transition state energies from **TS1** to **TS3** (**Fig. 1d**) explains why we often formed $C_{44}^{L(24,22)}$ with a single pulse from two adjacent $C_{22}$.

Using the described pulses for unmasking, movement and fusing rings, we fused $C_{22}$ with $C_{44}^{L(24,22)}$ (**Fig. 2a** and Supplementary **Figs. S13** and **S14**) and obtained the lemniscate $C_{66}^{L(24,24,22)}$ shown in **Fig. 2b** (see the calculated mechanism in Supplementary **Fig. S15**). Applying a pulse with $V_P$ = 4.5 V, we cleaved one bond, forming the $C_{66}^{L(46,22)}$ lemniscate (**Fig.

**2c**). By increasing $V_P$ to 4.7 V, ring opening was induced and yielded cyclocarbon $C_{66}$, see **Fig. 2d**.

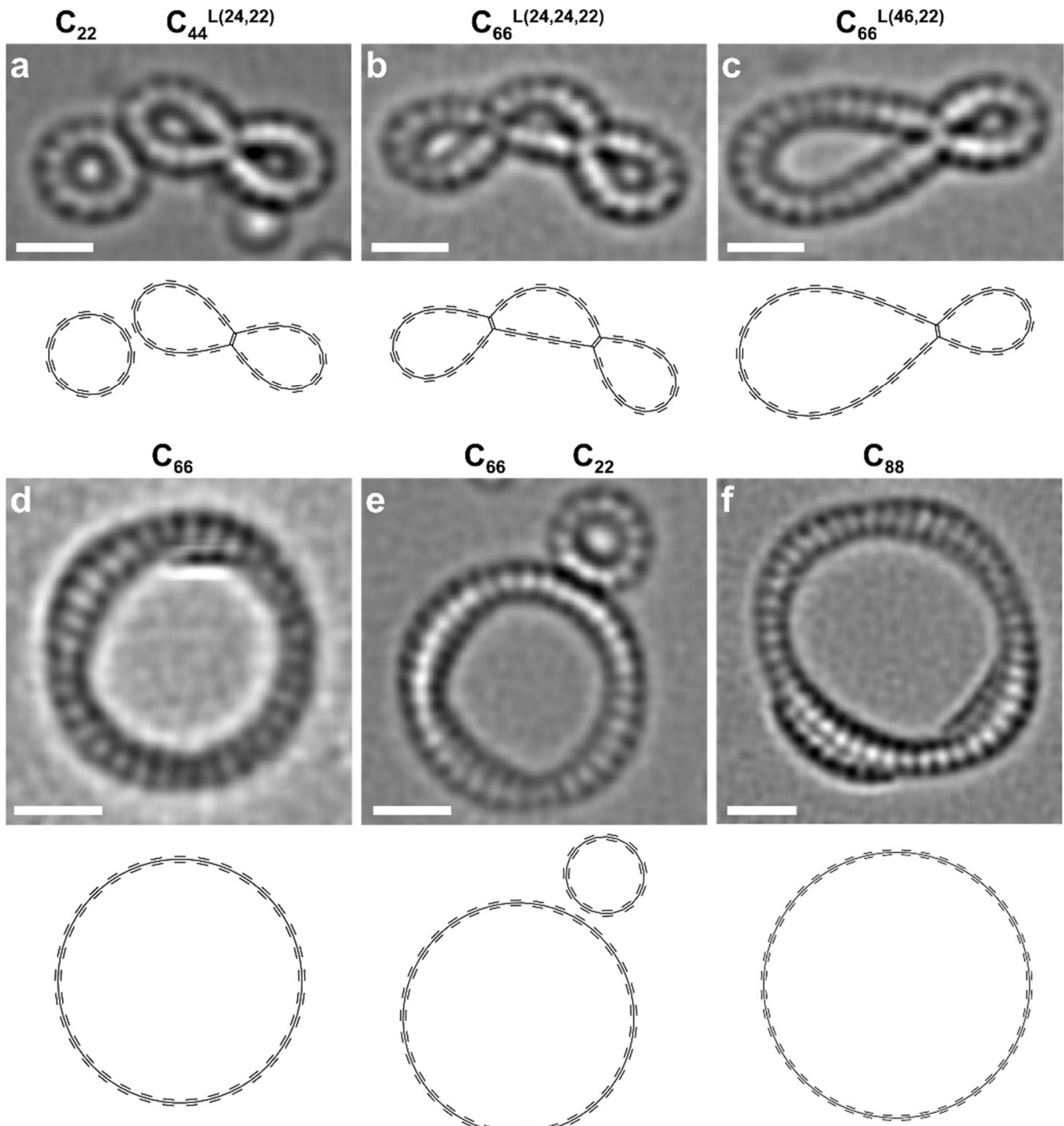

**Figure 2. On-surface synthesis of $C_{66}$ and $C_{88}$.** Sequence of Laplace-filtered AFM data (upper panels) and respective chemical structures (bottom panels). **a**, $C_{22}$ adjacent to $C_{44}^{L(24,22)}$. **b**, Lemniscate $C_{66}^{L(24,24,22)}$ formed by fusion (pulse parameters: $V_p$ = 4.5 V, $\Delta z$ = 8.5 Å, $I_P$ = 7 pA). **c**, Lemniscate $C_{66}^{L(46,22)}$ formed by ring-opening ($V_P$ = 4.5 V, $\Delta z$ = 8 Å, $I_P$ = 9 pA). **d**, $C_{66}$ formed by ring opening ($V_P$ = 4.7 V, $\Delta z$ = 10 Å, $I_P$ = 0.5 pA). **e**, $C_{22}$ moved to be adjacent to $C_{66}$ using lateral manipulation ($V_P$ = 6.5 V, $\Delta z$ = 19 Å, $I_P$ = 1 pA). **f**, $C_{88}$ formed by fusion of $C_{22}$ and $C_{66}$ ($V_P$ = 6.5 V, $\Delta z$ = 12 Å, $I_P$ = 100 pA). Reactions and lateral manipulations were induced using the STM tip by the abovementioned voltage pulse parameters with $\Delta z$ with respect to the STM setpoint $V$ = 0.2 V, $I$ = 0.3 pA. The predicted transition structures, energies and barriers are shown in Supplementary **Fig. S15**. Note that $C_{88}$ shows an apparent doubling of some segments of the rings in the AFM images, related to tip-induced lateral displacements during image acquisition (see Supplementary **Fig. S26** and **SI section 8** for adsorption site analysis of $C_{88}$). Scale bars: 1 nm. For AFM parameters and raw data see Supplementary **Fig. S13**.

Moving another $C_{22}$ towards $C_{66}$, see **Fig. 2e**, and fusing the molecules with a pulse laterally placed between the molecules, led to the formation of $C_{88}$, see **Fig. 2f.** We reproduced the fusing of cyclocarbons several times and formed 19 individual $C_{44}$, seven $C_{66}$ and two $C_{88}$. In some cases, other products were obtained, as shown in Supplementary **Fig. S16**.

Samples of the solution-phase synthesised $C_{22}$ precursor **1** contained traces of a byproduct **2**, that is, $C_{24}(CO)_8$ (about 1%, see **SI section 5**), which we found could be used as a precursor for $C_{24}$. Following a procedure similar as with **1**, we generated $C_{24}$ and fused it with $C_{22}$ to form $C_{46}$ (see Supplementary **Fig. S17**).

In addition, we used another custom synthesised precursor **3**, that is, $C_{20}(CO)_8$ [3] to form $C_{20}$ (see Supplementary **Fig. S18**). We sublimed precursors **1** and **3** onto the same sample, formed $C_{20}$ and $C_{22}$ and fused them into $C_{42}$ by applying a voltage pulse sequence similar as for the synthesis of $C_{44}$.

## Characterization of large cyclocarbons

All cyclo[*N*]carbons $C_N$ formed in this study, i.e., $C_{20}$, $C_{22}$, $C_{42}$, $C_{44}$, $C_{46}$, $C_{66}$, $C_{88}$, exhibit substantial BLA, as observed from the $N/2$ AFM-resolved bright features related to shorter bonds (see Supplementary **Figs. S19-S26**), in line with theoretical predictions [9] and resembling previous results of even-numbered cyclocarbons for $N \geq 16$ [1, 3-7]. For each cyclocarbon, we mapped the negative (NIR) and positive (PIR) ion resonances, which reflect the orbital densities associated with electron attachment (NIR) and detachment (PIR) [35, 36]. The NIR and PIR densities for both $N = 4n$ and $N = 4n+2$ cyclocarbons, show a particle-on-a-ring-like structure with $N/2$ lobes along the ring, located above the long (NIR) or short (PIR) bonds (see Supplementary **Figs. S27-S33**). The origin of these resonance densities is different for $N = 4n$ (anti-aromatic) [3] and $N = 4n+2$ (aromatic) [5] cyclocarbons, as discussed in **SI section 9**.

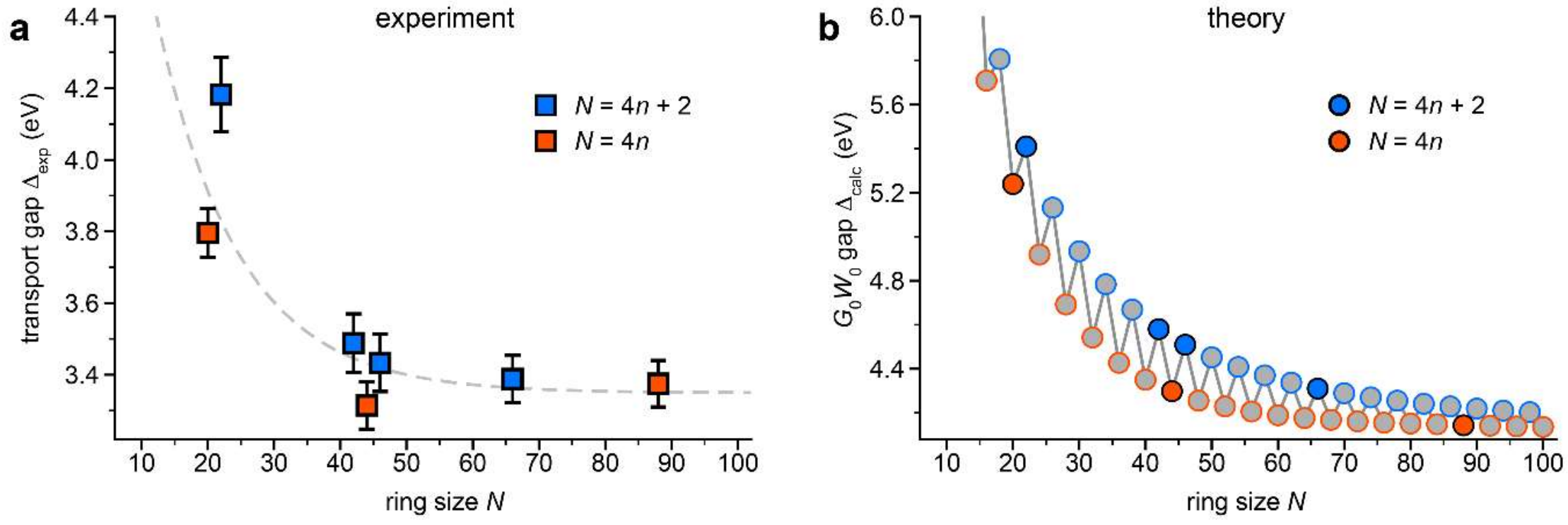

**Figure 3. Transport gaps of cyclocarbons. a**, STS measured gaps $\Delta_{exp}$, i.e., energy differences between the PIR and the NIR for the studied cyclocarbons on monolayer NaCl on Au(111) (for the STS spectra and estimation of errors, see Supplementary **Fig. S34**). The dashed line, which only serves as guidance, was obtained by fitting $\Delta_{exp} = ae^{-bN} + c$, with $a$, $b$, and $c$ as parameters. **b**, Gas-phase calculated IPEA gaps $\Delta_{calc}$ obtained using $G_0W_0$@OX-BLYP30/def2-QZVP.

For the different $C_N$ created in this study, we measured the transport gaps $\Delta_{exp}$ by scanning tunnelling spectroscopy (STS) on monolayer NaCl on Au(111) as the difference between the peaks of the PIR and the NIR resonances in d*I*/d*V* spectra, see **Fig. 3** and Supplementary **Fig. S34**.

Overall, the measured transport gaps decrease asymptotically with increasing *N*. However, $\Delta_{exp}(C_{20})$ is substantially smaller (by ~0.4 eV) than $\Delta_{exp}(C_{22})$, despite $C_{22}$ being larger than $C_{20}$. Moreover, the mean value of $\Delta_{exp}(C_{44})$ is smaller than the means of $\Delta_{exp}(C_{42})$ and $\Delta_{exp}(C_{46})$. At larger *N*, the differences between the gaps are below our experimental precision: $\Delta_{exp}(C_{66})$ is similar to $\Delta_{exp}(C_{46})$ and $\Delta_{exp}(C_{88})$.

## Theory

Most prior discussions of the electronic structure of cyclocarbons have been based on DFT results, which depend on the employed density functional approximation (DFA). In the case of cyclocarbons and other π-conjugated systems, DFT results are highly sensitive to the proportion of exact exchange (EE). High proportion of EE suppresses electron delocalisation, resulting in weaker aromaticity; however, at least about 35% short-range EE is necessary to reproduce the experimentally observed symmetry breaking in $C_{18}$ [8].

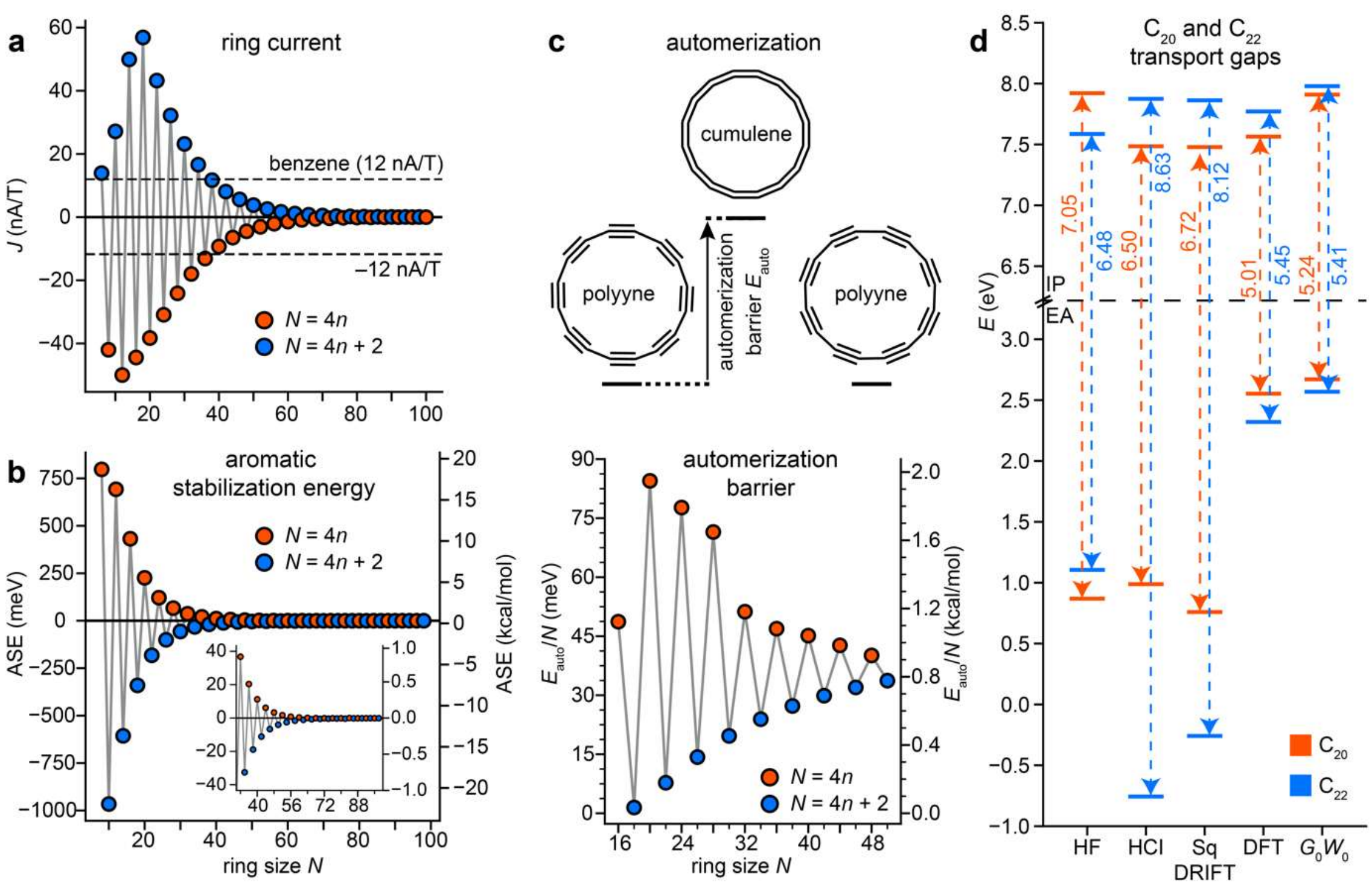

**Figure 4. Theory results for even cyclo[*N*]carbons up to *N* = 100. a**, Ring current. **b**, Aromatic stabilization energy (ASE). **c**, Automerization barrier. Calculations were performed with OX-BLYP30/def2-TZVP, further details in **SI sections 10–15**. **d**, Transport gaps $\Delta_{calc}$ of $C_{20}$ and $C_{22}$. Ionization potentials (IP), electron affinities (EA), and $\Delta_{calc}$ (IP – EA) were computed using Hartree-Fock (HF), heat bath CI (HCI), SqDRIFT executed on quantum hardware (see SI Section 16), OX-BLYP30/def2-TZVP (DFT), and $G_0W_0$@OX-BLYP30/def2-QZVP ($G_0W_0$).

Here, we build a finely-tuned range-separated DFA for cyclocarbons, OX-BLYP30 (Supplementary **Fig. S35**). Following our previous work on annulenes [12], porphyrin nanobelts [14, 15] and the $C_{48}$ catenane [13], we tuned OX-BLYP30 by benchmarking it against wavefunction theory methods (see Supplementary **Figs. S36** and **S37**). OX-BLYP30 has 30% EE in the short range, which is necessary to correctly reproduce the geometric features obtained for $N$ = 10–18 cyclocarbons by experiments [1-4] (see Supplementary **Fig. S38**).

We computed the transport gaps of cyclocarbons using OX-BLYP30 (Supplementary **Figs. S39** and **S40**) and the Green's function-based many-body $G_0W_0$ approach, [37-39] (**Fig. 3b**, see **Fig. S41** for other flavors of $GW$) using OX-BLYP30 as the starting point. We also used OX-BLYP30 to calculate magnetic (ring currents), energetic (aromatic stabilization energy and lowest optical excitation energy), and electronic (automerization barrier) signatures of aromaticity (**Fig. 4**, Supplementary **Figs. S42-S44**).

We find that all these properties, which are rooted in electronic delocalisation, oscillate between 4$n$ and 4$n$+2 cyclocarbons and attenuate asymptotically with increasing ring size. We note that the aromatic stabilisation energies (**Fig. 4b**) oscillate up to smaller $N$ than the transport gaps (**Fig. 3b**), ring currents (**Fig. 4a**), and automerization barriers (**Fig. 4c**). This difference in attenuation can be explained by recalling that the aromatic stabilisation energy is a pure ground-state property of a neutral molecule at equilibrium, whereas transport gaps, automerization barriers, and ring currents involve charged species, transition structures, and couplings with unoccupied states, respectively.

The OX-BLYP30 results suggest that aromaticity will be an important factor in the cyclocarbon electronic structure up to $N$ = 42, supporting our experimental findings. At this size, $C_{40}$ ($N$ = 4$n$, anti-aromatic) and $C_{42}$ ($N$ = 4$n$+2) cyclocarbons sustain paratropic and diatropic ring currents, respectively, with magnitude comparable to that in benzene (12 nA/T) (**Fig. 4a**), and the difference between their aromatic stabilisation energies are on the order of 1 kcal/mol (**Fig. 4b**).

A disadvantage of DFT, and to a lesser extent $G_0W_0$, is that their treatment of electron correlation is approximate. Configuration interaction (CI) treats electron correlation exactly, but it is only tractable in relatively small active spaces, typically correlating up to about 20 electrons and orbitals. To ascertain the origin of the transport gap differences in cyclocarbons, we performed approximate CI calculations to determine transport gaps in $C_{20}$ and $C_{22}$ in active spaces of around 30 π-electrons and orbitals (28 for $C_{20}$ and 32 for $C_{22}$, see SI Section 16). These large active space calculations were done using a semi-stochastic heat bath method (HCI) [40] and a quantum time evolution-based SqDRIFT algorithm [16] executed on quantum hardware (details in SI Section 16, Supplementary **Fig. S45**). Our results show that the uncorrelated Hartree-Fock (HF) method predicts a smaller transport gap for $C_{22}$ than $C_{20}$, as may be expected from size-based considerations [41], while adding π-correlation via HCI or SqDRIFT reverses this trend (**Fig. 4d**), giving qualitative agreement with DFT, $G_0W_0$, and experiment. Note that the EA values obtained with HF, HCI, and SqDRIFT are more negative compared to DFT and $G_0W_0$, due to the use of a small basis set [42].

## Discussion

Our experimental (STS) and computational ($G_0W_0$@OX-BLYP30) results are in qualitative agreement, showing that for $C_N$ with sizes up to $N$ = 42 the transport gaps of $N = 4n$ and corresponding $N = 4n\pm2$ cyclocarbons oscillate by about 5–10%, that is, on the order of a few 100 meV. However, the STS transport gaps measured on monolayer NaCl on Au(111) (**Fig. 3a**) quantitatively differ from the calculated (gas phase) transport gaps (**Fig. 3b**) because of the partial voltage drop across the NaCl layer [43] and the screening by the environment in the STS measurements [36, 44-48]. The latter results in decreased transport gaps on the surface with respect to isolated molecules in vacuum and is challenging to account for.

The larger gaps of (aromatic) $N = 4n\pm2$ cyclocarbons relative to their (anti-aromatic) $N = 4n$ counterparts (i.e. for a given $n$) indicate the coherent delocalisation of electrons over the whole molecular ring and persistent effects of aromaticity. If such delocalisation would be absent, the rings would behave similarly to linear polyynes [20, 21], without any oscillation between $N = 4n\pm2$ and $N = 4n$, and the gap would monotonously decrease with increasing $N$ towards the $N = \infty$ limit. However, both experiment and theory show a substantially smaller gap for $C_{20}$ relative to $C_{22}$, and indicate a smaller gap for $C_{44}$ relative to $C_{42}$ and $C_{46}$ (see **Fig. 3**). In addition, experiment and theory agree that with increasing $N$, the gaps and oscillations between $N = 4n\pm2$ and $N = 4n$ become smaller. The STS results confirm the validity of our theory, which predicts that aromaticity will be experimentally relevant at $N$ = 42.

## Conclusion

By tip-induced chemistry, we synthesized large cyclocarbons $C_N$ with different sizes, up to $C_{88}$. To access their aromaticity, we measured their transport gaps by scanning tunnelling spectroscopy. We observed oscillations of the transport gaps between aromatic $N = 4n+2$ and anti-aromatic $N = 4n$ cyclocarbons up to $N$ = 42, indicating that aromaticity persists for these cyclocarbons. In line, our theory using the cyclocarbon-tailored functional OX-BLYP30 predicts such oscillations of the transport gaps and substantial aromatic stabilisation energies and ring currents at $N$ = 42, supporting our estimate of the ring size at which aromaticity persists in cyclocarbons. Large active space approximate configuration interaction calculations executed on quantum hardware confirm that aromaticity originates from the correlation of π-electrons.

In the reaction pathway of fusing cyclocarbons we observed stable intermediates, amongst others carbon lemniscates, providing insights in the reaction mechanism. The persistence of aromaticity up to large sizes enables future studies where large cyclocarbons and carbon lemniscates could be model systems to study conductance, quantum interference, and the effects of (anti-)aromaticity in single atomic carbon wires and loops. For instance, STM break junction experiments of large cyclocarbons would allow the two leads (tip and sample) to be connected by two atomic wires of a single molecule, ideally suited to study interference effects on the single-molecule scale. Such effects might also be studied using near-term quantum hardware, as current quantum devices surpass routine conventional algorithms and rapidly approach best currently available conventional methods.

**Acknowledgements**: This work was funded by European Research Council grant no. 885606, ARO-MAT (H.L.A. and Y.G.); European Community Horizon 2020, grant project 101019310 CycloCarbonCatenane (Y.G. and H.L.A.); European Research Council Synergy grant no. 951519, MolDAM (L.S., F.P., F.A., J.E., L.-A.L. and L.G.); University of Manchester (I.R. and B.D.); Swiss National Science Foundation (SNSF) Swiss Postdoctoral Fellowships grant no. 233895, ISOTOPE (L.S.). I.R. acknowledges the assistance given by Research IT and the use of the Computational Shared Facility at The University of Manchester. M.V. acknowledges support from Charles University (Czechia) where he is enrolled as a PhD student, and from the Czech Science Foundation, via Grant No. 24-10982S. This work was supported by the Ministry of Education, Youth and Sports of the Czech Republic through the e-INFRA CZ (ID:90254). I.R. and B.D. acknowledge support from the EPSRC Materials for Quantum Network (EP/W037912/1).

**Author contributions**: Synthesis of the precursors: Y.G.; On-surface synthesis and STM and AFM measurements: L.S., F.P., F.A., J.E., L.-A.L., and L.G.; Theoretical analysis and computations: M.V., S.B., B.D., S.P., A.B., I.T., and I.R.; Writing: First draft L.S., F.P., I.R., and L.G. All authors commented on the manuscript and discussed the results.

**Competing interests**: The authors declare that they have no competing interests.

**Data and materials availability:** All experimental data are reported in the main text and supplementary materials.

### Methods

On-surface characterization and reactions were carried out with two home-built combined STM/AFMs, operated at a temperature $T$ = 5 K in ultra-high vacuum. The bias voltage $V$ was applied to the sample with respect to the tip. The Au(111) and Cu(111) single crystal surfaces were prepared by repeated cycles of sputtering with $Ne^+$ ions and annealing up to 820 K and 840 K, respectively. NaCl was evaporated onto the crystal surface with the sample held at 255 K – 330 K, resulting in partial coverage with (100)-terminated monolayer and bilayer NaCl islands. The precursors for $C_{22}$ (compound **1**) and $C_{20}$ (compound **3**), were thermally sublimed from an oxidized silicon wafer onto the cold ($T$ < 10 K) surface. If not stated otherwise all data shown was obtained on monolayer (one atomic layer thick) NaCl films on Au(111) with CO-functionalized tips [30]. AFM measurements were performed in non-contact mode using a qPlus sensor [49] operated in frequency-modulation mode [50]. The oscillation amplitude was kept constant at $A$ = 0.5 Å, if not stated otherwise. STM images were recorded at constant current. AFM images were obtained at $V$ = 0 V, at constant height, with the tip-height offset $\Delta z$ added to the STM controlled setpoint ($V$ = 0.2 V and $I$ = 0.3 pA, if not mentioned otherwise). Positive tip-height offsets $\Delta z$ correspond to an increase in tip-sample distance with respect to the setpoint. $I/V$ spectra were recorded at a fixed tip height with the feedback opened at the respective acquisition positions.

## Supplementary Information for

# On-surface synthesis and aromaticity of large cyclocarbons

**Authors:** Lisanne Sellies[1,#], Marco Vitek[2,#], Yueze Gao[3,#], Fabian Paschke[1,#], Florian Albrecht[1], Jakob Eckrich[1], Beren Dempsey[4], Stefano Barison[1,5], Leonard-Alexander Lieske[1], Samuele Piccinelli[1,6], Alberto Baiardi[1], Ivano Tavernelli[1], Harry L. Anderson[3,*], Igor Rončević[4,*], Leo Gross[1,*]

**Affiliations:**

[1]IBM Research Europe – Zurich, CH-8803 Rüschlikon, Switzerland

[2] Institute of Organic Chemistry and Biochemistry of the Czech Academy of Sciences, Flemingovo nám. 542/2, 160 00 Prague 6, Czechia

[3] Department of Chemistry, Oxford University, Chemistry Research Laboratory, Oxford, UK

[4] Department of Chemistry, University of Manchester, Manchester M13 9PL, UK

[5] Institute of Physics, Eidgenössische Technische Hochschule Zürich (ETHZ), CH-8092 Zurich, Switzerland

[6] Institute of Physics, École Polytechnique Fédérale de Lausanne (EPFL), CH-1015 Lausanne, Switzerland

[#] Equally contributing first authors

* Corresponding authors. Email: igor.roncevic@manchester.ac.uk; harry.anderson@chem.ox.ac.uk; lgr@zurich.ibm.com

## 1. Synthetic general methods:

Reagents were purchased reagent-grade from commercial suppliers and used without further purification. $MgSO_4$ was used as the drying reagent after the aqueous work-up. Petroleum ether was used having a boiling point range of 40–60 °C.

Thin layer chromatography (TLC) was carried out on aluminium-backed silica gel plates with 0.2 mm thick silica gel 60 F254 (Merck) and visualized via UV-light (254/364 nm).

Flash column chromatography was either carried out using flash silica gel 60 (230-400 mesh) obtained from Sigma-Aldrich, or on a Biotage Isolera One with a 200–400 nm UV detector.

$^{1}H$ and $^{13}C$ NMR spectra were recorded on Bruker AVIII HD 400 spectrometers at 400 MHz ($^{1}H$) and 101 MHz ($^{13}C$) and Bruker AVIII HD 500 spectrometers at 500 MHz ($^{1}H$) and 126 MHz ($^{13}C$), respectively, at 298 K. $^{1}H$ and $^{13}C$ NMR chemical shifts are reported in ppm relative to $SiMe_4$ ($\delta$ = 0) and were referenced internally with respect to residual solvent protons using the reported values ($^{1}H$: $CDCl_3$: 7.26 ppm; $^{13}C$: $CDCl_3$: 77.16 ppm). All chemical shifts are reported in ppm, coupling constants are reported in Hz and $^{1}H$ multiplicities are reported in accordance with the following: s = singlet; d = doublet; t = triplet; and m = multiplet.

High-resolution mass spectrometry measurements were carried out by electrospray ionization (ESI) mass spectrometry on a Thermo Scientific Q Exactive Hybrid Quadrupole-Orbitrap mass spectrometer by the mass spectrometry service at the University of Oxford.

UV-vis spectra were recorded in solution on a Perkin-Elmer Lambda 20 spectrometer at 21 °C, in fused silica cuvettes with a path length of 1 cm.

IR spectra were recorded as a thin film on a Bruker Tensor 27 spectrometer equipped with a Diamond ATR sample compartment.

Compounds **S1** (ref. 1), **S2** (ref. 2), and **S3** (ref. 2) were synthesized as previously reported. Compounds **$C_{16}$-ketal** (ref. 2), **$C_{18}$-ketal** (ref. 1), **$C_{20}$-ketal** (ref. 2), **$C_{24}$-ketal** (ref. 2), **$C_{20}(CO)_8$** (ref. 2), and **$C_{24}(CO)_8$** (ref. 2) have been reported in the literatures.

**$C_{16}$-ketal** **$C_{18}$-ketal** **$C_{20}$-ketal** **$C_{24}$-ketal**

**$C_{20}(CO)_8$** **$C_{24}(CO)_8$**

**2. Synthetic protocols:**

**S1** TBAF, THF/$H_2O$, 0 °C → **S2** × 2 + **S3**; CuCl, TMEDA, $CH_2Cl_2$, $O_2$, 25 °C → **$C_{22}$-ketal** (16%)

**$C_{22}$-ketal:** To a solution of compound **S1** (ref. 1) (80 mg, 0.177 mmol) in THF/$H_2O$ (6 mL/0.06 mL) was added tetrabutylammonium fluoride (TBAF, 1.0 M in THF, 390 µL, 0.39 mmol) at 0 °C. The solution was stirred for 1 h at 0 °C, then water (10 mL) and petroleum ether (20 mL) were added, the layers were separated, and the aqueous phase was extracted with petroleum ether (2 × 10 mL). The organic phases were combined and concentrated to ca. 2 mL in *vacuo*. Purification by passing through a silica plug (ethyl acetate/petroleum ether 1:4) resulted in a solution containing compound **S2** (ref. 2), which was then concentrated to ca. 2 mL. This resulting solution was used immediately in the next step, and diluted with $CH_2Cl_2$ (23 mL). Compound **S3** (ref. 2) (30 mg, 0.072 mmol) was then added to the solution. Then the Glaser-Hay catalyst was prepared by adding CuCl (7.1 mg, 0.072 mmol) to a solution of TMEDA (54 µL, 0.358 mmol) in $CH_2Cl_2$ (300 mL), and the solution was stirred for 5 min at 25 °C. To this resulting solution containing Hay-Glaser catalyst was added the solution of compound **S2** and **S3** in $CH_2Cl_2$ (25 mL) over 4 h via a syringe pump. Then the mixture was stirred for another 20 h. $H_2O$ (50 mL) was added, the layers were separated, and the aqueous phase was extracted with $CH_2Cl_2$ (2 × 20 mL). The organic phases were combined, washed with brine (20 mL), dried ($MgSO_4$), and filtered. Solvent removal and purification by column chromatography (silica gel, ethyl acetate/$CH_2Cl_2$/petrol ether 1:5:30 to 1:5:10) afforded compound **$C_{22}$-ketal** (10 mg, 16%) as a yellow-orange solid.
***$R_f$*** = 0.57 (ethyl acetate/petrol ether 1:1). **IR (ATR):** 2981 (s), 2889 (m), 1725 (w), 1252 (m), 1083 (m) $cm^{-1}$. **UV/Vis** ($CHCl_3$) $\lambda_{max}$ (ε): 314 (sh, 16900), 331 (sh, 28600), 353 (53800), 372 (94800), 396 (17100), 407 (sh, 11900), 423 (14500), 434 (20600), 456 (sh, 2770). **$^1$H NMR** (500 MHz, $CDCl_3$) δ 3.65 (s, 12H; $OCH_3$), 3.64 (s, 12H; $OCH_3$), 3.63 (s, 12H; $OCH_3$), 3.61 (s, 12H; $OCH_3$) (see **Fig. S1**). **$^{13}$C NMR** (125 MHz, $CDCl_3$) δ 138.1, 137.2, 135.6, 133.2, 109.4, 109.3, 109.24, 109.19, 95.2, 88.5, 88.2, 87.8, 79.4, 79.2, 78.4, 52.5, 52.40, 52.39, 52.2 (see **Fig. S2**). **ESI HRMS *m/z*:** calculated for $C_{46}H_{48}O_{16}Na^+$ ([M + Na]$^+$) 879.2830, found 879.2835.

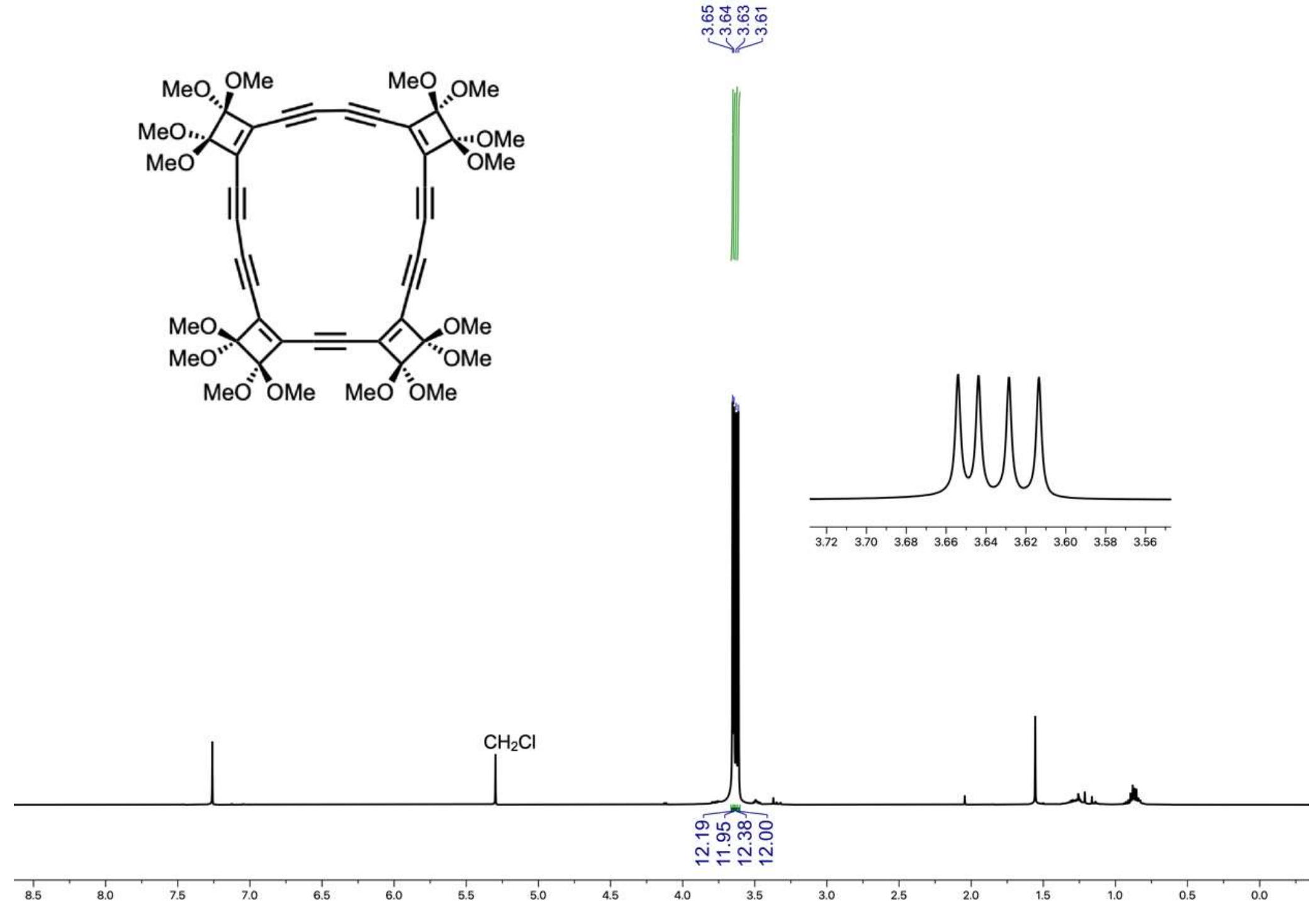

**Figure S1.** $^{1}$H NMR (500 MHz) spectrum of compound **$C_{22}$-ketal** in $CDCl_3$.

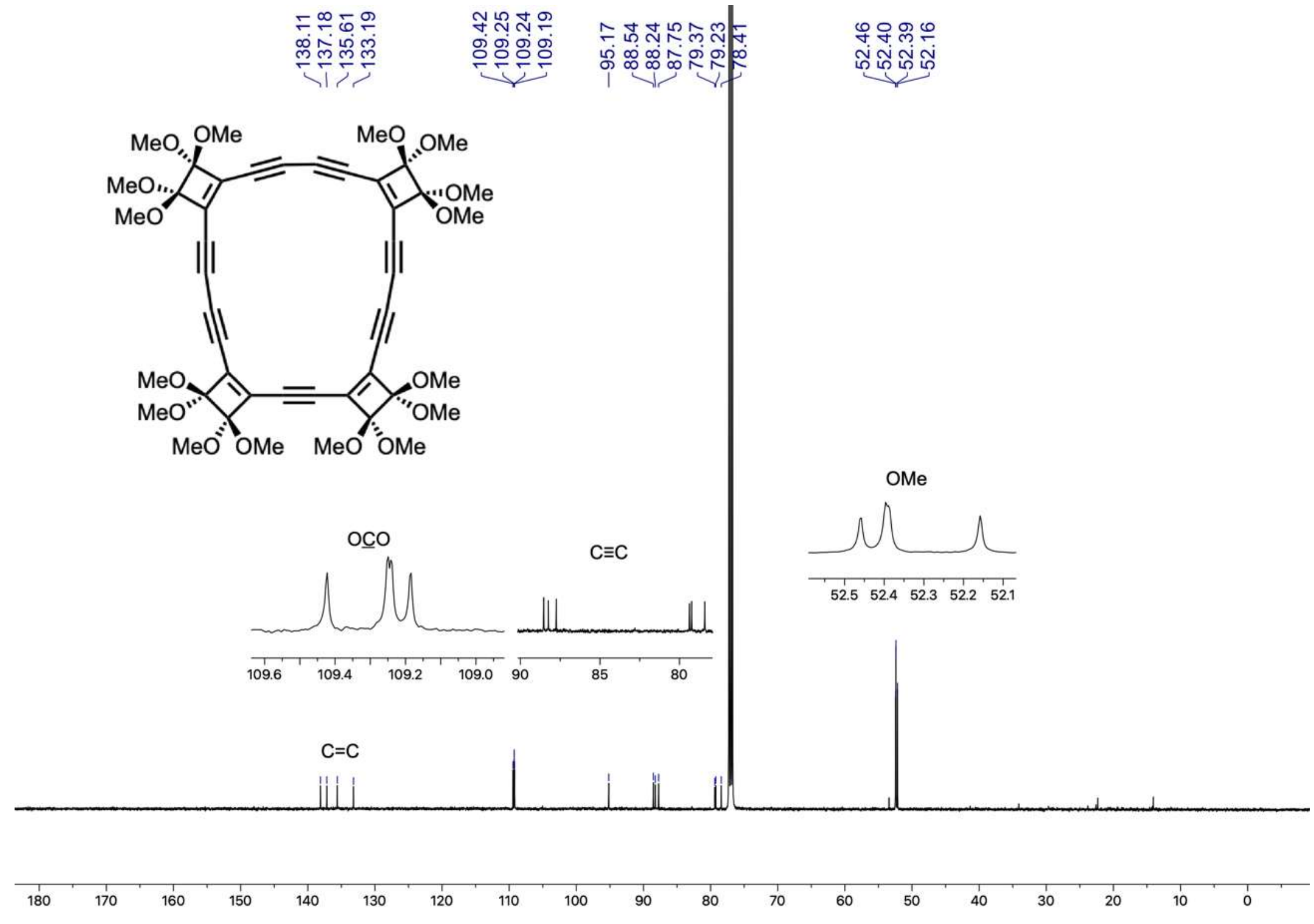

**Figure S2.** $^{13}$C NMR (125 MHz) spectrum of compound **$C_{22}$-ketal** in $CDCl_3$.

**Compound $C_{22}(CO)_8$**: Trifluoroacetic acid (TFA, 1.00 mL) and $H_2O$ (20.0 µL, 1.12 mmol) were added to **$C_{22}$-ketal** (6 mg, 0.007 mmol) in a vial. The solution was stirred for 3.5 h at 21 °C while wrapped in aluminium foil to avoid light. TFA was removed by evaporation under a flow of $N_2$ gas and then under high vacuum to afford **$C_{22}(CO)_8$** (3.3 mg, 97%) as a grey-red solid. Since the solid form of **$C_{22}(CO)_8$** was light sensitive and slowly decomposed, the compound was stored in a diluted $CHCl_3$ solution shielding from light at –20 °C. **UV/Vis** ($CHCl_3$) $\lambda_{max}$ (minimum values of ε): 320 (32000), 352 (33100), 373 (45600), 403 (72200), 465 (16100). **$^{13}$C NMR** (125 MHz, $CDCl_3$) δ 190.64, 190.61, 190.59, 190.0, 179.7, 178.2, 175.8, 173.0, 107.2, 107.0, 106.8, 105.9, 82.1, 81.69, 81.67 (see **Fig. S3**).

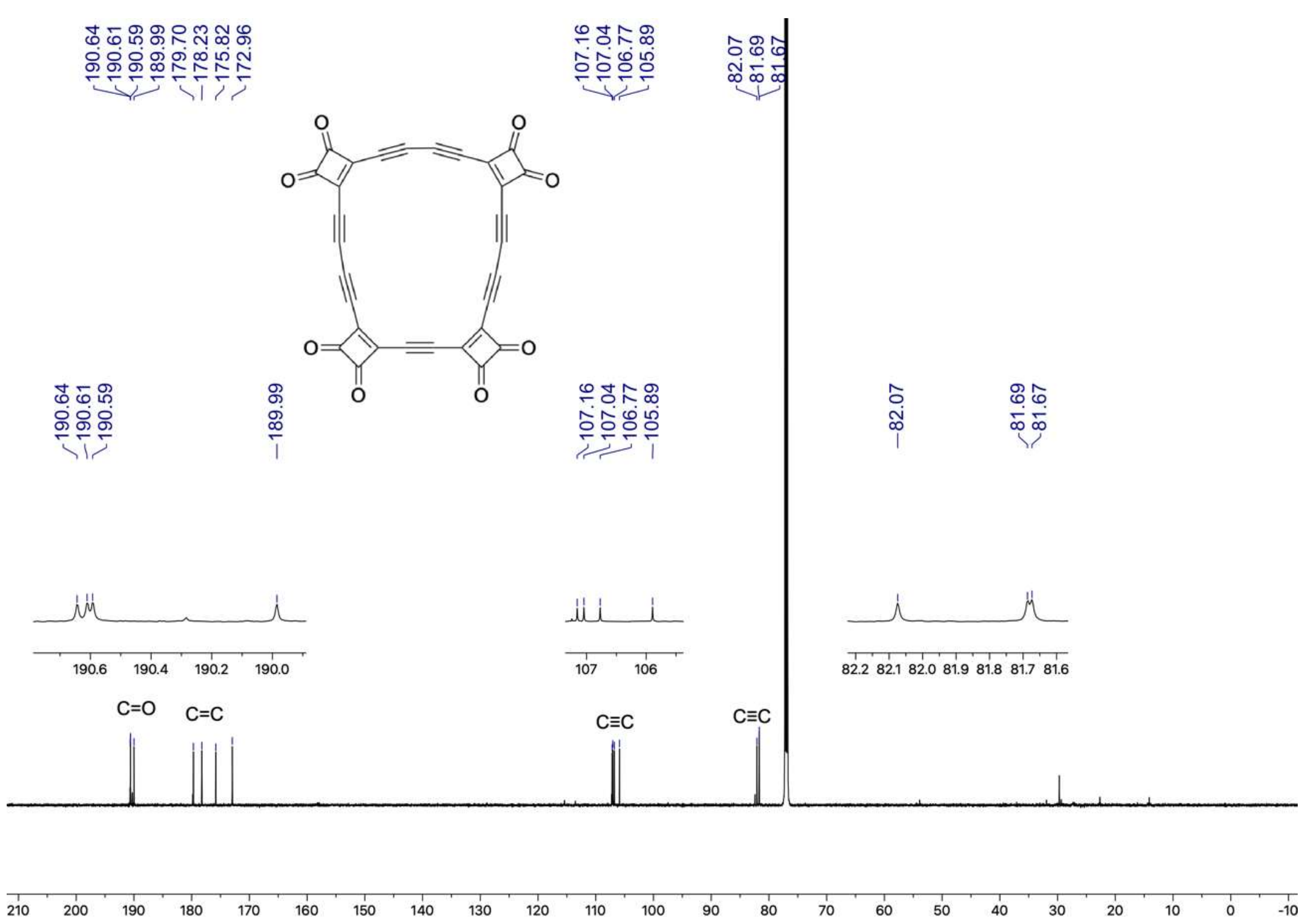

**Figure S3.** $^{13}$C NMR (151 MHz) spectrum of compound **$C_{22}(CO)_8$** in $CDCl_3$.

## 3. NMR comparison:

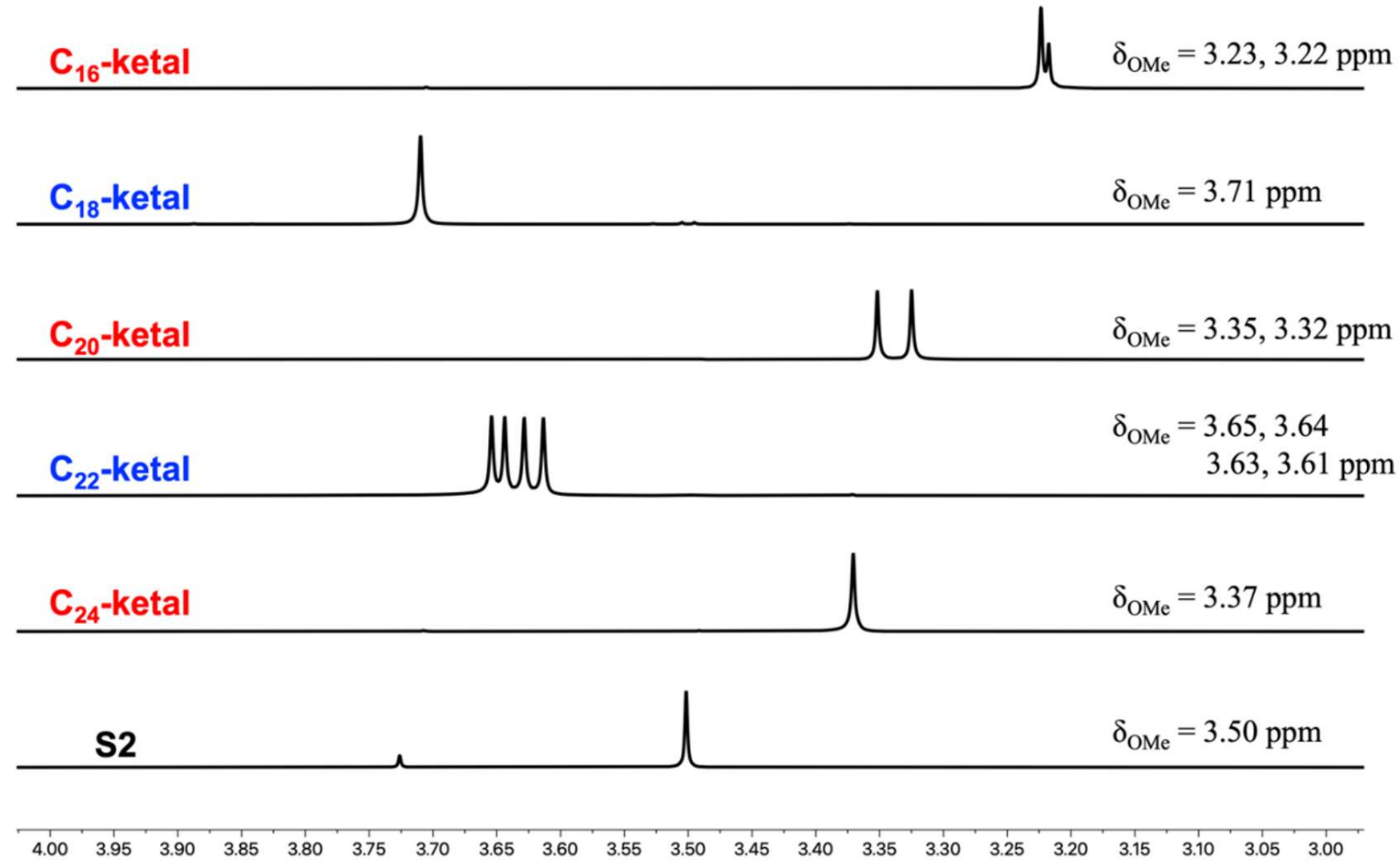

**Figure S4.** Partial $^{1}$H NMR spectra of even **C$_{N}$-ketals** ($N$ = 16 – 24) and **S2** (400 MHz, $CDCl_3$) showing evidence for an anti-aromatic ring current in compound **C$_{20}$-ketal** and an aromatic ring current in **C$_{22}$-ketal**.

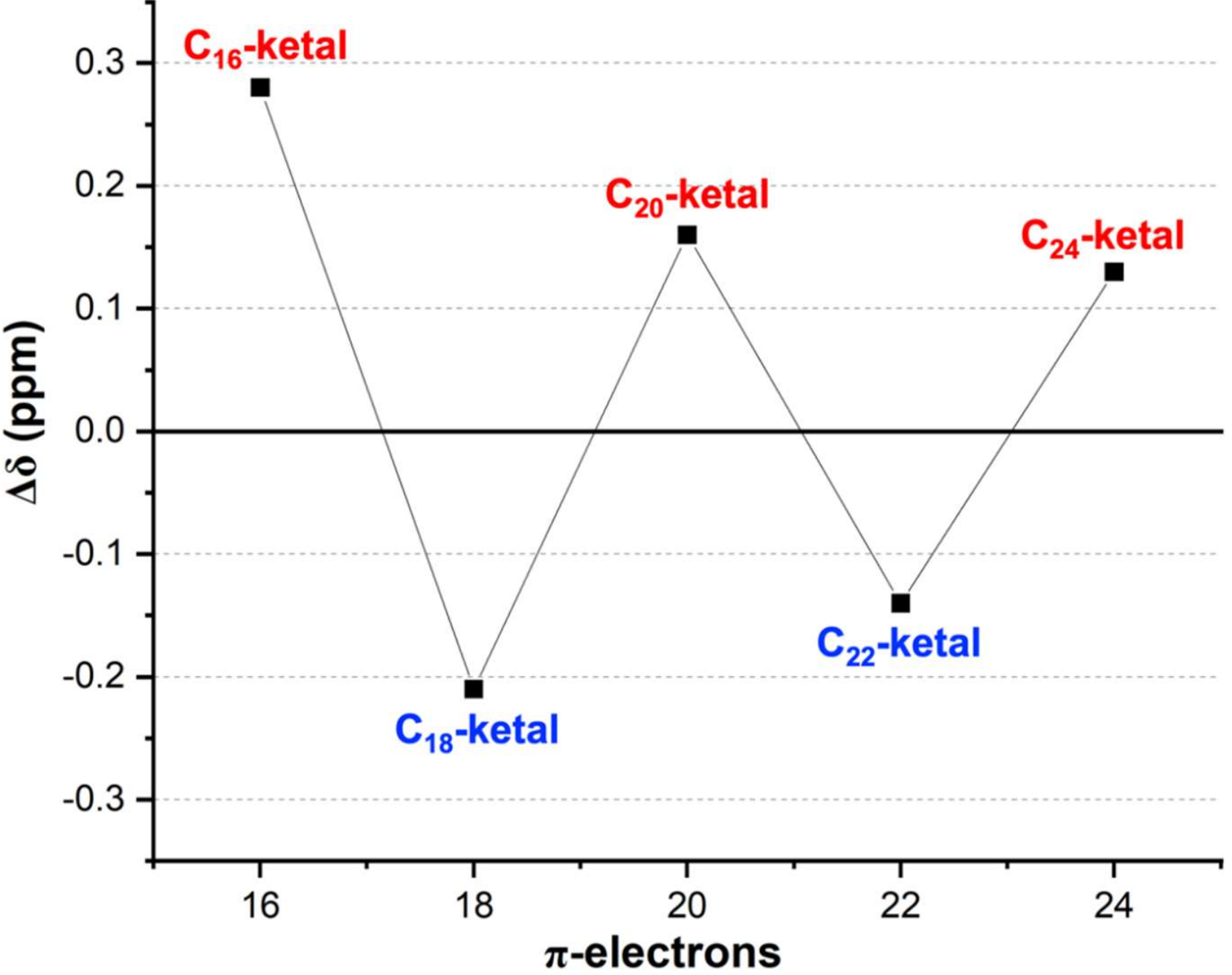

**Figure S5.** Summary of the shielding and deshielding of the methoxy groups across the five different rings, relative to compound **S2**. A similar alternation between aromatic and antiaromatic shielding effects in dehydroannulenes was reported in ref. 2 and ref. 3.

## 4. Stability tests:

The thermal decomposition of compound $C_{22}(CO)_8$ at 25 °C was monitored by UV-vis spectroscopy, as measured in $CHCl_3$ (**Fig. S6b**). To perform these experiments, concentrated solutions of $C_{22}(CO)_8$ in $CHCl_3$ were prepared, such that diluting 100 µL of the concentrated solution into 2.5 mL of $CHCl_3$ in a 10-mm pathlength cuvette gave a UV-vis absorption spectrum with an optical density of ca. 0.8 at $\lambda_{max}$. Solid-state samples were prepared, by quickly evaporating 100 µL of the concentrated solution in each vial under a flow of $N_2$. These samples were wrapped with aluminium foil to avoid light exposure. After keeping the solid-state sample in the dark for a certain time, $CHCl_3$ (2.5 mL) was added, the solution was placed in a cuvette, and the UV-vis spectrum was recorded. The results show that $C_{22}(CO)_8$ decomposes with a half-life of about 9 minutes in the solid state at room temperature (**Fig. S6c**). However, it is much more stable at low temperature and can be stored at –80 °C.

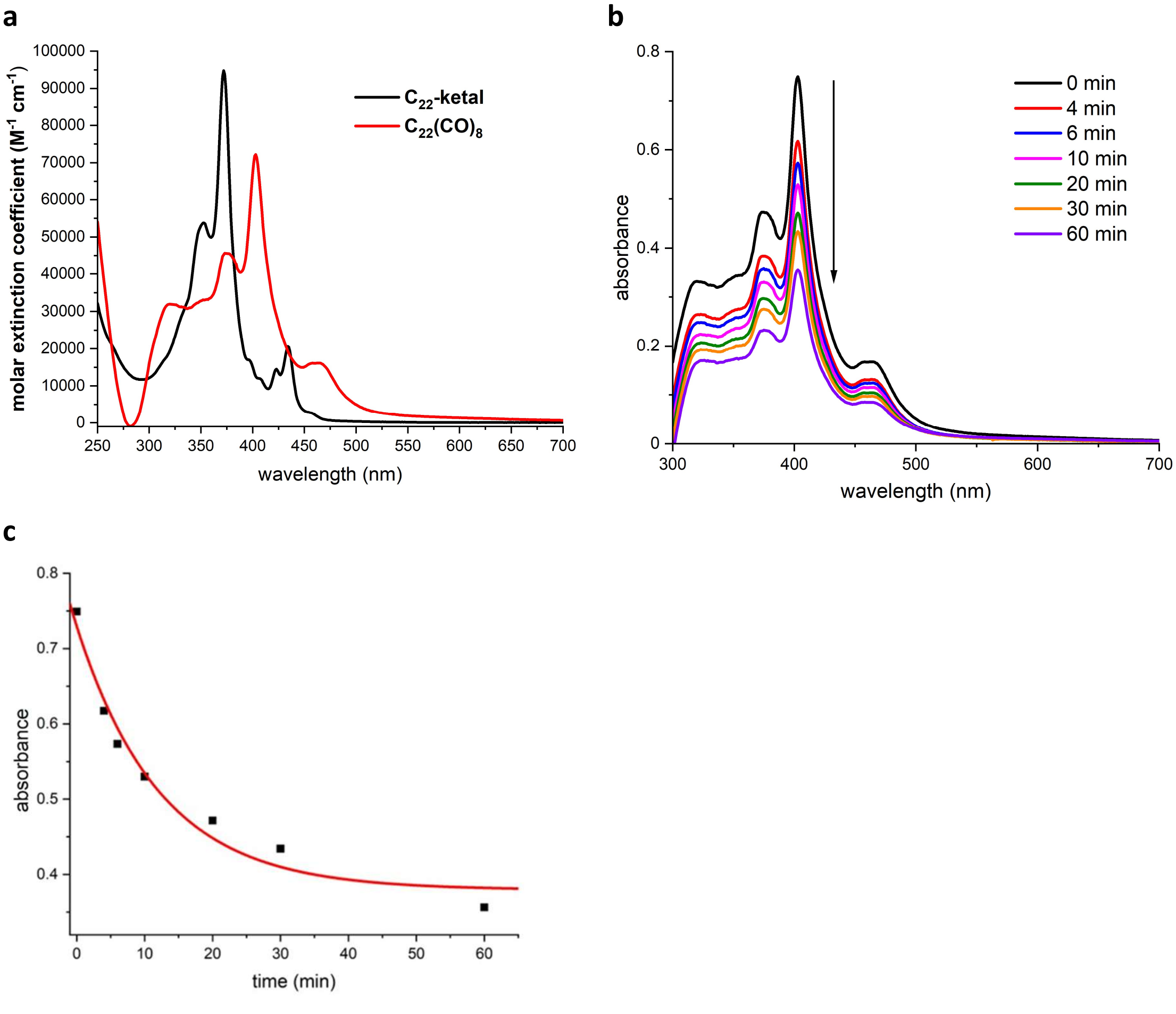

**Figure S6.** (**a**) Quantitative UV-vis spectra of **$C_{22}$-ketal** and $C_{22}(CO)_8$. (**b**) Stability test for $C_{22}(CO)_8$ as a solid state as monitored by UV-vis spectroscopy in $CHCl_3$ at 25 °C shielded from light. (**c**) Plot of UV-vis absorbance decay of $C_{22}(CO)_8$ extracted from **Fig. S3b**, monitored at $\lambda_{max}$ = 403 nm. UV-vis data were fitted to a first-order decay model, $A(t) = a*\exp(-kt)+b$, with $a$ = 0.35; $b$ = 0.38; $k$ = 0.08 min$^{-1}$, giving $t_{1/2}$ = 9 min.

## 5. $C_{24}$ by-product:

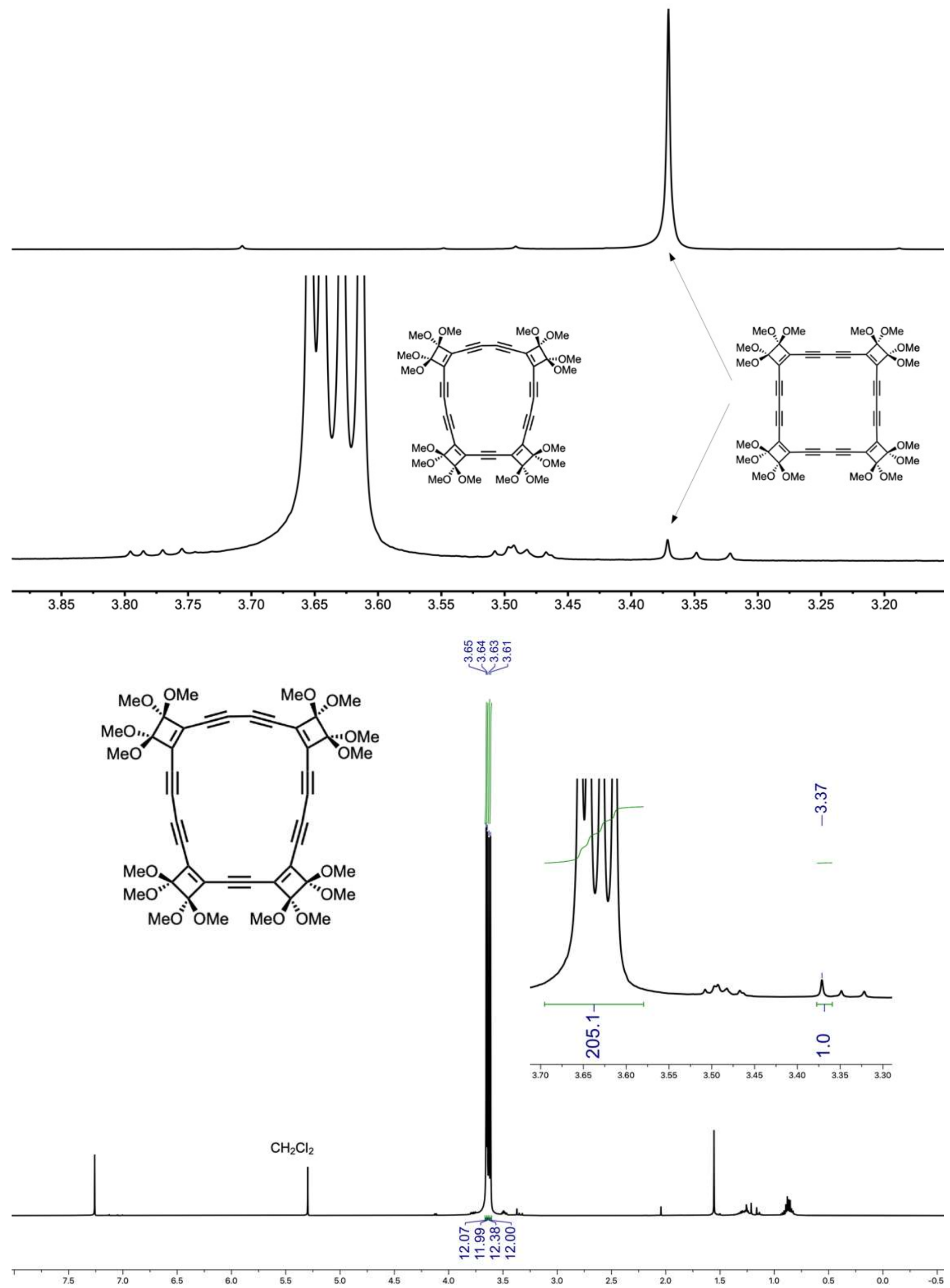

**Figure S7.** $^{1}$H NMR (500 MHz) spectra of **$C_{22}$-ketal** showing the presence of **$C_{24}$-ketal** by-product in $CDCl_3$. The top trace shows the $^{1}$H NMR (500 MHz) spectrum of the **$C_{24}$-ketal**.[2]

During the synthesis of **$C_{22}$-ketal** product, a small amount of the **$C_{24}$-ketal** by-product is also formed. Most of this **$C_{24}$-ketal** impurity can be removed by silica gel chromatography, though approximately 0.5–1.0% typically remains, as indicated by the integrals in **Fig. S7**.

## 6. Generation of large cyclocarbons:

We formed 172 individual $C_{22}$ molecules. We determined the yield for the conversion of precursor **1** into $C_{22}$ from 45 precursor molecules **1** on the NaCl surface, of which we successfully converted 33 into $C_{22}$, corresponding to a yield of 73%. We formed 15 individual $C_{20}$ molecules from precursor molecules **3**. Note that some precursors were found partly demasked on the surface after the preparation; likely such demasking took place during the thermal sublimation from the Si wafer onto the surface.

In total, by fusing $C_{22}$, we formed 19 individual $C_{44}$, seven $C_{66}$ and two $C_{88}$. By fusing a $C_{22}$ with a $C_{24}$ (generated from **2,** a rare side product of **1**), we formed one $C_{46}$. By fusing $C_{20}$ with $C_{22}$, we formed two $C_{42}$.

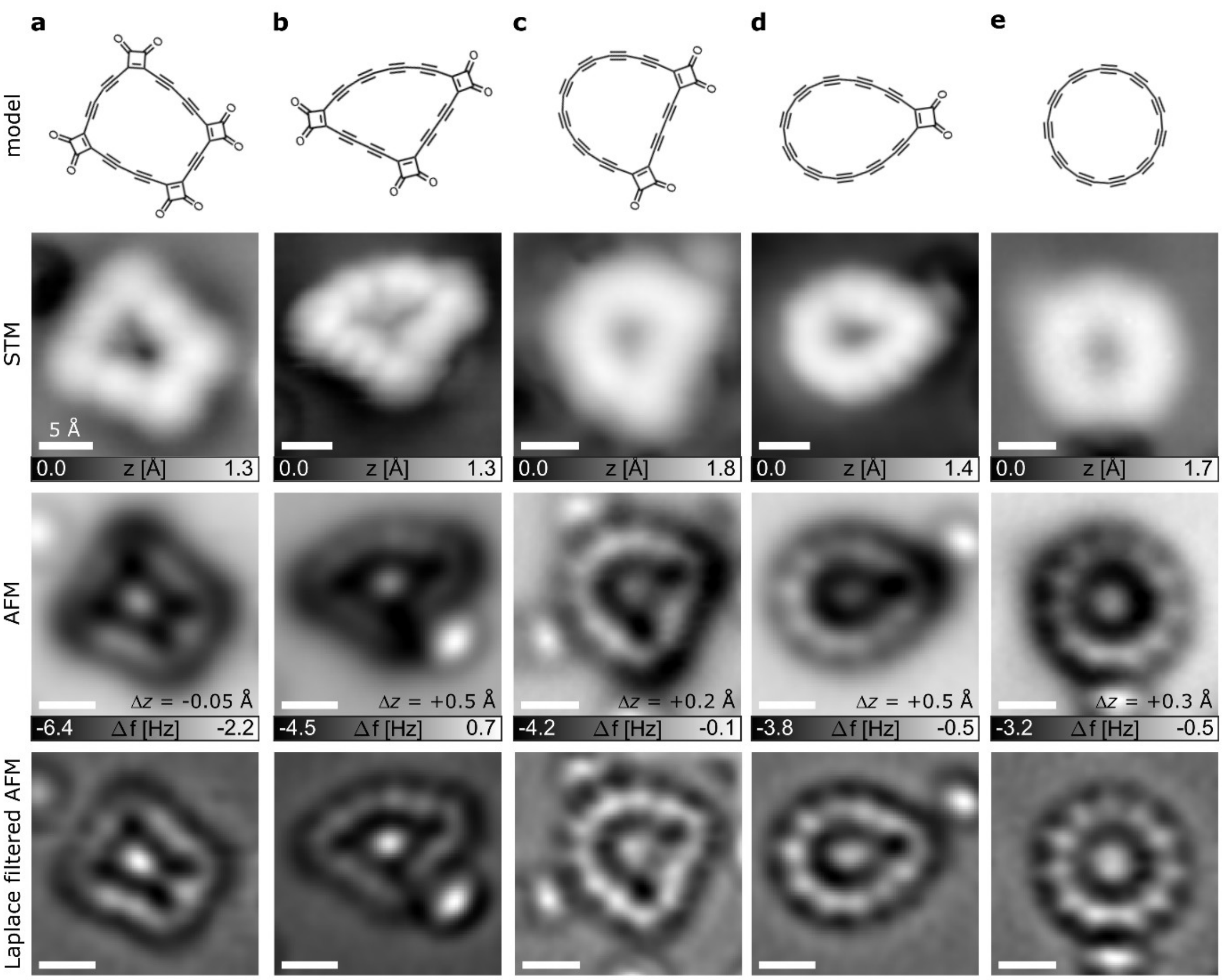

**Figure S8. On-surface synthesis of $C_{22}$.** Chemical structures (1st row), STM data (2nd row), low-pass filtered AFM data (3rd row) and Laplace-filtered AFM data (4th row). (**a**) Precursor **1** of $C_{22}$, (**b-d**), Intermediates observed after applying voltage pulses on **1**, (**e**) cyclocarbon $C_{22}$. All molecules are adsorbed on bilayer NaCl on Cu(111). STM parameters, $V$ = 0.2 V, $I$ = 0.5 pA. AFM parameters, the indicated $\Delta z$ are the tip-height offsets with respect to the STM setpoint: $V$ = 0.2 V, $I$ = 0.5 pA. Scalebars, 5 Å.

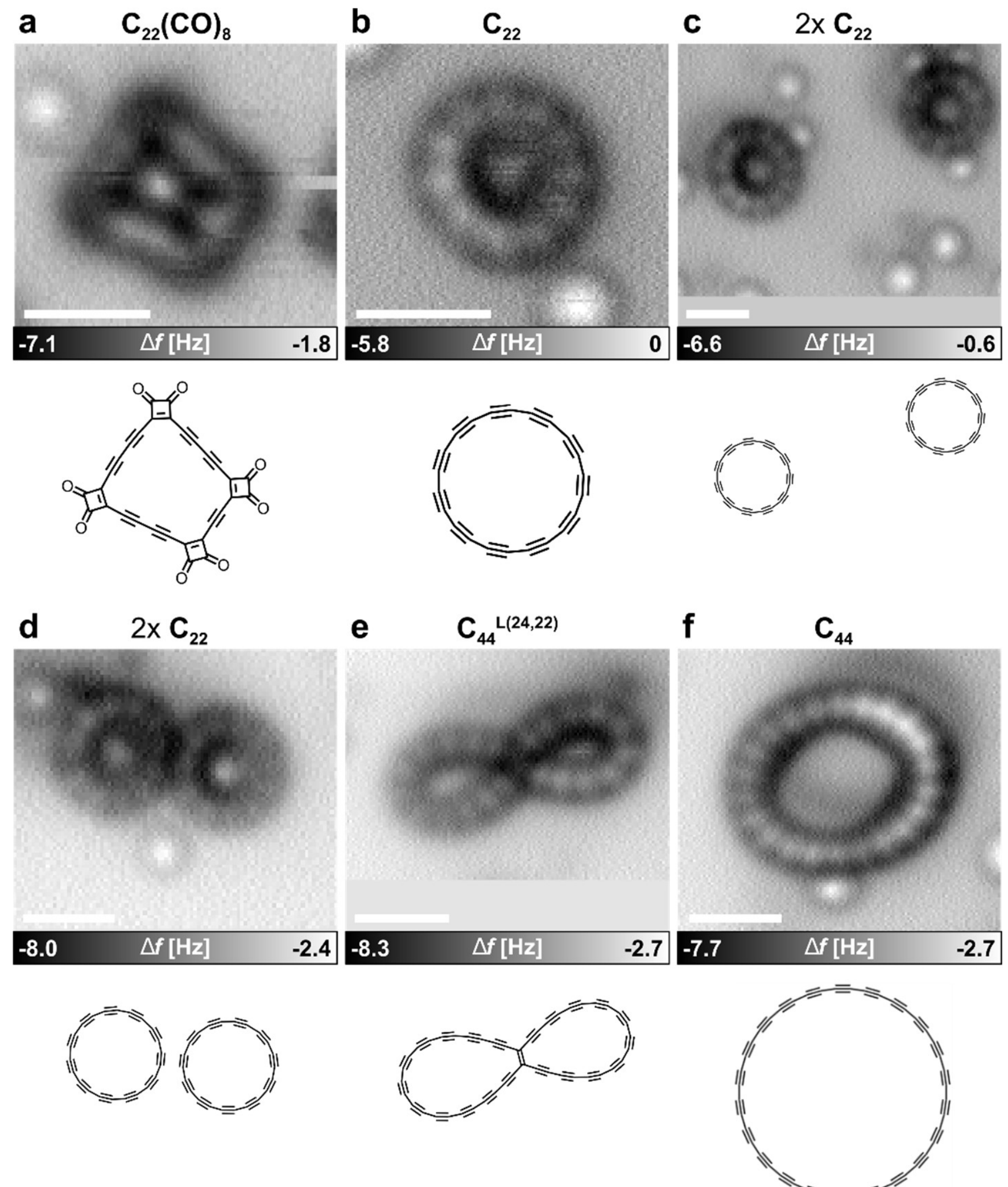

**Figure S9. On-surface synthesis of $C_{22}$ and $C_{44}$, AFM raw data.** AFM raw data (upper panels) of the Laplace-filtered data shown in **Fig. 2a-c**, **e**, **h** and **i** of the main text, respectively. Corresponding chemical structures (bottom panels). (**a**) was measured on bilayer NaCl on Cu(111), while (**b-f**) were measured on monolayer NaCl on Au(111). CO molecules appear as round bright features. AFM parameters: $A$ = 0.5 Å (**a**, **c-f**) or $A$ = 0.7 Å (**b**), tip-height offsets $\Delta z$ with respect to the STM setpoint: $V$ = 0.2 V, $I$ = 0.5 pA: (**a**) $\Delta z$ = -0.05 Å; with respect to the STM setpoint of $V$ = 0.2 V, $I$ = 0.8 pA: (**b**) $\Delta z$ = -0.9 Å; and with respect to an STM setpoint of $V$ = 0.2 V, $I$ = 0.3 pA: (**c**) $\Delta z$ = -1.2 Å, (**d**) $\Delta z$ = -1.2 Å, (**e**) $\Delta z$ = -1.2 Å, (**f**) $\Delta z$ = -1.4 Å. Scalebars, 1 nm.

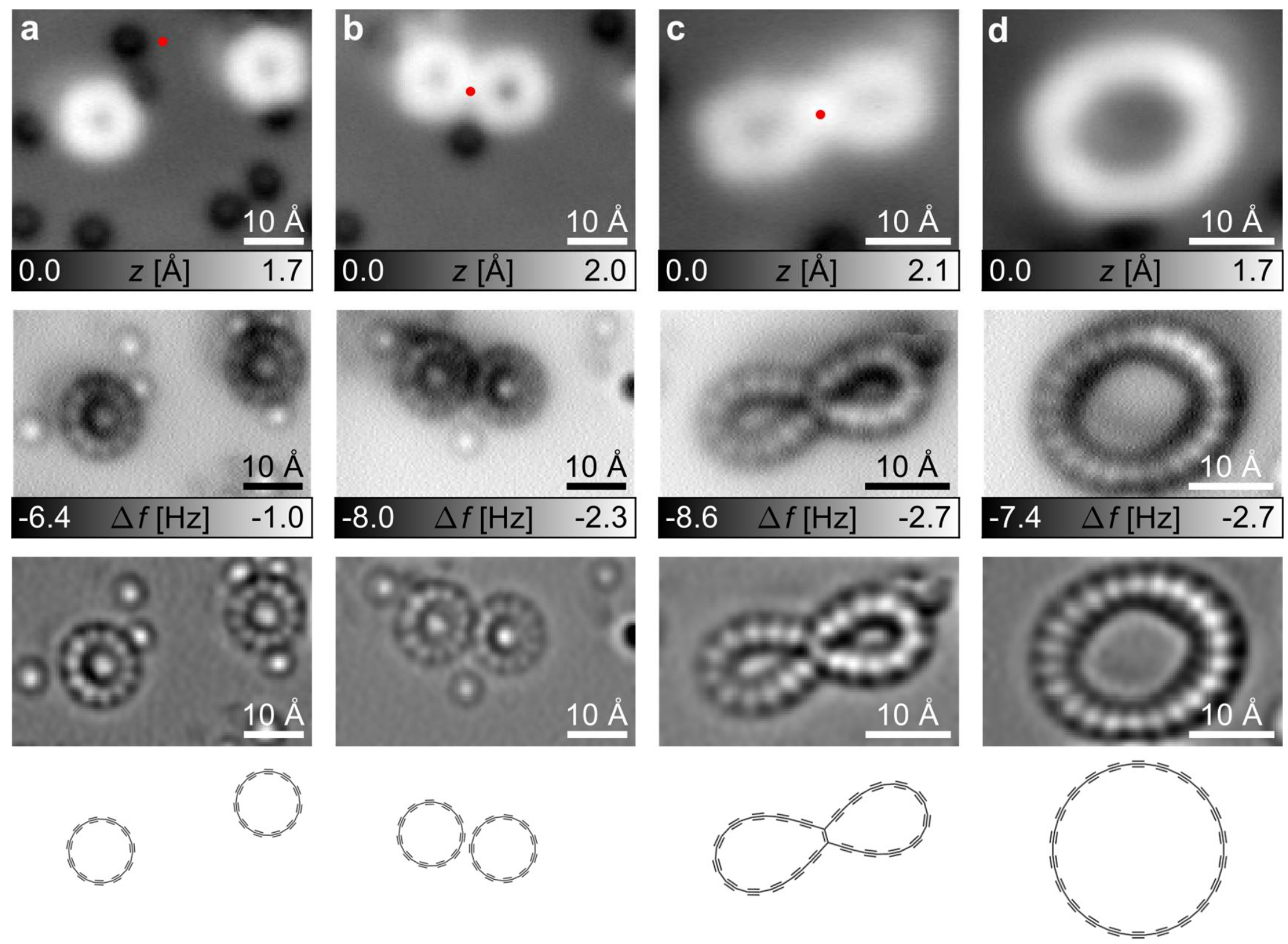

**Figure S10. On-surface synthesis of $C_{44}$, STM and AFM data**. STM data (1st row), AFM raw data (2nd row), Laplace-filtered AFM data (3rd row) and chemical structures (4th row). (**a**) Two $C_{22}$ about 5 nm apart, (**b**) two adjacent $C_{22}$, (**c**) lemniscate $C_{44}^{L(24,22)}$, (**d**) cyclocarbon $C_{44}$. Red dots mark the lateral position of the tip, at which pulses were applied. The tip was retracted from the STM setpoint: $V$ = 0.2 V, $I$ = 0.3 pA by the tip-height offset $\Delta z$ and the voltage $V_p$ was applied for 0.1 s. Parameters for the pulse indicated in (**a**), that is, the pulse that was applied between recording images (**a**) and (**b**) to laterally move the molecules: $V_p$ = 6 V, $\Delta z$ = 17 Å. Pulse indicated in (**b**) to fuse the molecules: $V_p$ = 4.5 V, $\Delta z$ = 8 Å. Pulse indicated in (**c**) to break the C=C bond of the lemniscate: $V_p$ = 4.5 V, $\Delta z$ = 8 Å. STM parameters for all images: $V$ = 0.2 V, $I$ = 0.3 pA. AFM parameters: tip-height-offsets (**a**, **b**) $\Delta z$ = -1.2 Å and (**c**, **d**) $\Delta z$ = -1.4 Å.

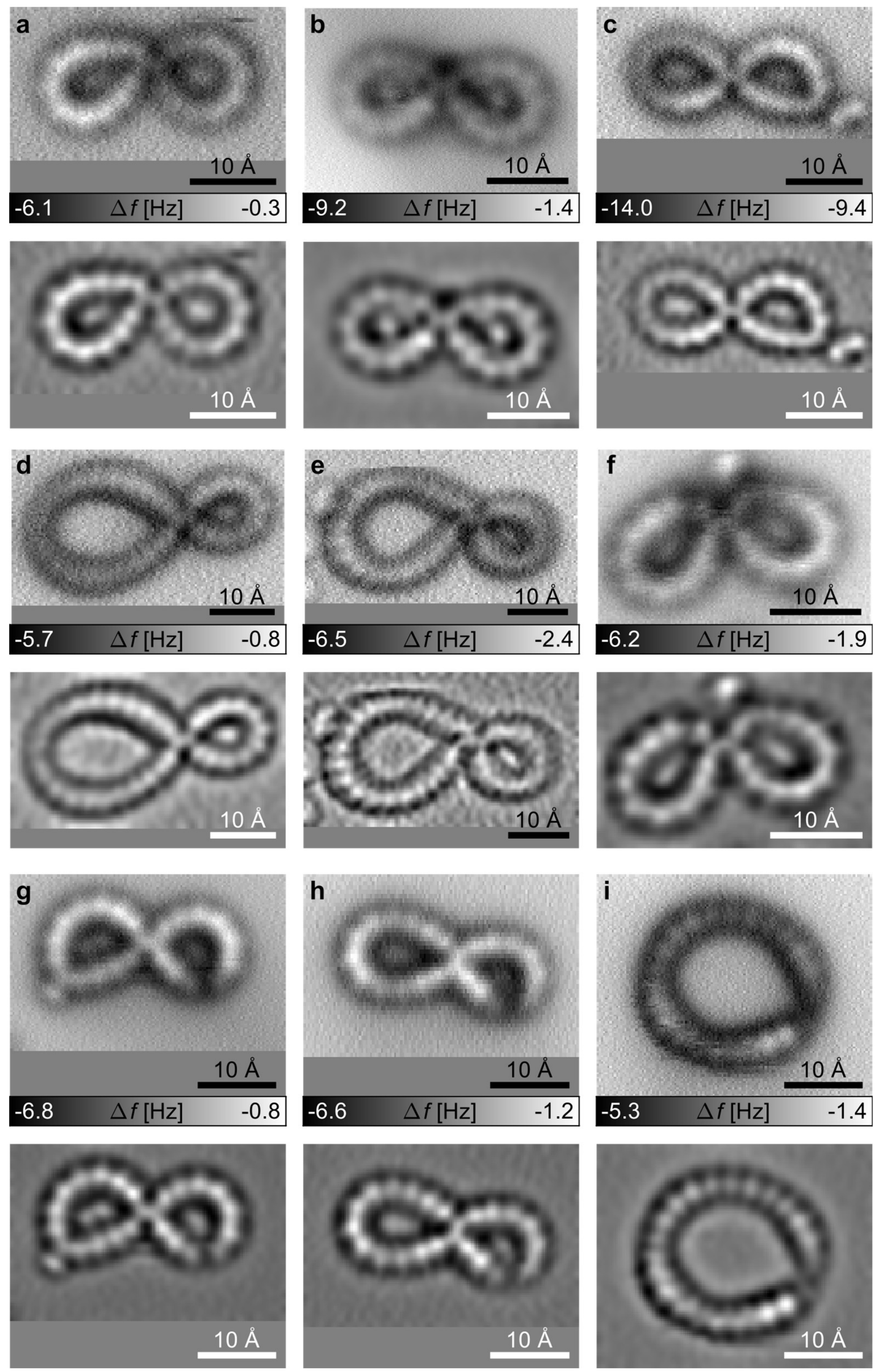

**Figure S11. Carbon rings connected by a single bond.** AFM raw data in top panels, Laplace-filtered AFM data below. (**a-c**, **f**) Structures featuring a single C-C bond between carbon rings observed after fusing two $C_{22}$. (**d**, **e**) Structure observed after fusing a $C_{22}$ and a $C_{44}$. (**g**) Structure formed from two $C_{22}$ precursors. The left carbon ring is connected to an (unknown) side group. The right ring, which exhibits a dark feature within the ring, likely includes an (unknown) defect. The two rings are connected by a C-C bond. (**h**) Structure formed after applying a voltage pulse above the molecule shown in (**g**) with $V_P$ = 6.0 V, Δ*z* = 12 Å from the STM setpoint *V* = 0.2 V, *I* = 0.3 pA . As a result, the (unknown) side group was dissociated from the left ring, and, instead of being connected by a single C-C bond, the rings shared a C=C bond after the pulse forming a lemniscate. (**i**) Structure after scanning with *V* = 5.0 V, *I* = 0.3 pA above the molecule shown in (**h**), resulting in ring opening. The unknown defect at the bottom right of the ring remained. Parameters: (**a**, **c-f**) were measured on monolayer NaCl on Au(111), while (**b**, **g-h**) were measured on bilayer NaCl on Au(111). AFM parameters: *A* = 0.5 Å (**b**, **f-i**) or *A* = 1.0 Å (**a**, **c-e**), tip-height offsets Δ*z* with respect to the STM setpoint: *V* = 0.2 V, *I* = 0.3 pA: (**a**) Δ*z* = -1.4 Å, (**b**) Δ*z* = 0.1 Å, (**c**) Δ*z* = -1.4 Å, (**d**) Δ*z* = -1.2 Å, (**e**) Δ*z* = -1.1 Å, (**f**) Δ*z* = -1.2 Å, (**g**) Δ*z* = -0.1 Å, (**h**) Δ*z* = 0.0 Å, (**i**) Δ*z* = 0.4 Å.

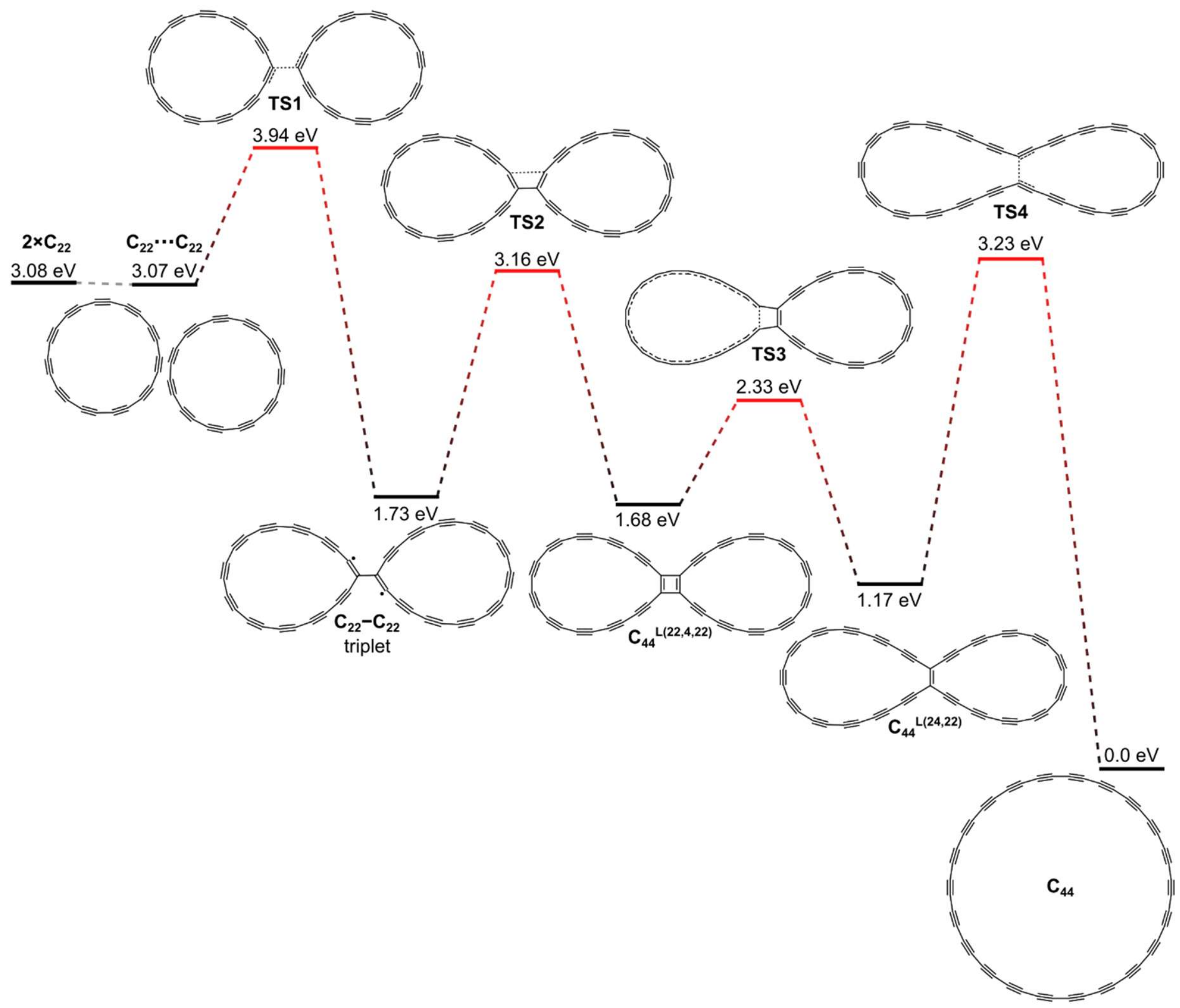

**Figure S12.** Reaction-energy profile for the formation of $C_{44}$ from two $C_{22}$, showing relative energies of intermediates and transition states and their chemical structures, computed at the OX-BLYP30/def2-TZVP level of theory, including zero-point vibrational energy (see **section 11** for details).

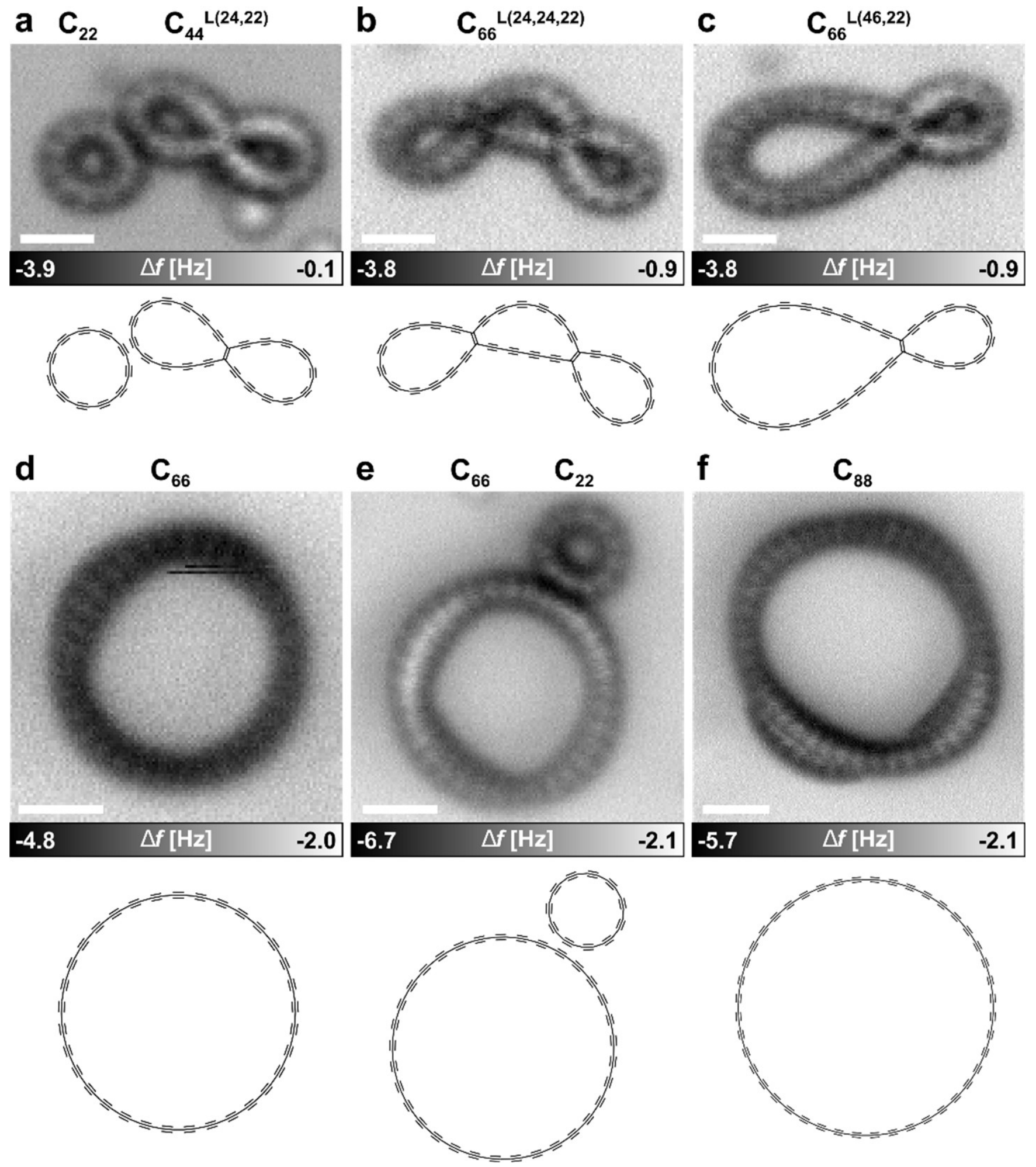

**Figure S13. On-surface synthesis of $C_{66}$ and $C_{88}$, AFM raw data.** AFM raw data (upper panels) of the Laplace filtered data shown in **Fig. 2** of the main text. Corresponding chemical structures (bottom panels). Tip-height offsets $\Delta z$ with respect to an STM setpoint of $V$ = 0.2 V, $I$ = 0.3 pA: (**a**) $\Delta z$ = -1.0 Å, (**b**) $\Delta z$ = -0.8 Å, (**c**) $\Delta z$ = -0.8 Å, (**d**) $\Delta z$ = -0.7 Å, (**e**) $\Delta z$ = -1.0 Å, (**f**) $\Delta z$ = -1.2 Å. Scalebars, 1 nm.

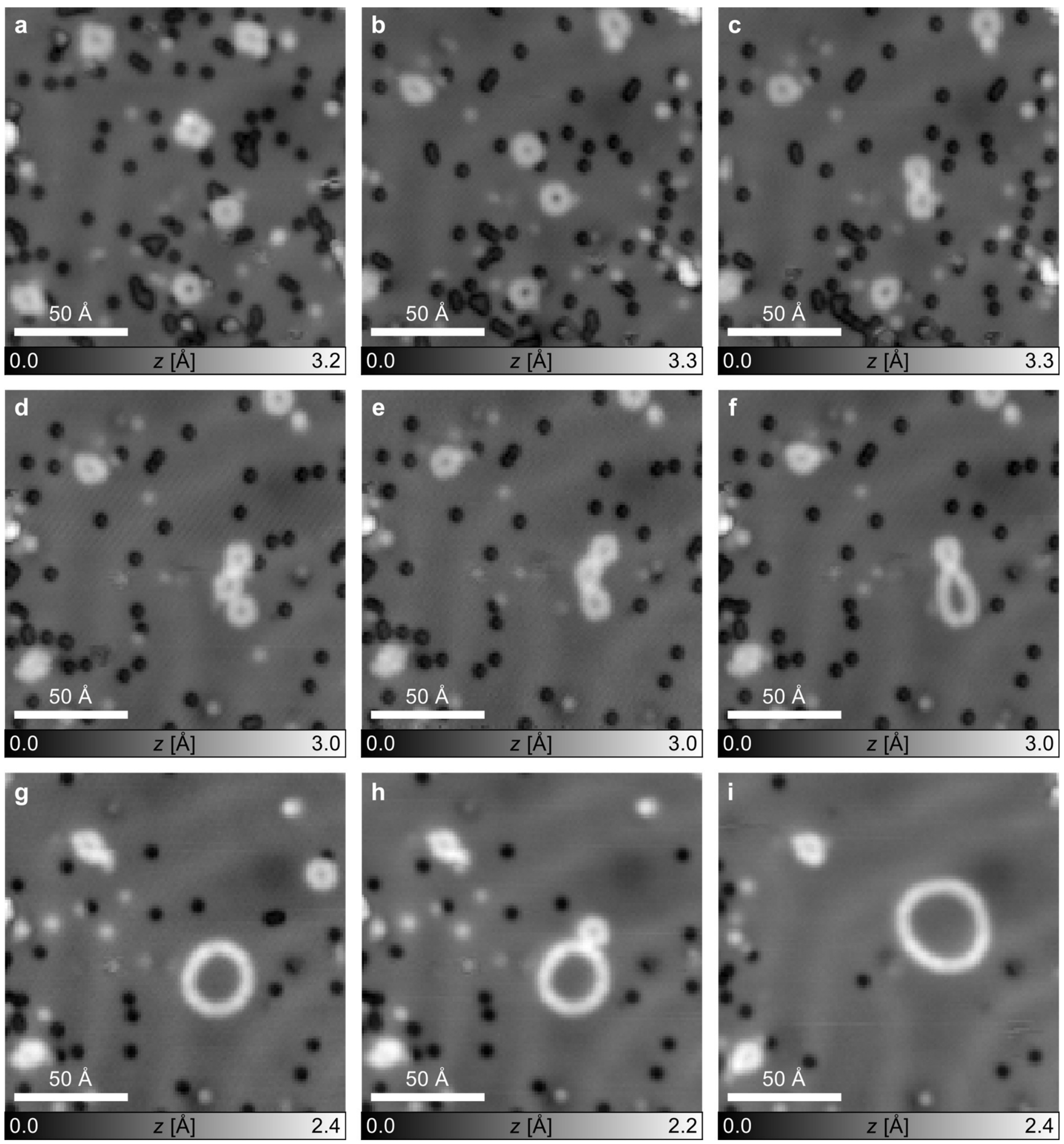

**Figure S14. Synthesis of $C_{88}$, sequence of STM overview images.** (**a-i**) Sequence of STM images acquired during the tip-induced synthesis of $C_{88}$ from four $C_{22}$ precursors **1** on monolayer NaCl. Pulses were applied to induce chemical reactions. To that end, the tip was retracted from the STM setpoint ($V$ = 0.2 V, $I$ = 0.3 pA) by the tip-height offset $\Delta z$ and the voltage $V_p$ was applied for about 0.1 s. Parameters for the pulses: (**a**) that is, the pulses that were applied between recording images (**a**) and (**b**) to demask precursors **1**, $V_p$ = 4.5 V, $\Delta z$ = 13 Å, 12 Å and 11 Å; (**b**) to move $C_{22}$ and fuse two $C_{22}$ to a $C_{44}$ lemniscate, $V_p$ = 6 V, $\Delta z$ = 17 Å; (**c**) to move a $C_{22}$, $V_p$ = 6 V, $\Delta z$ = 17 Å; (**d**) to fuse a $C_{22}$ and a $C_{44}$ lemniscate, $V_p$ = 4.5 V, $\Delta z$ = 8.5 Å; (**e**) to break one of the two C=C bonds in the $C_{66}$ lemniscate, $V_p$ = 4.5 V, $\Delta z$ = 8 Å; (**f**) to break the other C=C bond in the $C_{66}$ lemniscate, $V_p$ = 4.7 V, $\Delta z$ = 10 Å; (**g**) to move a $C_{22}$, $V_p$ = 6.5 V, $\Delta z$ = 19 Å; (**h**) to fuse a $C_{22}$ and a $C_{66}$ to cyclocarbon $C_{88}$, $V_p$ = 6.5 V, $\Delta z$ = 12 Å. STM parameters, $V$ = 0.2 V, $I$ = 0.3 pA. Constant-height AFM data corresponding to panels (**d-i**) are shown in **Figs. 2** and **S13**. Note that the voltage pulses also led to desorption of CO molecules, which are imaged as small circular depressions by STM. The Au(111) herringbone reconstruction is faintly observed through the NaCl monolayer.

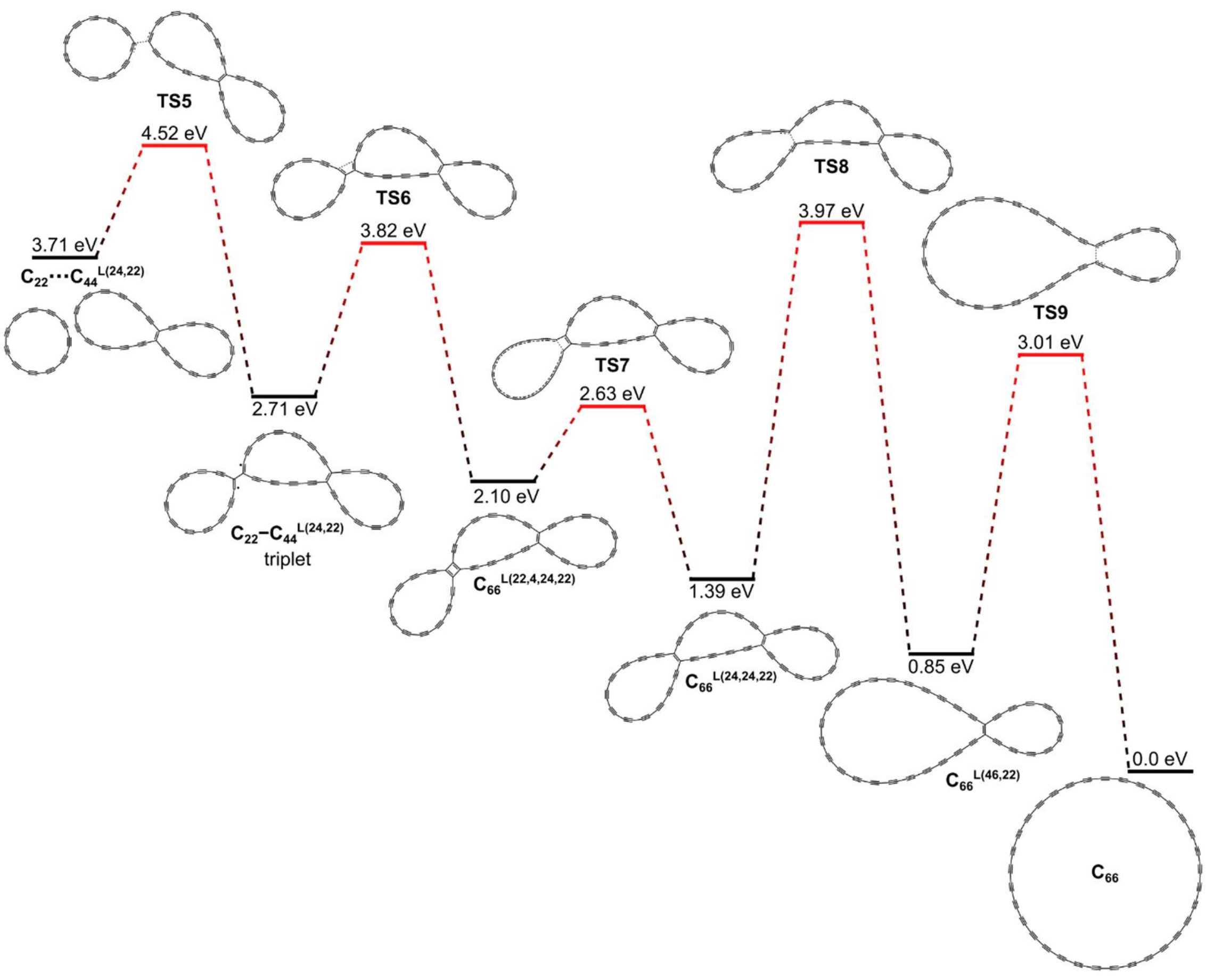

**Figure S15.** Reaction-energy profile for the formation of $C_{66}$ from $C_{22}$ and $C_{44}^{L(24,22)}$, showing relative energies of intermediates and transition states and their chemical structures, computed at the OX-BLYP30/def2-TZVP level of theory, including zero-point vibrational energy (see **section 11** for details).

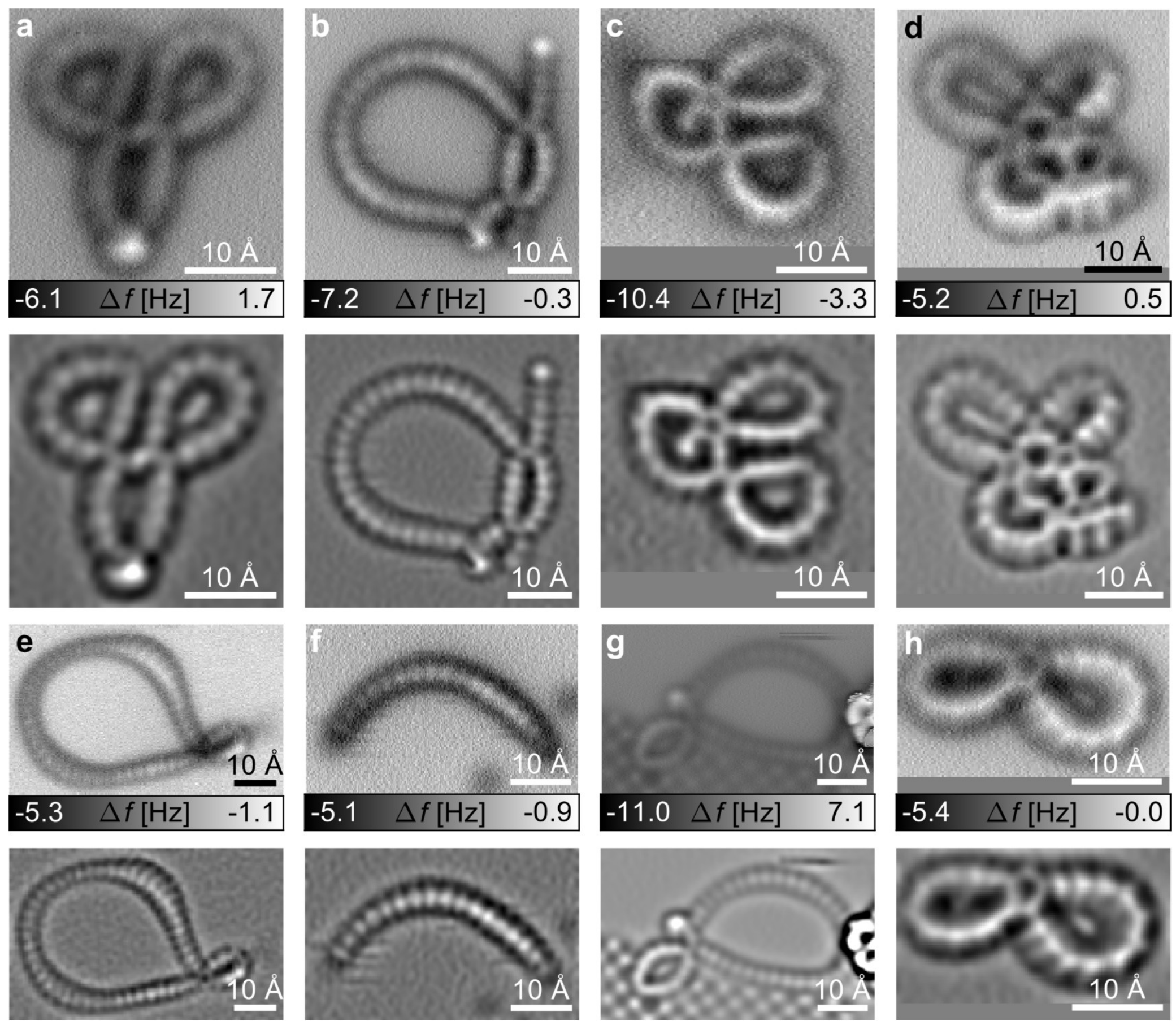

**Figure S16. Other products formed by fusing cyclocarbons.** (**a-h**) AFM raw data (top) and Laplace-filtered AFM data (bottom). All structures shown were formed by voltage pulses from several precursors **1** of $C_{22}$. None of these structures could be transformed into a cyclocarbon by further applied voltage pulses. (**a-c**, **e**, **f**, **h**) on monolayer NaCl on Au(111), (**d**, **g**) on bilayer NaCl on Au(111). Parameters: $A$ = 0.5 Å (**a**, **b**, **d-g**) or $A$ = 1.0 Å (**c**, **h**), tip-height offsets $\Delta z$ with respect to an STM setpoint of $V$ = 0.2 V, $I$ = 1.0 pA: (**a**) $\Delta z$ = -1.1 Å, (**b**) $\Delta z$ = -0.75 Å, (**f**) $\Delta z$ = -0.3 Å; and with respect to an STM setpoint of $V$ = 0.2 V, $I$ = 0.3 pA: (**c**) $\Delta z$ = -1.05 Å, (**d**) $\Delta z$ = 0.4 Å, (**e**) $\Delta z$ = -1.1 Å, (**g**) $\Delta z$ = 0.3 Å, (**h**) $\Delta z$ = -0.3 Å.

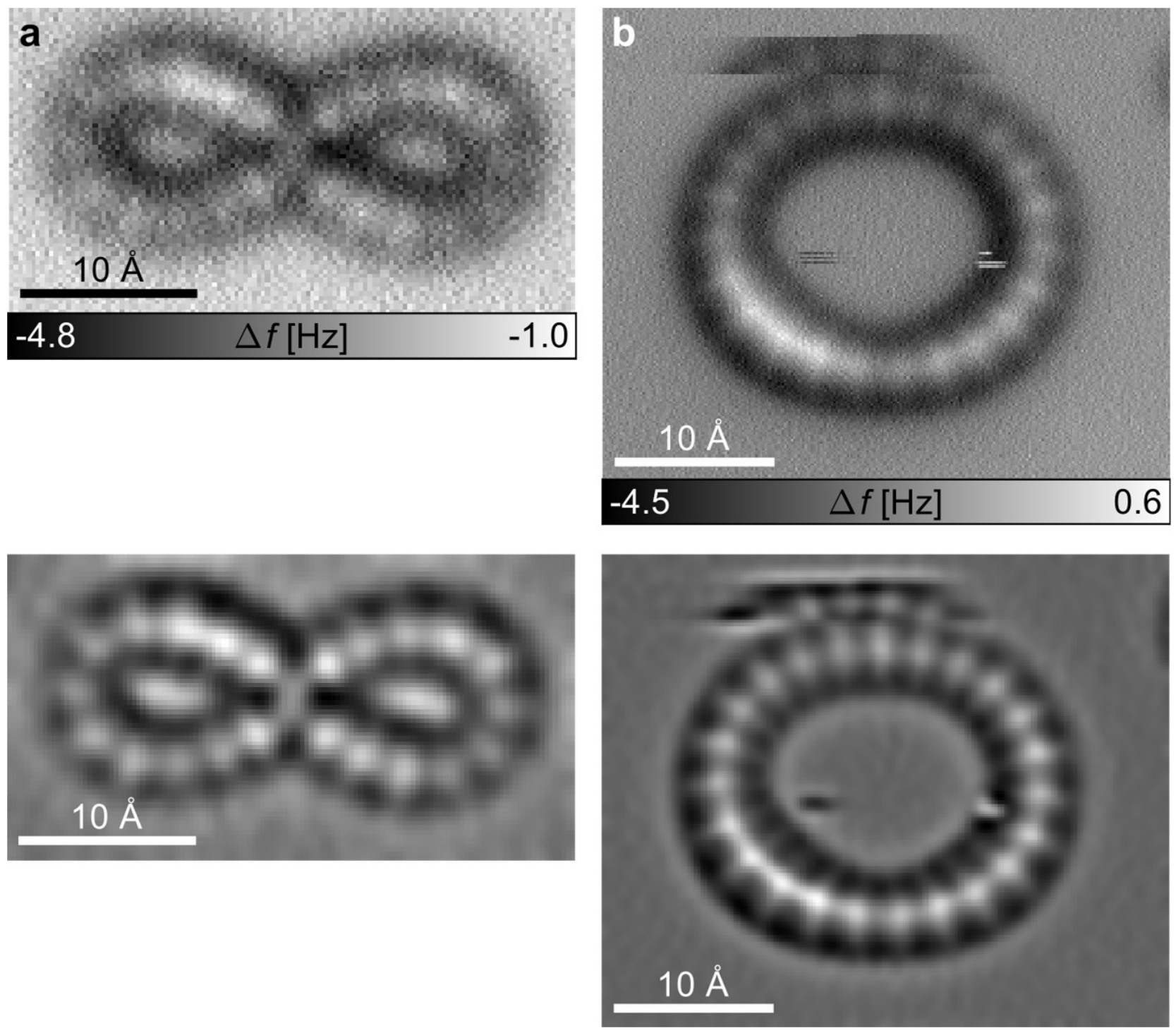

**Figure S17. Formation of $C_{46}$.** AFM raw data (top) and Laplace-filtered AFM data (bottom). (**a**) Lemniscate $C_{46}^{L(24,24)}$ formed by atom manipulation. (**b**) $C_{46}$ formed from the lemniscate shown in (**a**) by atom manipulation. Parameters for the pulses: the pulse that was applied before recording image (**a**) to form the lemniscate: $V_p$ = 4.5 V, $\Delta z$ = 8 Å; and before recording image (**b**) to break the C=C bond of the lemniscate and thus form $C_{46}$: $V_p$ = 6.5 V, $\Delta z$ = 18 Å; AFM parameters: $A$ = 1.0 Å, tip-height offsets with respect to the STM setpoint of $V$ = 0.2 V, $I$ = 0.3 pA: (**a**) $\Delta z$ = -1.45 Å and (**b**) $\Delta z$ = -1.3 Å.

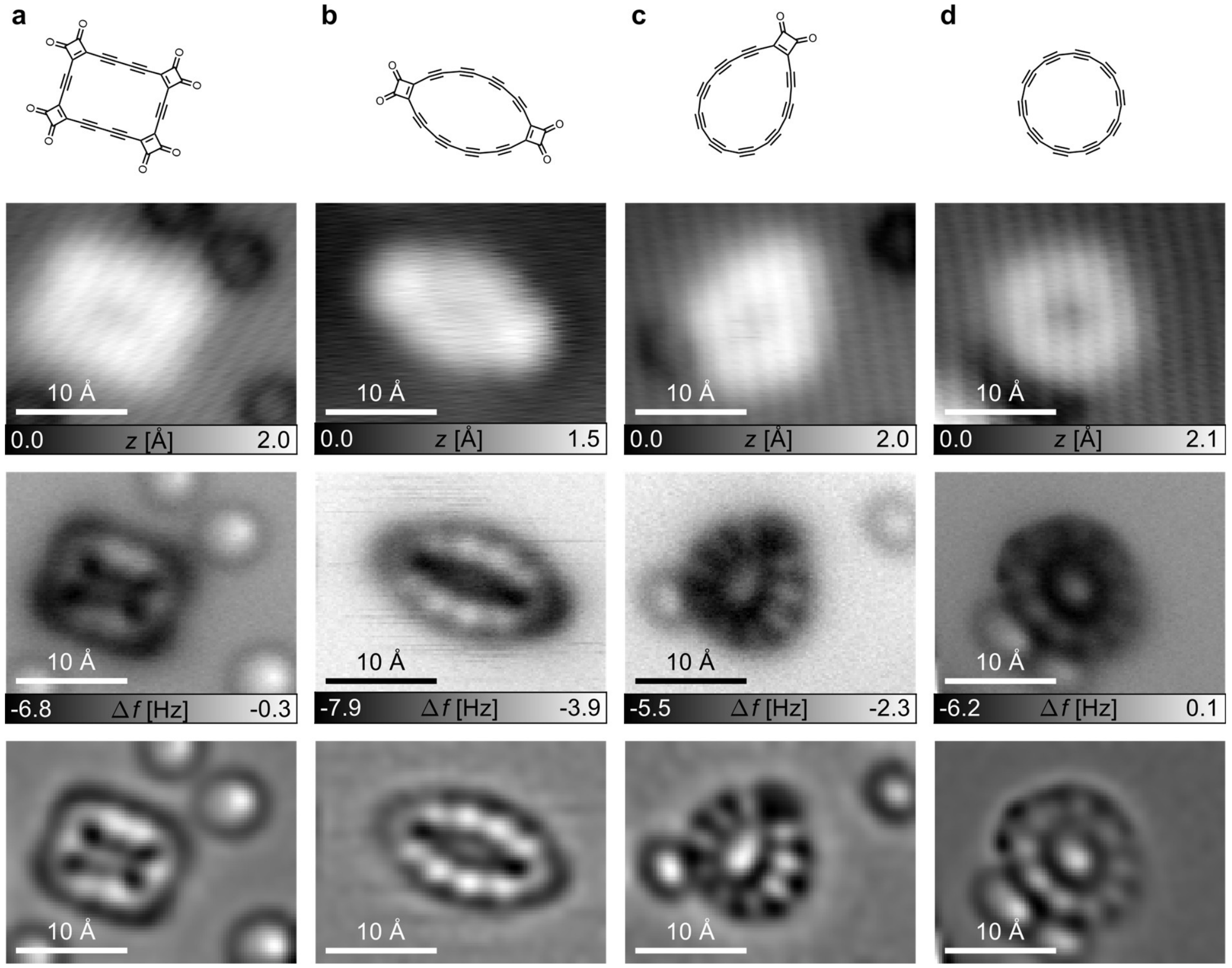

**Figure S18. On-surface synthesis of $C_{20}$.** Chemical structures (1st row), STM data (2nd row), AFM raw data (3th row) and Laplace-filtered AFM data (4th row). (**a**) Precursor of $C_{20}$, that is, compound **3**. (**b**, **c**) Intermediates observed after applying voltage pulses on precursors **3**, (**d**) cyclocarbon $C_{20}$. STM parameters, $V$ = 0.2 V, $I$ = 0.5 pA (**a**, **c**) or $I$ = 0.3 pA (**b**, **d**). AFM parameters, tip-height offsets $\Delta z$ with respect to an STM setpoint of $V$ = 0.2 V, $I$ = 0.5 pA: (**a**) $\Delta z$ = -0.5 Å, (**c**) $\Delta z$ = -0.25 Å and (**d**) $\Delta z$ = -0.25 Å; and with respect to an STM setpoint of $V$ = 0.2 V, $I$ = 0.3 pA: (**b**) $\Delta z$ = -1.0 Å.

## 7. AFM characterization of large cyclocarbons:

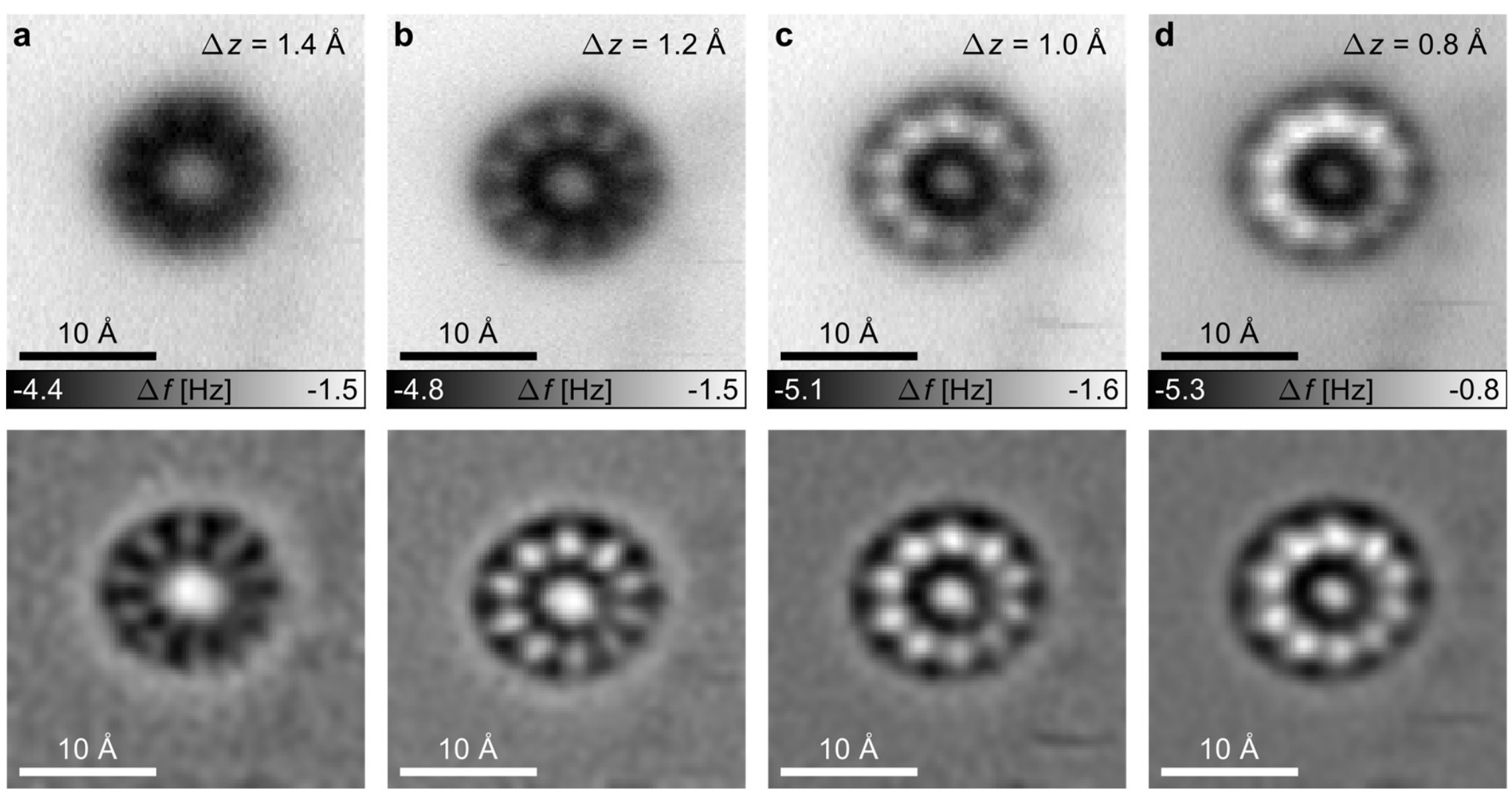

**Figure S19. Additional AFM data on $C_{20}$.** (**a-d**) AFM raw data (top) and Laplace-filtered AFM data (bottom). The molecule is on bilayer NaCl on Au(111). Tip-height offsets Δ*z* with respect to the STM setpoint: *V* = 0.2 V, *I* = 0.3 pA.

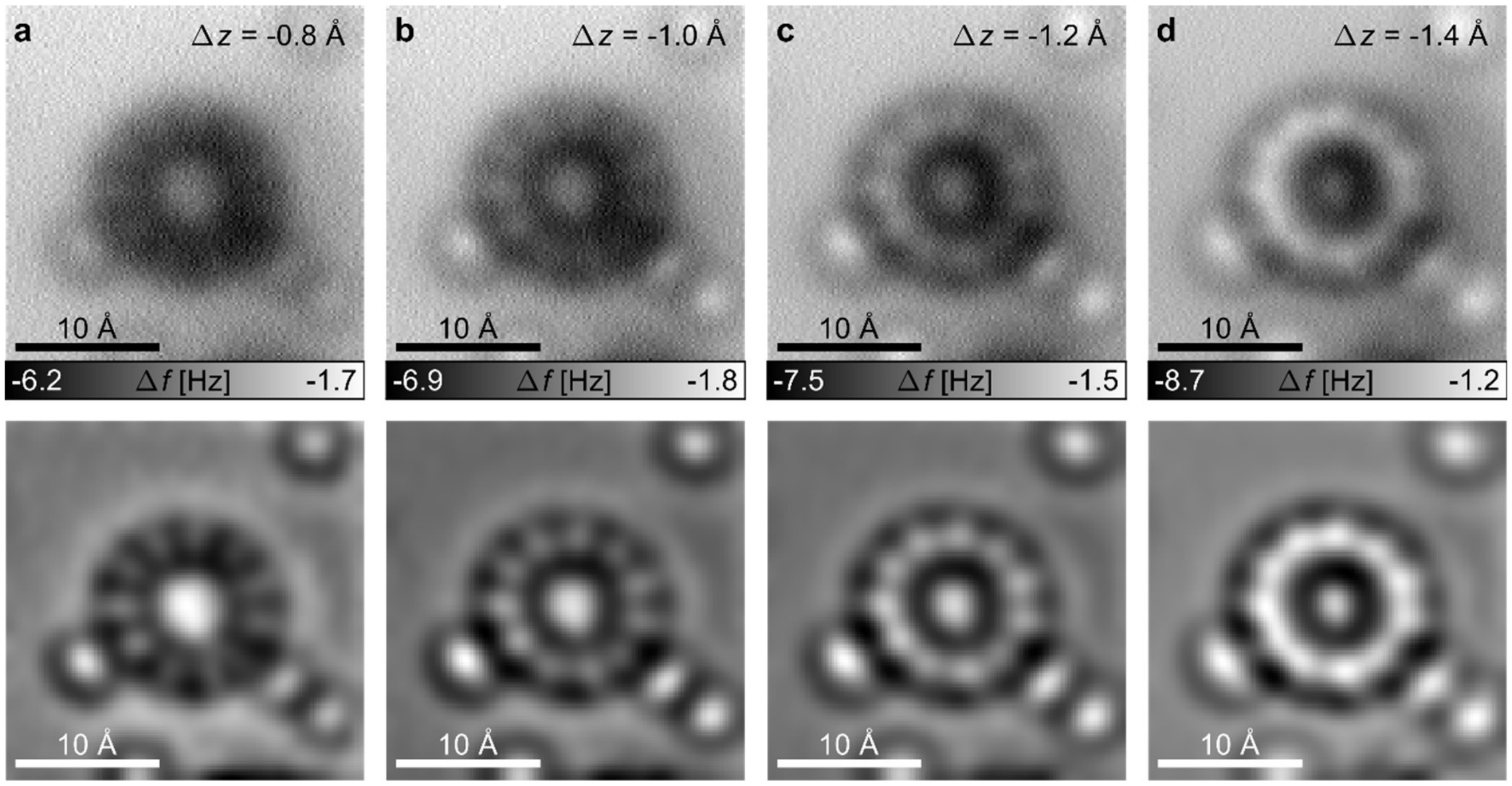

**Figure S20. Additional AFM data on $C_{22}$.** (**a-d**) AFM raw data (top) and Laplace-filtered AFM data (bottom). Tip-height offsets Δ*z* with respect to the STM setpoint: *V* = 0.2 V, *I* = 0.3 pA.

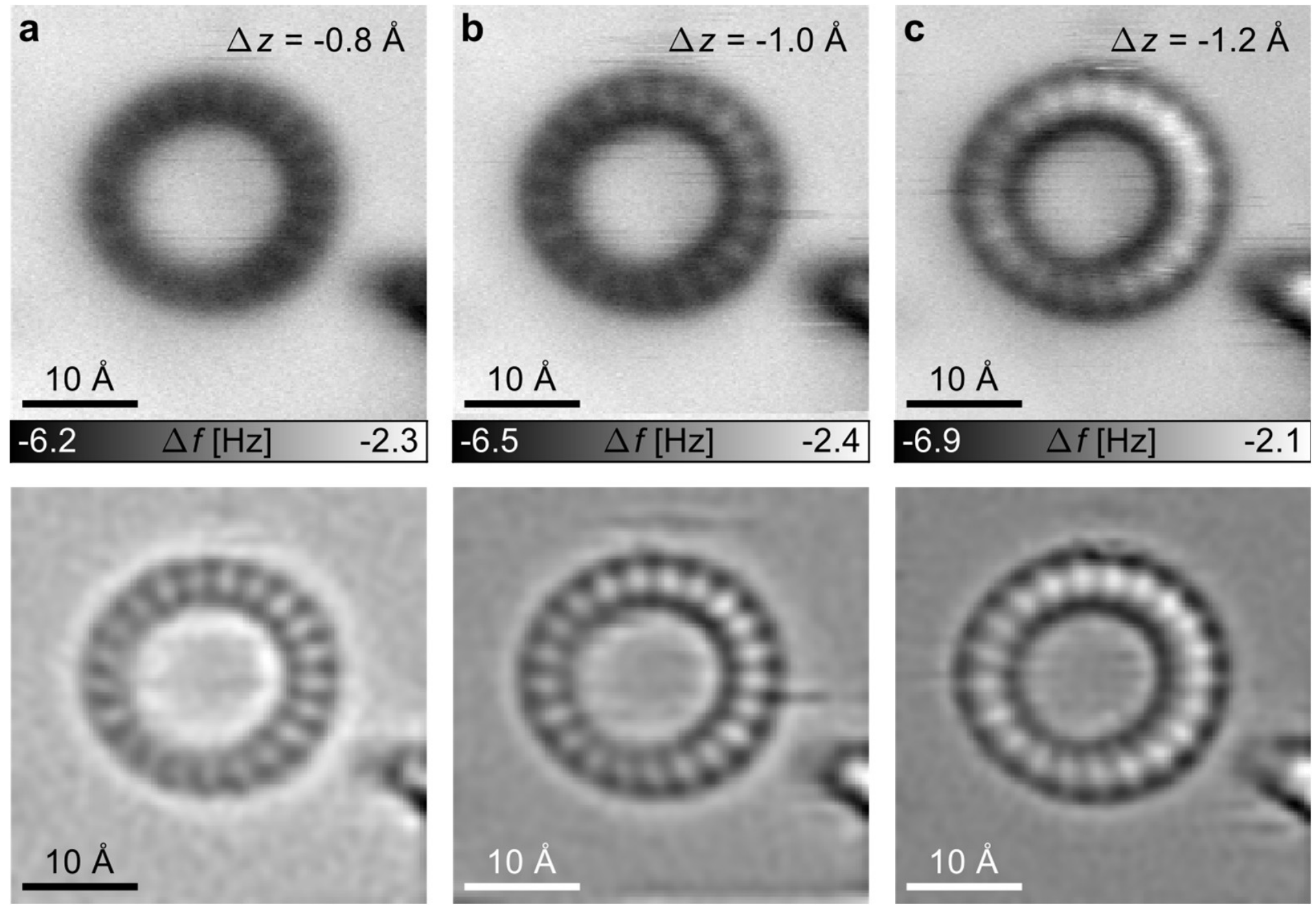

**Figure S21. AFM data on $C_{42}$.** (**a-d**) AFM raw data (top) and Laplace-filtered AFM data (bottom). Tip-height offsets Δ*z* with respect to the STM setpoint: *V* = 0.1 V, *I* = 0.3 pA.

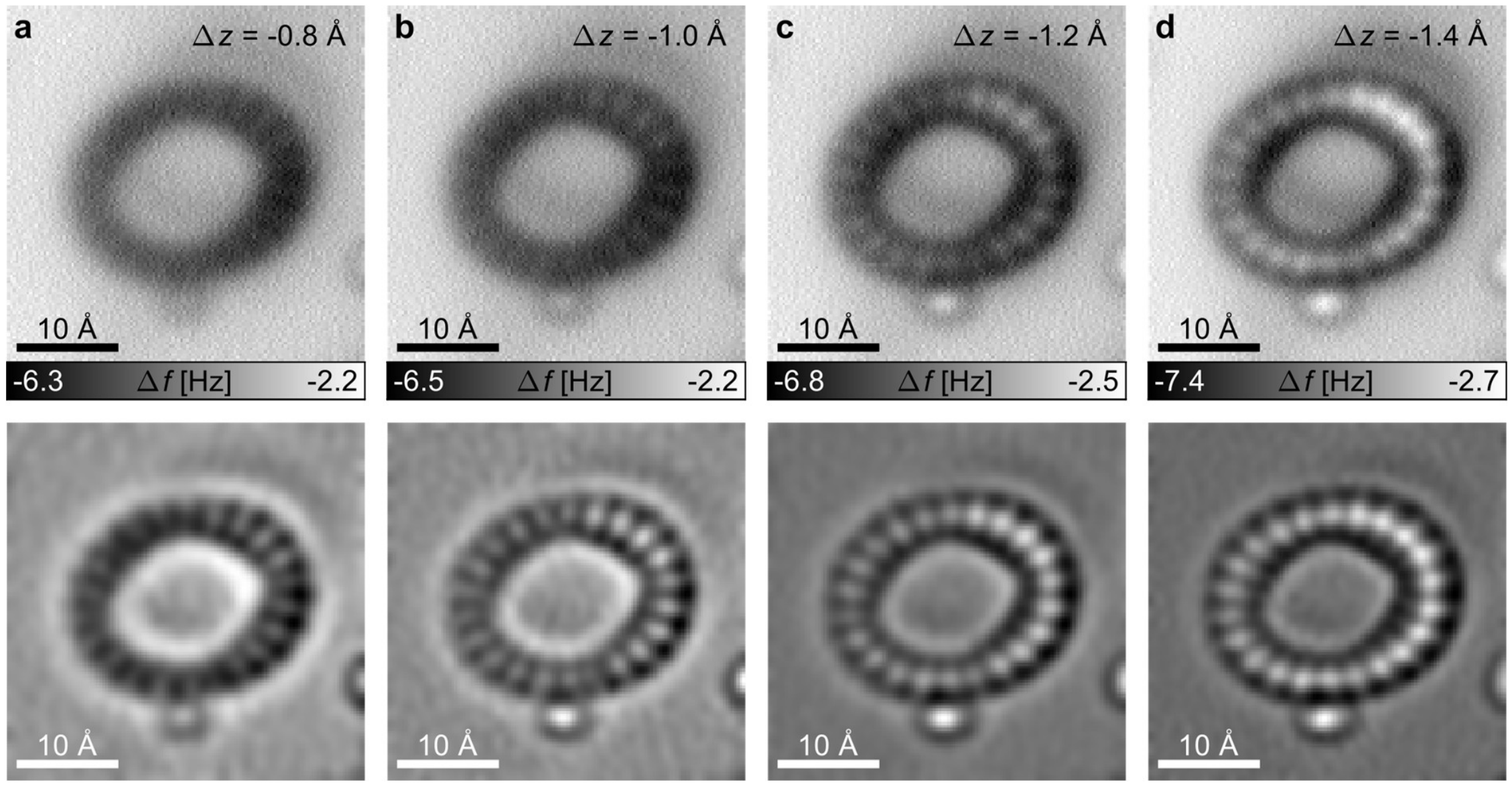

**Figure S22. Additional AFM data on $C_{44}$.** (**a-d**) AFM raw data (top) and Laplace-filtered AFM data (bottom). Tip-height offsets Δ*z* with respect to the STM setpoint: *V* = 0.2 V, *I* = 0.3 pA.

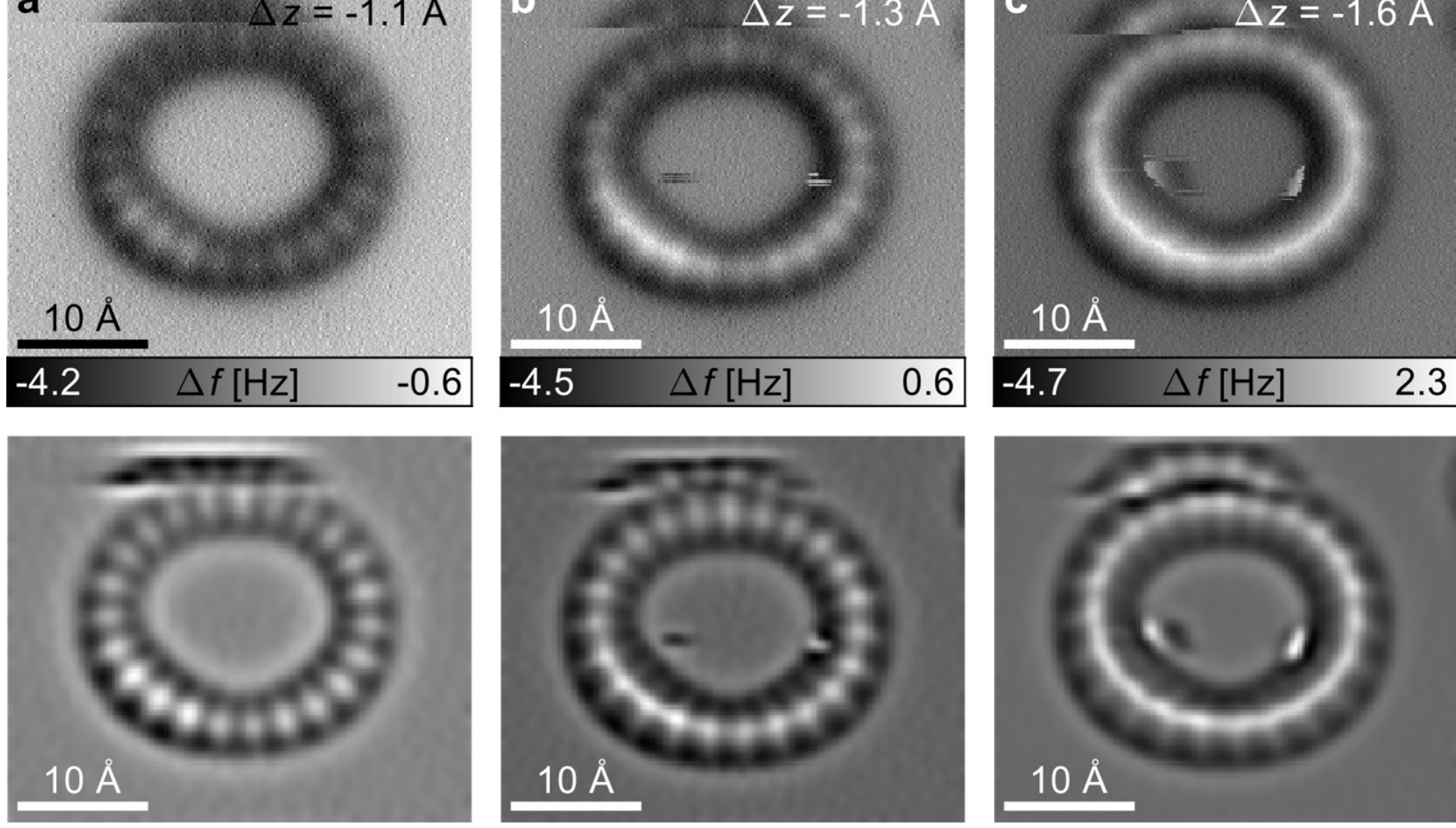

**Figure S23. Additional AFM data on $C_{46}$.** (**a-c**) AFM raw data (top) and Laplace-filtered AFM data (bottom). The doubled appearance of the ring segment in the upper part of the AFM measurement is due to the molecule being displaced by interaction with the tip during scanning. AFM parameters: $A$ = 1.0 Å, tip-height offsets $\Delta z$ with respect to the STM setpoint: $V$ = 0.2 V, $I$ = 0.3 pA, are indicated.

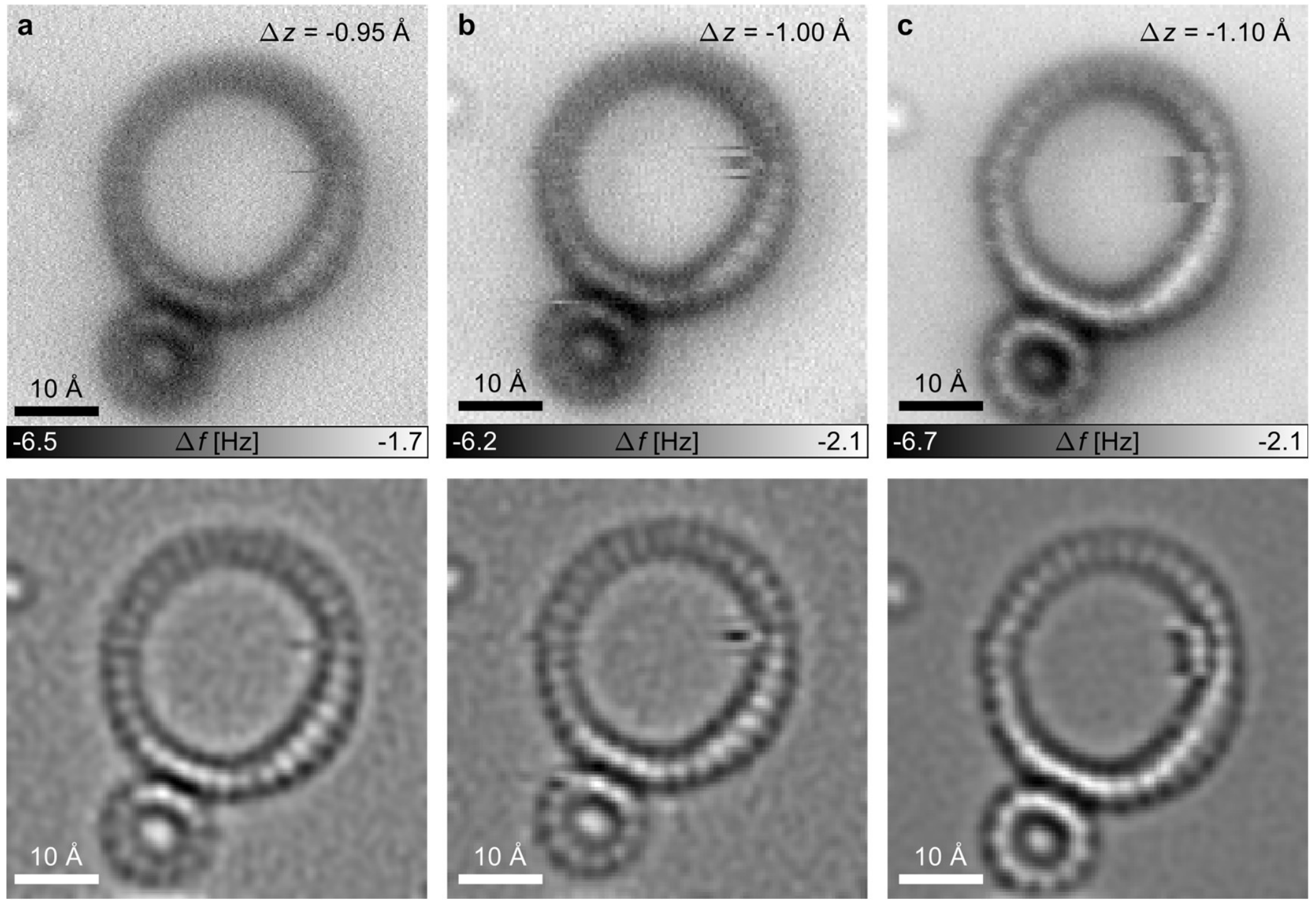

**Figure S24. Additional AFM data on $C_{66}$.** (**a-c**) AFM raw data (top) and Laplace-filtered AFM data (bottom). Some segments of the ring appear doubled in the AFM images, as explained in **section 8**. A $C_{22}$ is adsorbed next to $C_{66}$. Tip-height offsets $\Delta z$ with respect to the STM setpoint: $V$ = 0.2 V, $I$ = 0.3 pA.

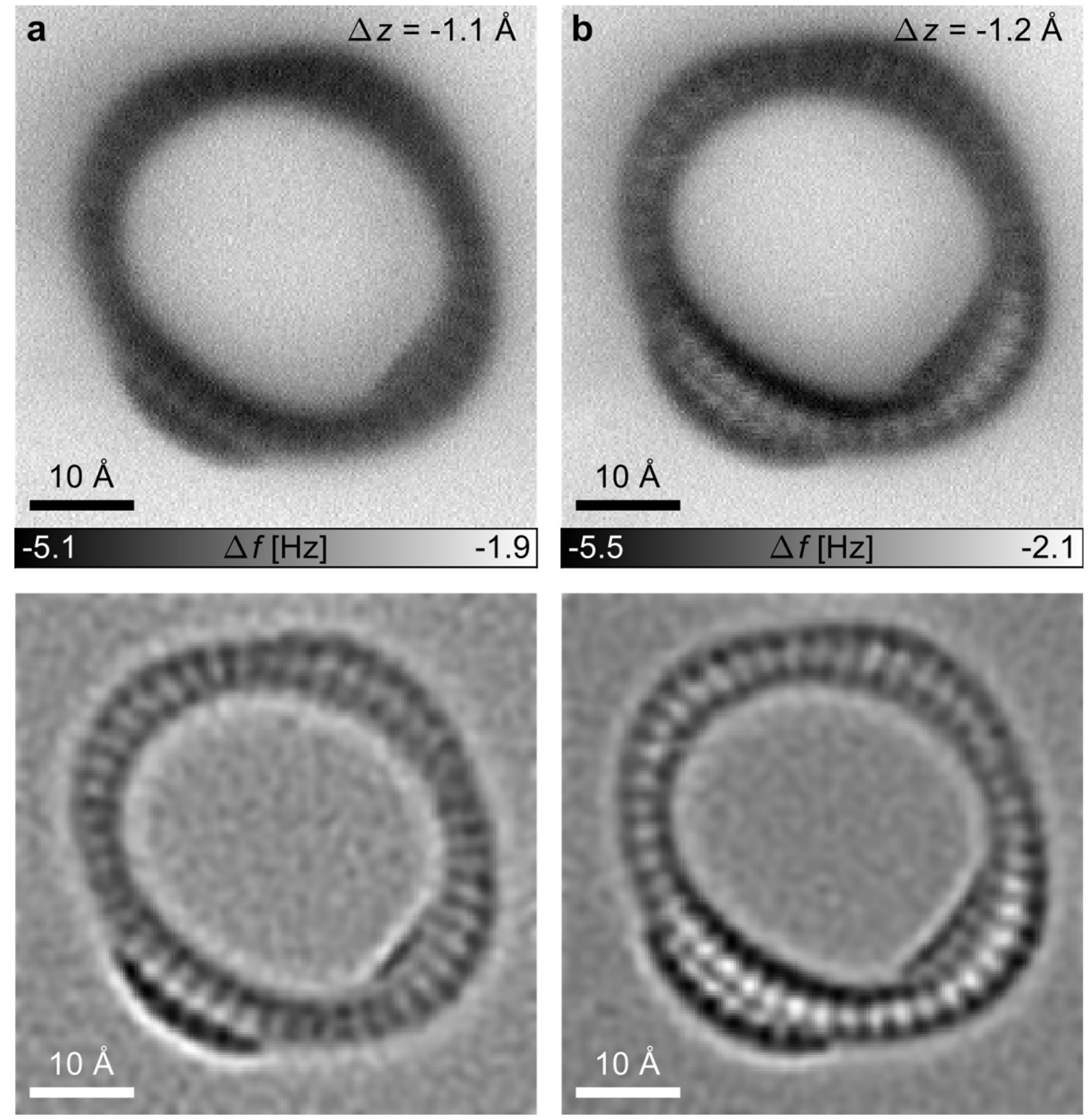

**Figure S25. Additional AFM data on $C_{88}$.** (**a**, **b**) AFM raw data (top) and Laplace-filtered AFM data (bottom). Some segments of the ring appear doubled in the AFM images, as explained in **section 8**. Tip-height offsets Δ*z* with respect to the STM setpoint: *V* = 0.2 V, *I* = 0.3 pA.

## 8. Doubling of ring segments of large cyclocarbons and intensity modulation of STM orbital density maps:

In AFM images, the Cl sites of NaCl are observed as bright features, from which we can deduce the crystallographic orientation of the NaCl surface and the adsorption site of the molecule. For large rings ($C_{46}$, $C_{66}$ and $C_{88}$) we often observe that segments of the ring appear doubled in AFM images, see **Figs. S23, S24c** and **S25**. By resolving the substrate, we find that the segments that appear doubled are aligned to the non-polar <110> directions of the NaCl surface, see **Fig. S26**. The doubled appearance of these segments we explain by a low energy barrier for different adsorption geometries of the segments along the <110> NaCl directions. By attractive interaction with the AFM tip, the segments locally move between neighbouring <110> oriented rows of Na and Cl atoms. This is similar to other molecules with small diffusion barriers on NaCl, e.g. triangulene, which changes its orientation on the surface, and moves under the AFM tip.[4]

For the atomically resolved NaCl monolayer in **Fig. S26**, we observe a modulation of the intensity of the Cl atoms. The modulation can be attributed to the Moiré pattern, resulting from the incommensurate growth of NaCl on Au(111). For monolayer NaCl, the contrast of the Moiré pattern is more pronounced compared to bilayer NaCl. The Moiré pattern might also affect the energy potential landscape for molecular adsorption. In addition, the Moiré pattern might be the reason for the intensity modulation of the STM orbital density maps and in-gap images of large cyclocarbons, e.g. as seen in **Figs. S32** and **S33**.

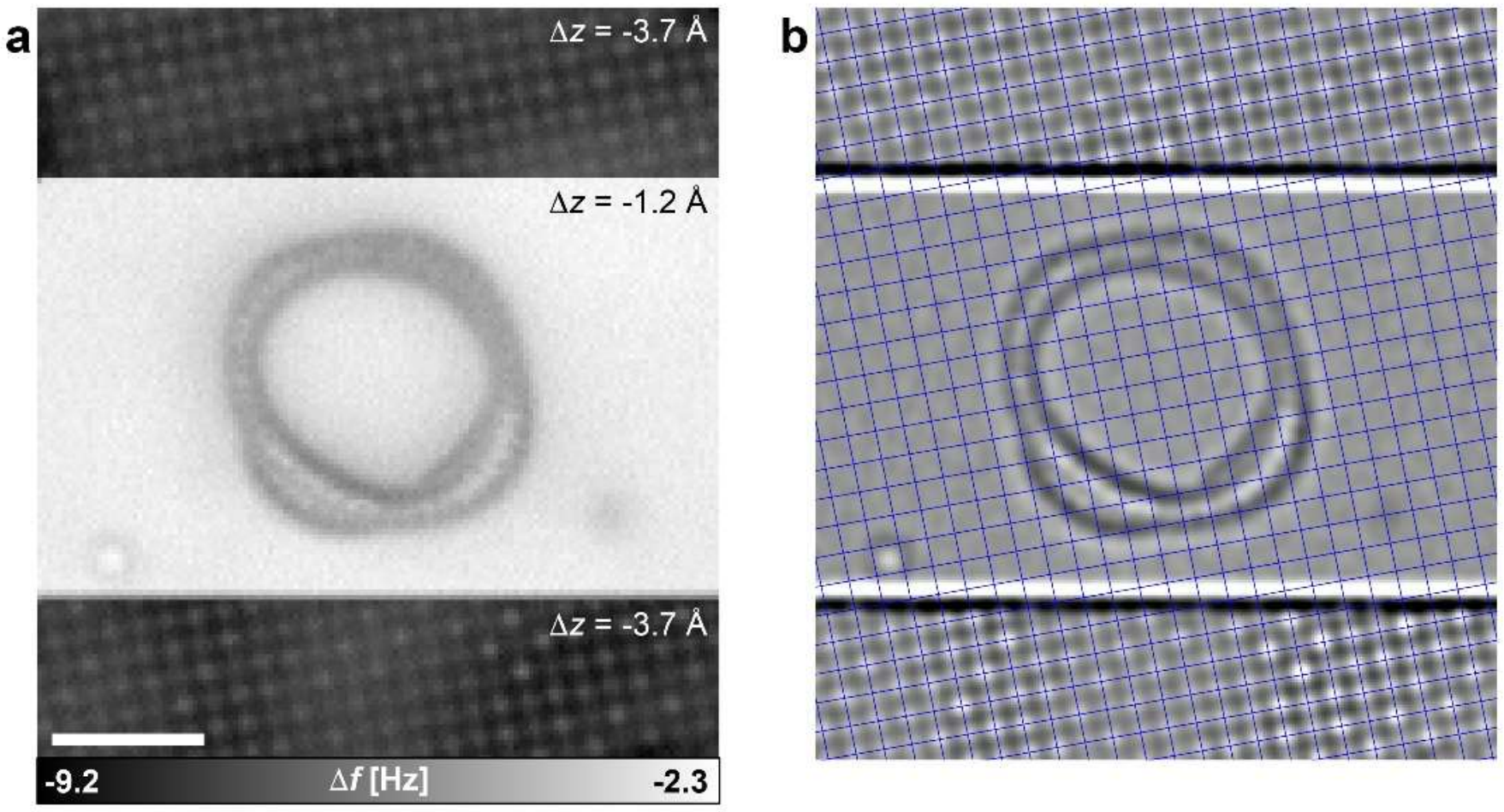

**Figure S26. Adsorption-site determination of $C_{88}$.** AFM raw data (**a**) and Laplace filtered AFM data (**b**) of $C_{88}$ on monolayer NaCl. To resolve the NaCl film atomically, the tip-height offset $\Delta z$ was changed within the image, to be 2.5 Å smaller (closer to the surface) in the top and bottom part of the image, compared to the middle part. The blue lines in (**b**) indicate the NaCl lattice, with vertices indicating the positions of the $Cl^-$ ions of the NaCl monolayer. Tip-height offsets $\Delta z$ with respect to the STM setpoint $I$ = 0.3 pA, $V$ = 0.2 V (at the top centre of the image).

## 9. STM characterization of large cyclocarbons:

For each cyclocarbon formed in this study, we mapped by STM the negative (NIR) and positive (PIR) ion resonances, which reflect the orbital densities associated with electron attachment (NIR) and detachment (PIR)[5,6]. The NIR and PIR densities for both $N = 4n$ (**Figs. S27**, **S30** and **S33**) and $N = 4n+2$ cyclocarbons (**Figs. S28**, **S29**, **S31** and **S32**), show a particle-on-a-ring-like structure with $N/2$ lobes along the ring. The lobes are located above the long bonds in case of the NIR and located above the short bonds in case of the PIR. However, the origin of these resonance densities is different for anti-aromatic $N = 4n$ (ref. 2) and aromatic $N = 4n+2$ (ref. 7) cyclocarbons, as can be rationalized by the condition, that the π-orbitals (except the lowest ones in energy) all have an even number of lobes along the ring, because the sign of the orbital switches between each two lobes. In $N = 4n$ (anti-aromatic) cyclocarbons, the number of observed lobes, that is, $N/2$, is even and the NIR density corresponds to one lowest unoccupied molecular orbital (LUMO), while the PIR corresponds to one highest occupied molecular orbital (HOMO). In the case of $N = 4n+2$ (aromatic) cyclocarbons, the number of observed lobes, that is $N/2$, is odd, and the PIR density results from a superposition of two degenerate HOMOs, with $(N/2 – 1)$ lobes each. Similarly, the NIR density results from two degenerate LUMOs with $(N/2 + 1)$ lobes[7].

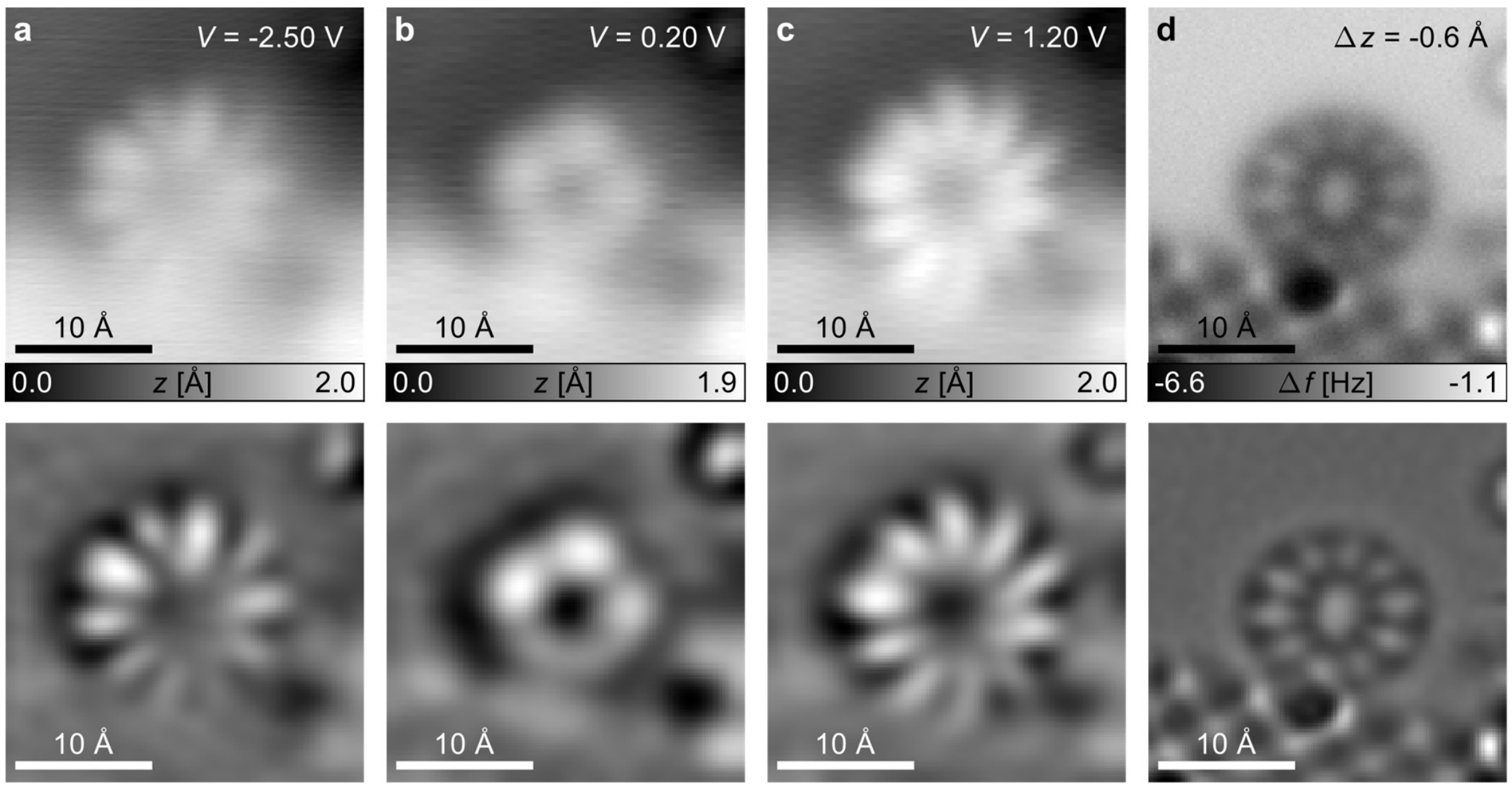

**Figure S27. STM orbital density maps of $C_{20}$.** (**a-c**) STM raw data (top) and Laplace-filtered STM data (bottom), at indicated sample voltages *V*. (**a**) At *V* = -2.50 V, corresponding to the PIR; (**b**) at *V* = 0.20 V, i.e., in-gap; (**c**) at *V* = 1.20 V, corresponding to the NIR. (**d**) AFM raw data (top) and Laplace-filtered AFM data (bottom), for comparison to the orbital density maps (**a**) and (**c**). The molecule is on monolayer NaCl on Au(111), next to a bilayer NaCl island. STM parameters, *I* = 0.7 pA, *V* as indicated. AFM parameters, tip-height offset Δ*z* with respect to the STM setpoint: *V* = 0.2 V, *I* = 0.3 pA.

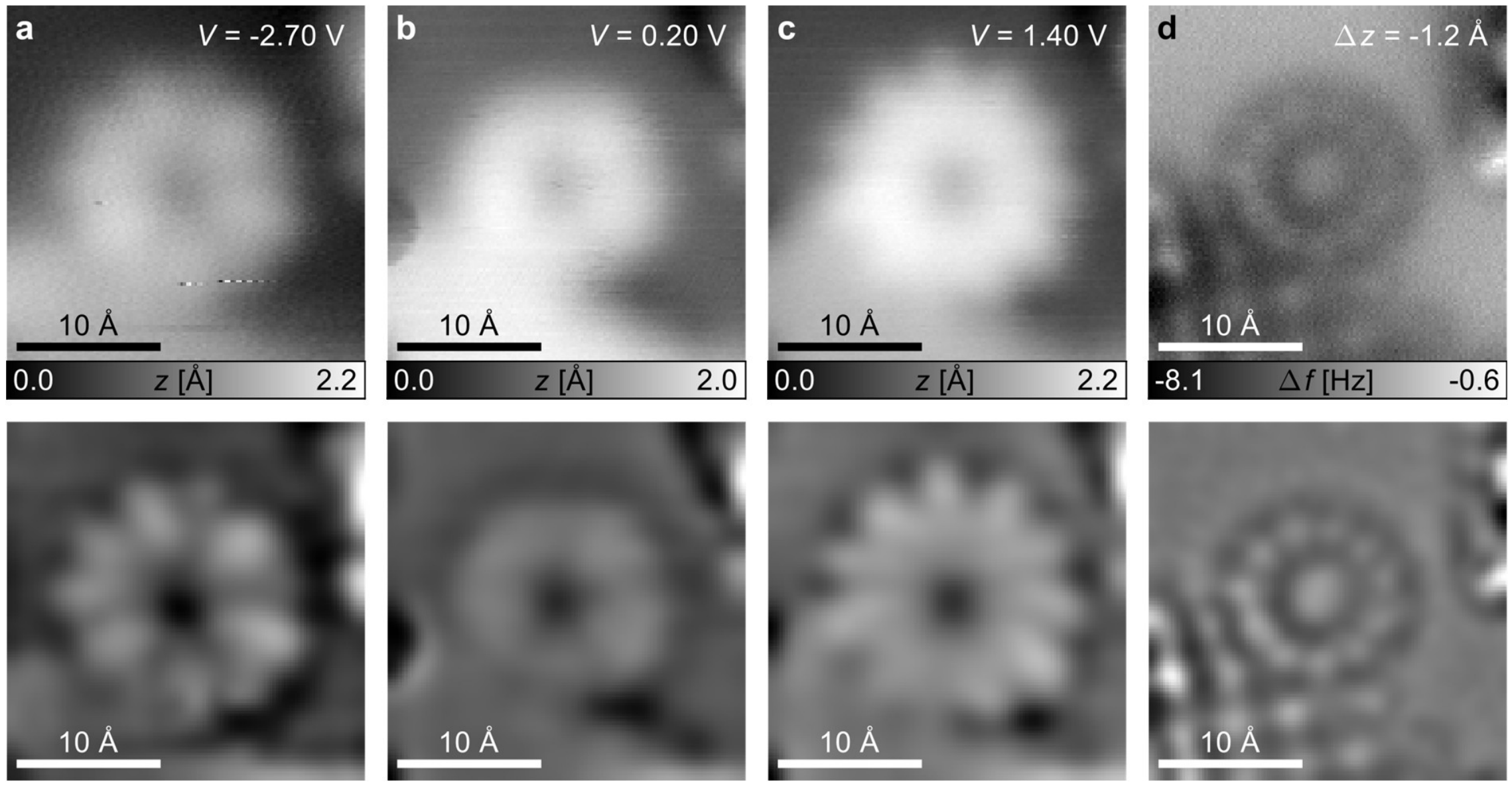

**Figure S28. STM orbital density maps of $C_{22}$.** (**a-c**) STM raw data (top) and Laplace-filtered STM data (bottom), at indicated sample voltages *V*. (**a**) At *V* = -2.70 V, corresponding to the PIR; (**b**) at *V* = 0.20 V, i.e., in-gap; (**c**) at *V* = 1.40 V, corresponding to the NIR. (**d**) AFM raw data (top) and Laplace-filtered AFM data (bottom), for comparison to the orbital density maps (**a**) and (**c**). The molecule is on monolayer NaCl on Au(111), next to a bilayer NaCl island. STM parameters, *I* = 0.3 pA, *V* as indicated. AFM parameters, tip-height offset Δ*z* with respect to the STM setpoint: *V* = 0.2 V, *I* = 0.3 pA.

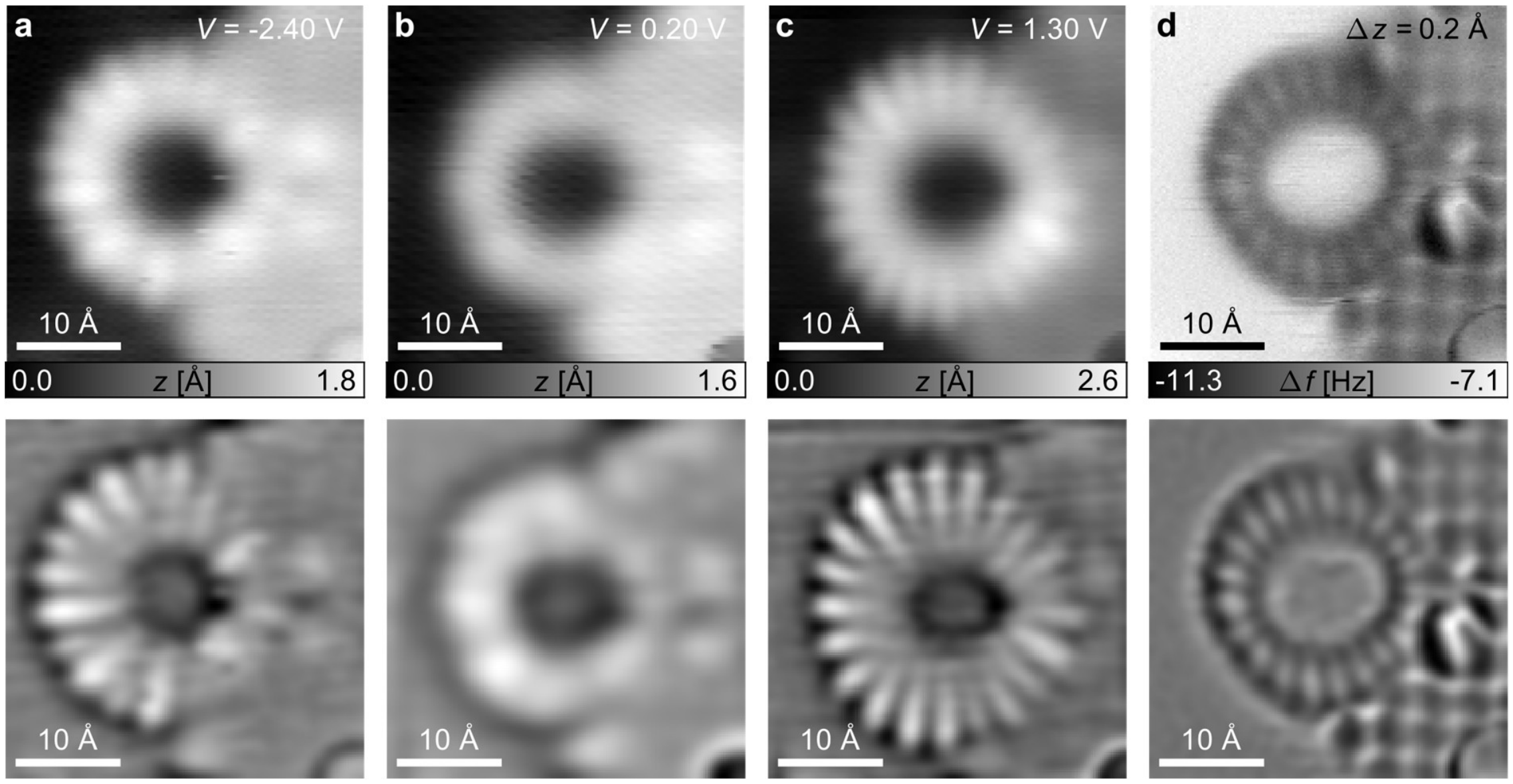

**Figure S29. STM orbital density maps of $C_{42}$.** (**a-c**) STM raw data (top) and Laplace-filtered STM data (bottom), at indicated sample voltages *V*. (**a**) At *V* = -2.40 V, corresponding to the PIR; (**b**) at *V* = 0.20 V, i.e., in-gap; (**c**) at *V* = 1.30 V, corresponding to the NIR. (**d**) AFM raw data (top) and Laplace-filtered AFM data (bottom), for comparison to the orbital density maps (**a**) and (**c**). The molecule is on monolayer NaCl on Au(111), next to a bilayer NaCl island. STM parameters, *I* = 0.3 pA (**a**, **b**) or 0.5 pA (**c**), *V* as indicated. AFM parameters, tip-height offset Δ*z* with respect to the STM setpoint: *V* = 0.2 V, *I* = 0.3 pA.

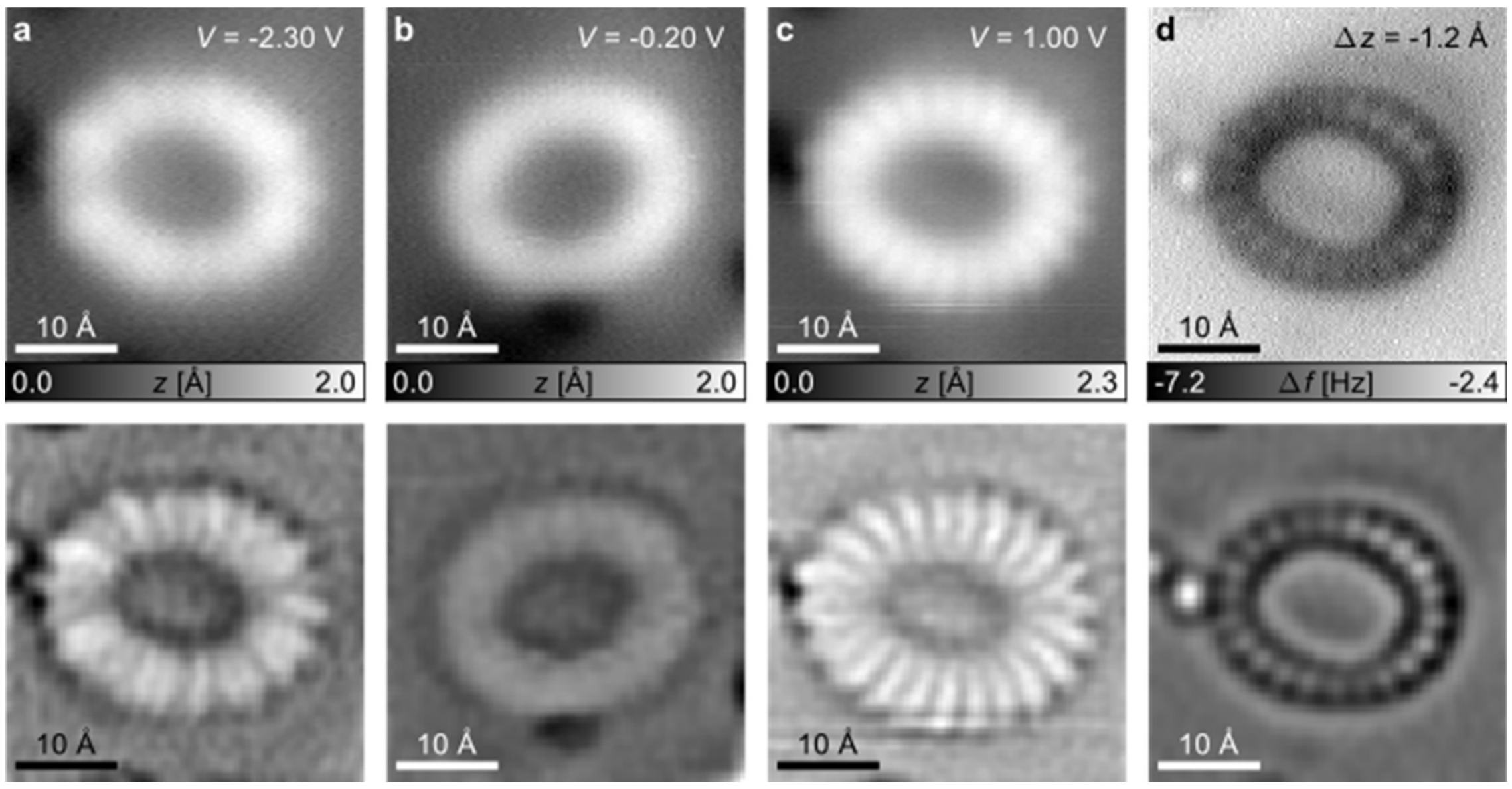

**Figure S30. STM orbital density maps of $C_{44}$.** (**a-c**) STM raw data (top) and Laplace-filtered STM data (bottom), at indicated sample voltages *V*. (**a**) At *V* = -2.30 V, corresponding to the PIR; (**b**) at *V* = -0.20 V, i.e., in-gap; (**c**) at *V* = 1.00 V, corresponding to the NIR. (**d**) AFM raw data (top) and Laplace-filtered AFM data (bottom), for comparison to the orbital density maps (**a**) and (**c**). The molecule was not moved between panels (**a**, **c**, **d**), but it was moved to a different adsorption site before taking the in-gap STM image (**b**). STM parameters, *I* = 0.3 pA, *V* as indicated. AFM parameters, tip-height offset Δ*z* with respect to the STM setpoint: *V* = 0.2 V, *I* = 0.3 pA.

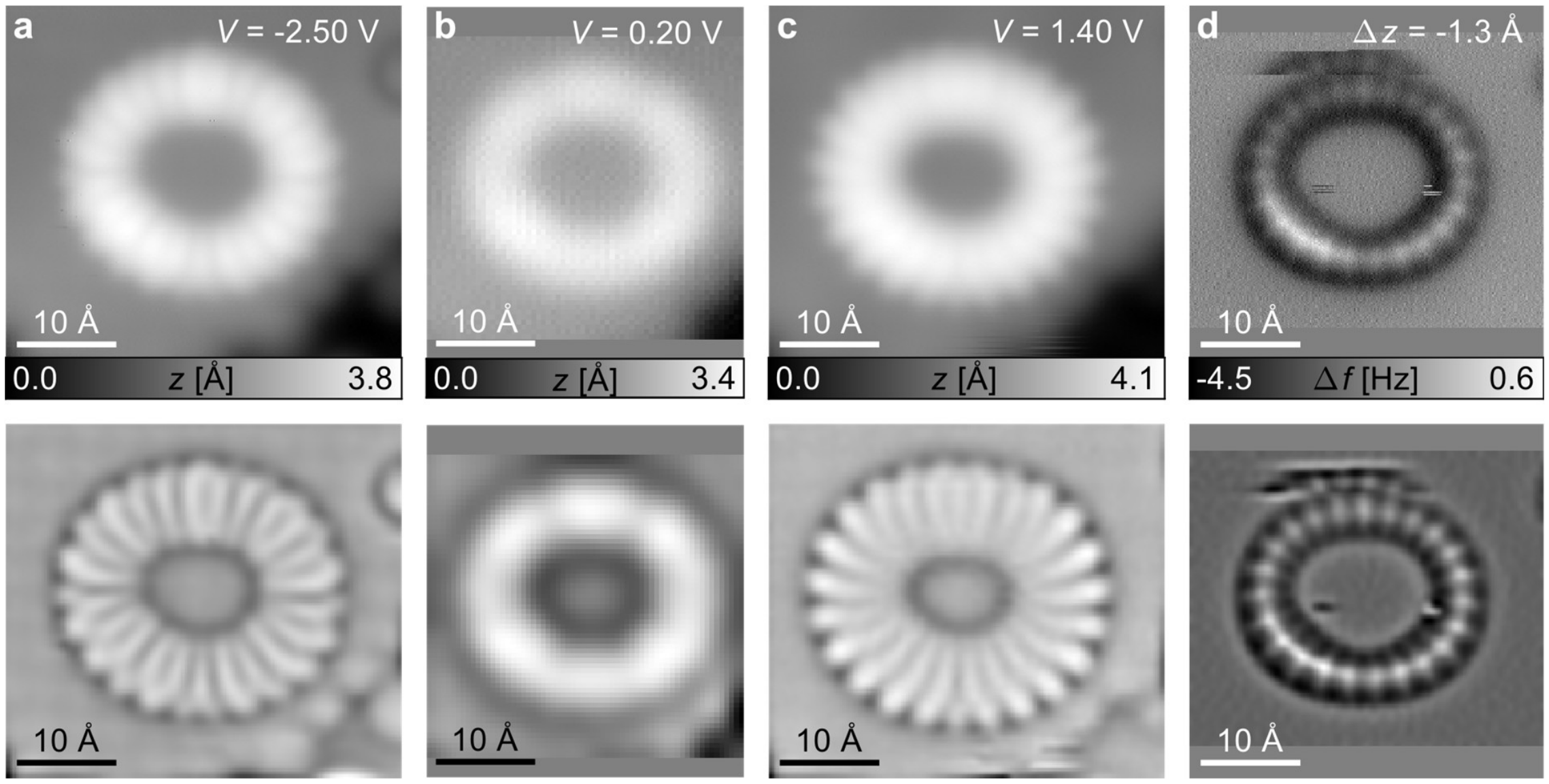

**Figure S31. STM orbital density maps of $C_{46}$.** (**a-c**) STM raw data (top) and Laplace-filtered STM data (bottom), at indicated sample voltages $V$. (**a**) At $V$ = -2.50 V, corresponding to the PIR; (**b**) at $V$ = 0.20 V, i.e., in-gap; (**c**) at $V$ = 1.40 V, corresponding to the NIR. (**d**) AFM raw data (top) and Laplace-filtered AFM data (bottom), for comparison to the orbital density maps (**a**, **c**). The doubled appearance of the ring segment in the upper part of the AFM measurement is due to the molecule being displaced by interaction with the tip during scanning. STM parameters, $I$ = 1.0 pA (**a**, **c**) and 0.3 pA (**b**), $V$ as indicated. AFM parameters: $A$ = 1.0 Å, tip-height offset $\Delta z$ with respect to the STM setpoint: $V$ = 0.2 V, $I$ = 0.3 pA.

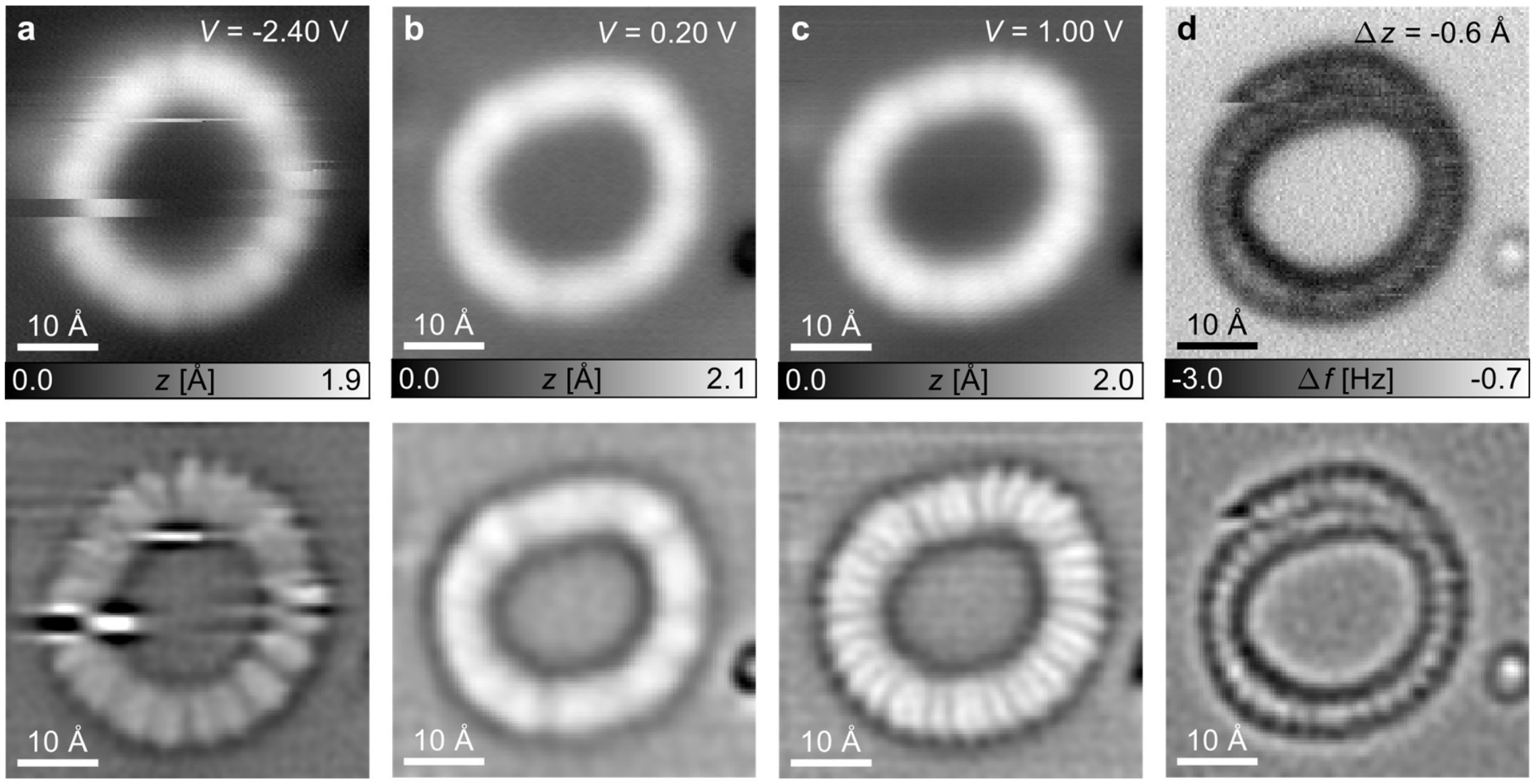

**Figure S32. STM orbital density maps of $C_{66}$.** (**a-c**) STM raw data (top) and Laplace-filtered STM data (bottom), at indicated sample voltages *V*. (**a**) At *V* = -2.40 V, corresponding to the PIR; (**b**) at *V* = 0.20 V, i.e., in-gap; (**c**) at *V* = 1.00 V, corresponding to the NIR. (**d**) AFM raw data (top) and Laplace-filtered AFM data (bottom), for comparison to the orbital density maps (**a**) and (**c**). The molecule was not moved between panels (**b-d**), but was moved to a different adsorption site before taking the PIR image (**a**). STM parameters, *I* = 0.3 pA (**a**, **b**) and *I* = 0.5 pA (**c**), *V* as indicated. AFM parameters, tip-height offset Δ*z* with respect to the STM setpoint: *V* = 0.2 V, *I* = 0.3 pA.

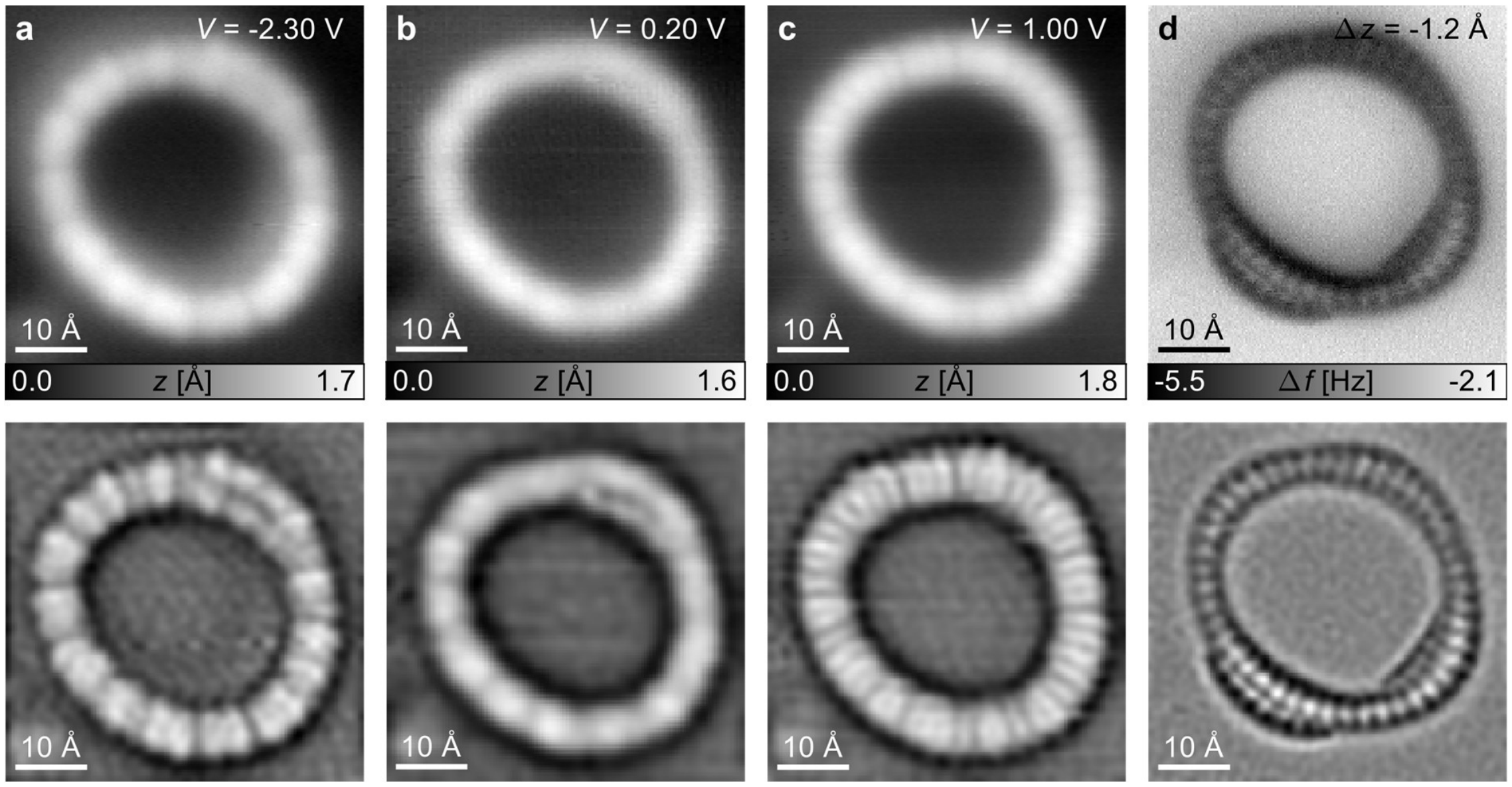

**Figure S33. STM orbital density maps of $C_{88}$.** (**a-c**) STM raw data (top) and Laplace-filtered STM data (bottom), at indicated sample voltages $V$. (**a**) At $V$ = -2.30 V, corresponding to the PIR; (**b**) at $V$ = 0.20 V, i.e., in-gap; (**c**) at $V$ = 1.00 V, corresponding to the NIR. (**d**) AFM raw data (top) and Laplace-filtered AFM data (bottom), for comparison to the orbital density maps (**a**) and (**c**). STM parameters, $I$ = 0.5 pA (**a**, **c**) and $I$ = 0.3 pA (**b**), $V$ as indicated. AFM parameters, tip-height offset $\Delta z$ with respect to the STM setpoint: $V$ = 0.2 V, $I$ = 0.3 pA.

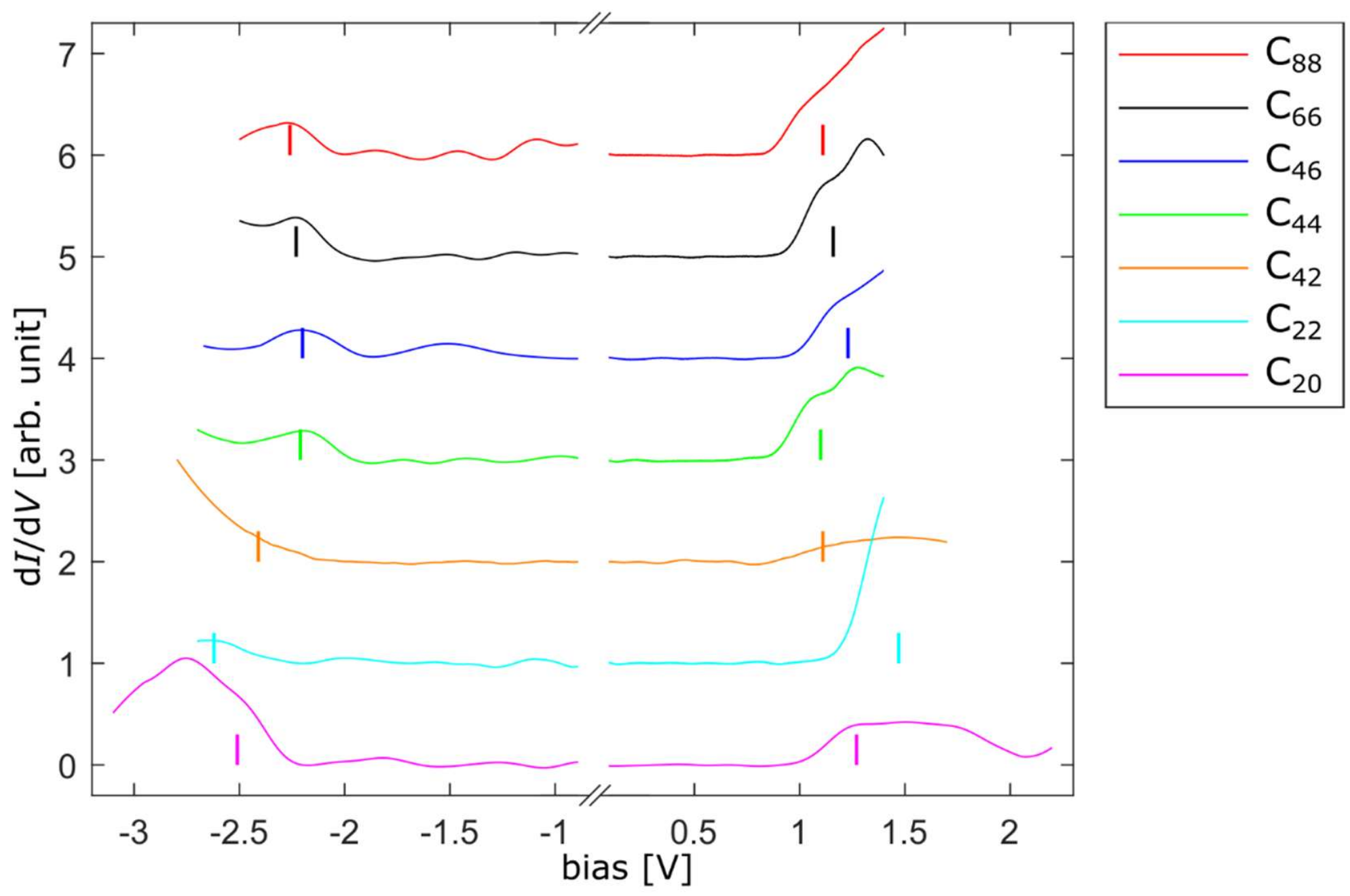

**Figure S34. STS of cyclocarbon molecules.** Representative d*I*/d*V* spectra of cyclocarbon molecules derived from *I*/*V* spectra obtained at constant height. Note that the tip heights were chosen to result in maximum currents of only about 1 pA, because larger currents typically resulted in displacements of the molecules on the surface. For every microtip used, background *I*/*V* spectra were taken. After numerical deriving the d*I*/d*V* spectra, the respective background signal was subtracted. Spectra are vertically offset. The peaks at positive and negative bias were fitted to a minimal set of Gaussians of the same width, corresponding to the full width at half maximum reported for pentacene[5] that is 0.3 V. Vertical markers represent the positions of the fitted Gaussians at lowest positive and negative voltages. The transport gaps shown in **Fig. 3a** in the main text correspond to the peak-to-peak spacing of these Gaussians. STM maps of the ion resonances (**Figs. S27** to **S33**) confirmed that the peaks assigned in the d*I*/d*V* spectra correspond to the respective PIR and NIR. The errors displayed in **Fig. 3a** stem from two contributions: i) For all derived values an error of ±65 mV to take the scatter for different microtips into account. ii) For ring sizes with several individual molecules probed ($C_{20}$, $C_{22}$, $C_{42}$, $C_{44}$) the standard deviation of the experimentally derived gap values. But for $C_{46}$, $C_{66}$ and $C_{88}$, for which we recorded *I*/*V* data for only one individual molecule, the 95% confidence intervals of the Gaussian fits are used as second contribution to the error.

### 10. Input cards for OX-BLYP30:

GAUSSIAN 16:

```
# Single point energy calculation with OX-BLYP30 (ω=0.025, E_X^HF: 30% => 100%) in Gaussian16

%nproc=36
%mem=72GB
#p def2tzvp cam-b3lyp
IOp(3/107=0025000000,3/108=0025000000,3/119=0700000000,3/120=0700000000,3/130=03000,3/131=03000)
```

ORCA 6.1:

```
# Single point energy calculation with OX-BLYP30 (ω=0.025, E_X^HF: 30% => 100%) in ORCA 6.1
! CAM-B3LYP def2-TZVP def2-TZVP/C def2/J RIJCOSX TightSCF

%method
   Method      DFT
   ScalHFX     0.30
   ScalDFX     0.00
   RangeSepEXX  True
   RangeSepMu   0.025
   RangeSepScal  0.70
End
```

TURBOMOLE:

```
# Single point energy calculation with OX-BLYP30 (ω=0.025, E_X^HF: 30% => 100%) in Turbomole

$dft
functional cam-b3lyp_own 0.30 0.70 0.025
```

### 11. Computational Methods:

All geometry optimizations were performed using the def2-TZVP basis set[8] within **GAUSSIAN 16**.[9] For functional tuning, geometry optimizations were carried out using the smaller def2-SVP basis set.[8] Harmonic frequency calculations were performed to confirm that the optimized geometries correspond to either minima (all positive frequencies) or first-order saddle points (a single imaginary frequency). In a few cases, very small imaginary frequencies ($<50$ cm$^{-1}$) related to out-of-plane torsion were obtained; these were disregarded as the calculations were done in the gas phase, but surface adsorption would enforce planarization. Local domain-based pair natural orbital coupled-cluster with single and double excitations and perturbative triples correction DLPNO-CCSD(T1)-F12[10] and fully internally contracted complete active space second-order perturbation theory FIC-CASPT2[11] single-point energies were evaluated with **ORCA 5.0.3**.[12] FIC-CASPT2 calculations were performed with the def2-TZVP basis set[8], whereas DLPNO-CCSD(T1)-F12 calculations employed F12-optimized basis sets. $S_1$ energies were computed by time-dependent DFT (TDDFT) using **GAUSSIAN 16**[9]. Ring current calculations were performed using **SYSMOIC**,[13] and $G_0W_0$[14,15] IPEA gaps were obtained using **TURBOMOLE 7.6**[16] with the def2-QZVP basis set[8]. Zero point vibrational energy (ZPVE) corrections to reaction barriers in **Figs. S12** and **S15** were computed using **Goodvibes**.[17]

## 12. Functional tuning:

This work follows our recently introduced strategy for fine-tuning DFAs.[18] Briefly, a single-point calculation with the ideal DFA will give the exact ground state energy $E_{exact}$. A geometry optimization with an ideal DFA will yield the correct equilibrium geometry and exact ground state energy at equilibrium, $E_{exact,eq}$. If we assume that high-level wavefunction theory methods such as CASPT2 and CCSD(T) give a reliable single-point reference energy $E_{ref}$, such that $E_{ref} \approx E_{exact}$, then we can tune DFAs by (1) performing a geometry optimization using a chosen DFA, (2) evaluating $E_{ref}$ using a high-level method, and (3) finding DFA parameters which minimize $E_{ref}$, i.e. which approach the correct equilibrium geometry and $E_{exact,eq}$.

In this work, we test the DFAs shown in **Fig. S35a**,**b** on the model systems shown in **Fig. S35c**, using FIC-CASPT2 (**section 12a**) and DLPNO-CCSD(T1) (**section 12b**) to evaluate $E_{ref.}$ Following our previous work, we tune the proportion of EE in the short range (at $r_{12} = 0$, see **Fig. S35a,b**) and the range-separation parameter $\omega$, which controls how quickly EE changes with the inter-electronic distance $r_{12}$.

As a starting point for functional tuning, we use the OX-B3LYP DFA (red in **Fig. S35b**), which was originally developed for edge-fused porphyrin nanobelts.[18] Its predictions were recently validated by $^{1}$H NMR spectroscopy, demonstrating quantitative agreement between theory and experiment.[19] For comparison, we also include DFAs often used to describe cyclocarbons (ωB97XD, BHLYP, and M06-2X; teal, purple, and orange, respectively, in **Fig. S35a**). Both FIC-CASPT2 and DLPNO-CCSD(T1) tuning indicates that OX-BLYP30 (dark blue in **Fig. S35b**) is the best DFA for describing cyclocarbons (see **sections 12a** and **12b**, respectively). OX-BLYP30 inherits the $\omega = 0.025\ a_0^{-1}$ (where $a_0$ is the Bohr radius, $a_0 = 0.529$ Å) range-separation parameter from OX-B3LYP, and includes 30% EE in the short-range. The 30% EE in the short range, assuming no range separation, is the minimum amount required to correctly reproduce the geometric features obtained for $N = 10–18$ cyclocarbons by experiments.[1,2,20,21,22]

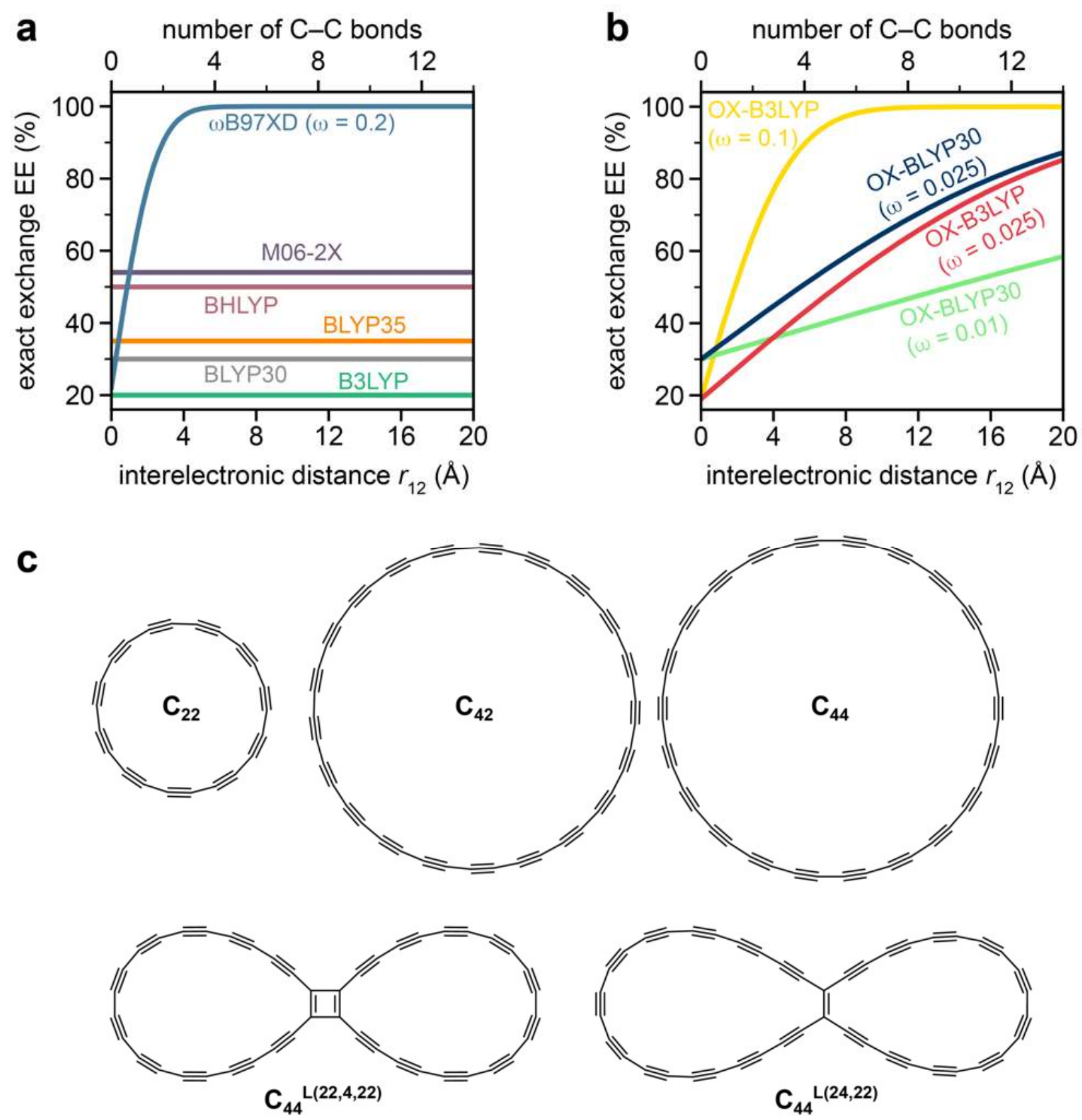

**Figure S35. DFA tuning.** (**a, b**) Variation of EE with interelectron distance ($r_{12}$) in popular hybrid functionals. In (**a**) ωB97XD (blue), M06-2X (plum), BHLYP (rose), BLYP35 (orange), BLYP30 (grey), B3LYP (dark green), and in (**b**) an OX-B3LYP variant with $\omega$ = 0.1 $a_0^{-1}$ (yellow), OX-BLYP30 (dark blue), OX-B3LYP (red), and OX-BLYP30 with $\omega$ = 0.01 $a_0^{-1}$ (light green). (**c**) Structures of the five cyclocarbon systems used for functional tuning, i.e., $C_{22}$, $C_{42}$, $C_{44}$, $C_{44}^{L(22,4,22)}$, $C_{44}^{L(24,22)}$.

12a. DFA tuning using FIC-CASPT2:

For each system in the test set (**Fig. 35c**), the relative energy $\Delta E_{ref}$ was defined relative to the lowest obtained value of $E_{ref}$ with any of the tested functionals. **Figure S36** shows the average of these relative energies for the whole test set, indicating that OX-BLYP30 ($\omega$ = 0.025 $a_0^{-1}$) and an OX-BLYP30 variant with $\omega$ = 0.01 $a_0^{-1}$ are most appropriate DFAs for cyclocarbons.

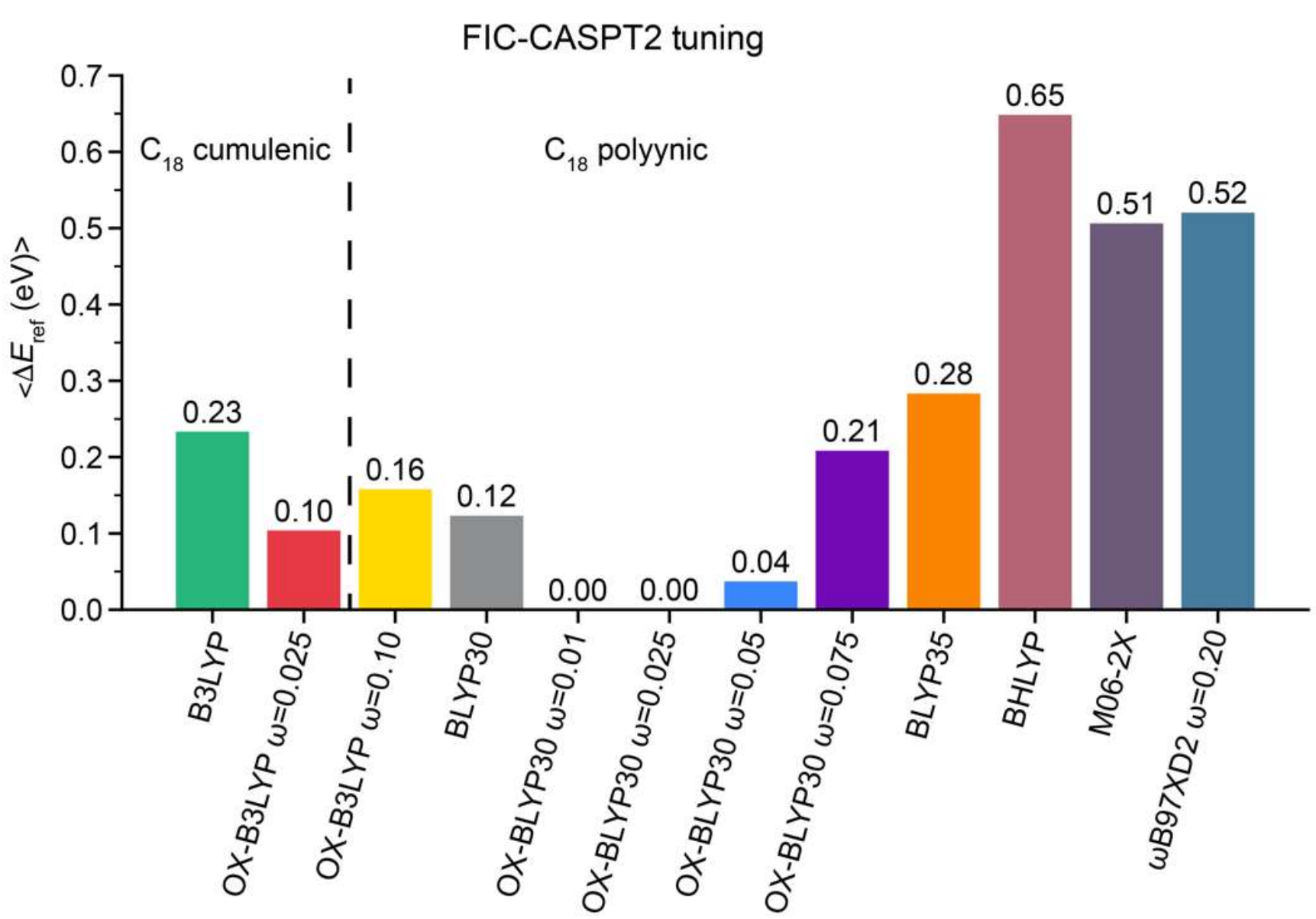

**Figure S36.** Average of relative single-point FIC-CASPT2 energies $\Delta E_{ref}$ evaluated on DFA-optimized geometries of the molecules in the training set (**Fig. S35c**). Energies are shown relative to the lowest obtained value. The dashed line indicates the boundary between functionals that yield a cumulenic geometry for $C_{18}$ (left) and those that stabilize the experimentally observed polyynic structure with bond-length alternation (right). For $C_{44}$ we used the (12,12) active space, i.e. correlated 12 electrons in 12 orbitals, while for the other molecules of the training set we used the (8,8) active space.

12b. DFA tuning using DLPNO-CCSD(T1):

For coupled clusters-based tuning, we limited our training set to $C_{42}$ and $C_{44}$, as lemniscates and smaller cyclocarbons showed substantial multireference character with the $T_1$ diagnostic >0.02. The results in **Fig. S37** show that OX-BLYP30 ($\omega$ = 0.025 $a_0^{-1}$) and an OX-BLYP30 variant ($\omega$ = 0.050 $a_0^{-1}$) are the best performers.

Combining results from **Fig. S36** and **Fig. S37**, OX-BLYP30 ($\omega$ = 0.025 $a_0^{-1}$) is the overall best-performing functional tested.

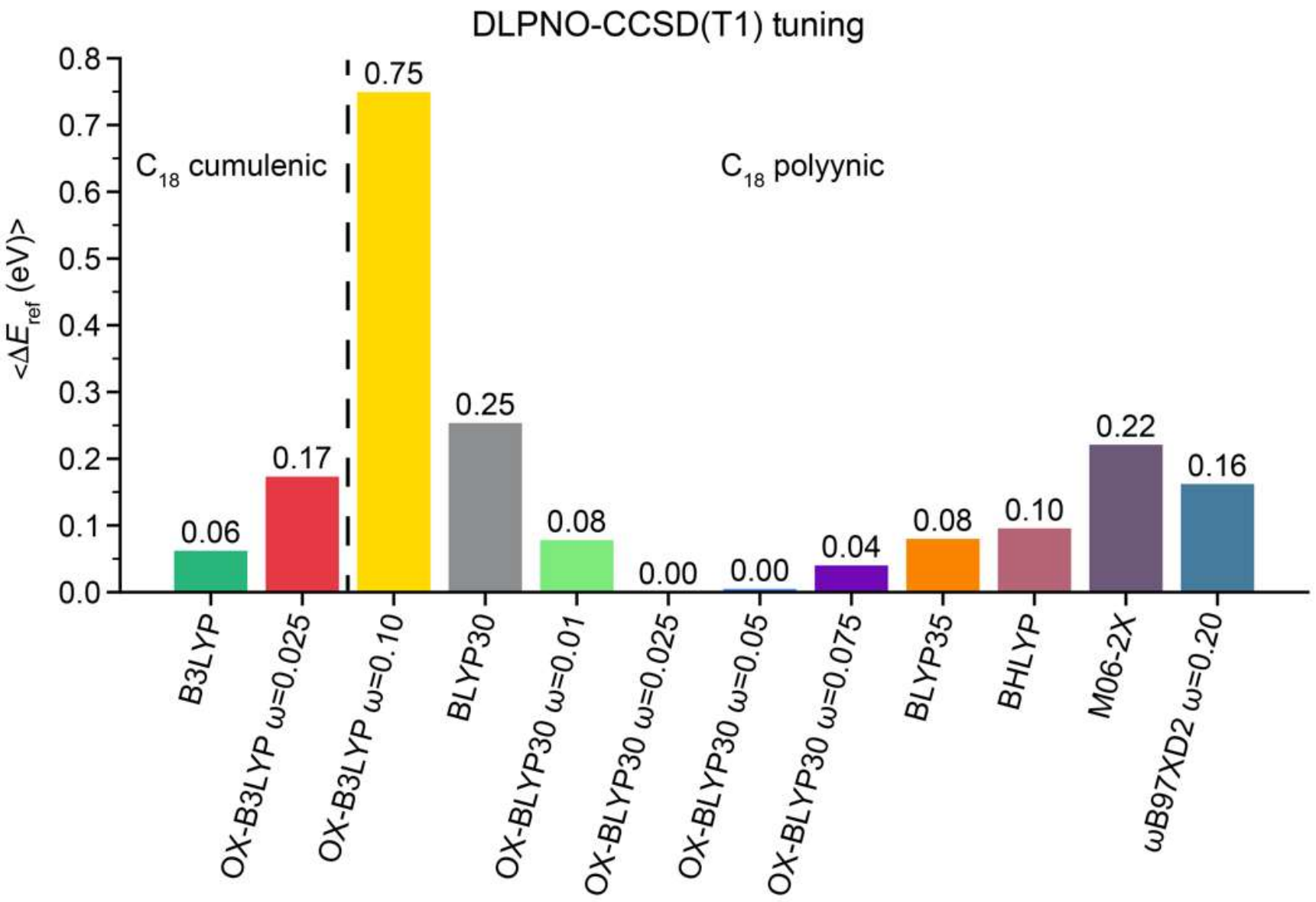

**Figure S37.** Average of relative single-point DLPNO-CCSD(T1) energies $\Delta E_{ref}$ evaluated on DFA-optimized geometries of $C_{42}$ and $C_{44}$. The explicitly correlated CC-PVDZ-F12 basis set was used. Energies are shown relative to the lowest obtained value. The dashed line indicates the boundary between functionals that yield a cumulenic geometry for $C_{18}$ (left) and those that stabilize the experimentally observed polyynic structure with bond-length alternation (right).

### 13. Bond-length alternation in cyclocarbons:

To compute the BLA in the optimized cyclo[*N*]carbon geometries, we identified all carbon–carbon bonds and sorted the bond lengths in ascending order. BLA was defined as the difference between the average of the longer half and the shorter half of these sorted bond lengths:

$$\mathrm{BLA} = \langle d_{\mathrm{long}} \rangle - \langle d_{\mathrm{short}} \rangle \quad (1)$$

where the averages are computed as:

$$\langle d_{\mathrm{short}} \rangle = \frac{2}{n} \sum_{i=1}^{n/2} d_i \quad (2)$$

$$\langle d_{\mathrm{long}} \rangle = \frac{2}{n} \sum_{i=n/2+1}^{n} d_i \quad (3)$$

with *n* being the number of C–C bond lengths (equal to the number of carbon atoms in the ring), and $d_i$ representing the sorted bond lengths in ascending order. This metric captures the extent of BLA along the ring and reflects the degree of cumulenic or polyynic character in the structure. **Fig. S38** shows the calculated BLA for even-*N* cyclocarbons with *N* = 6–100 using OX-BLYP30.

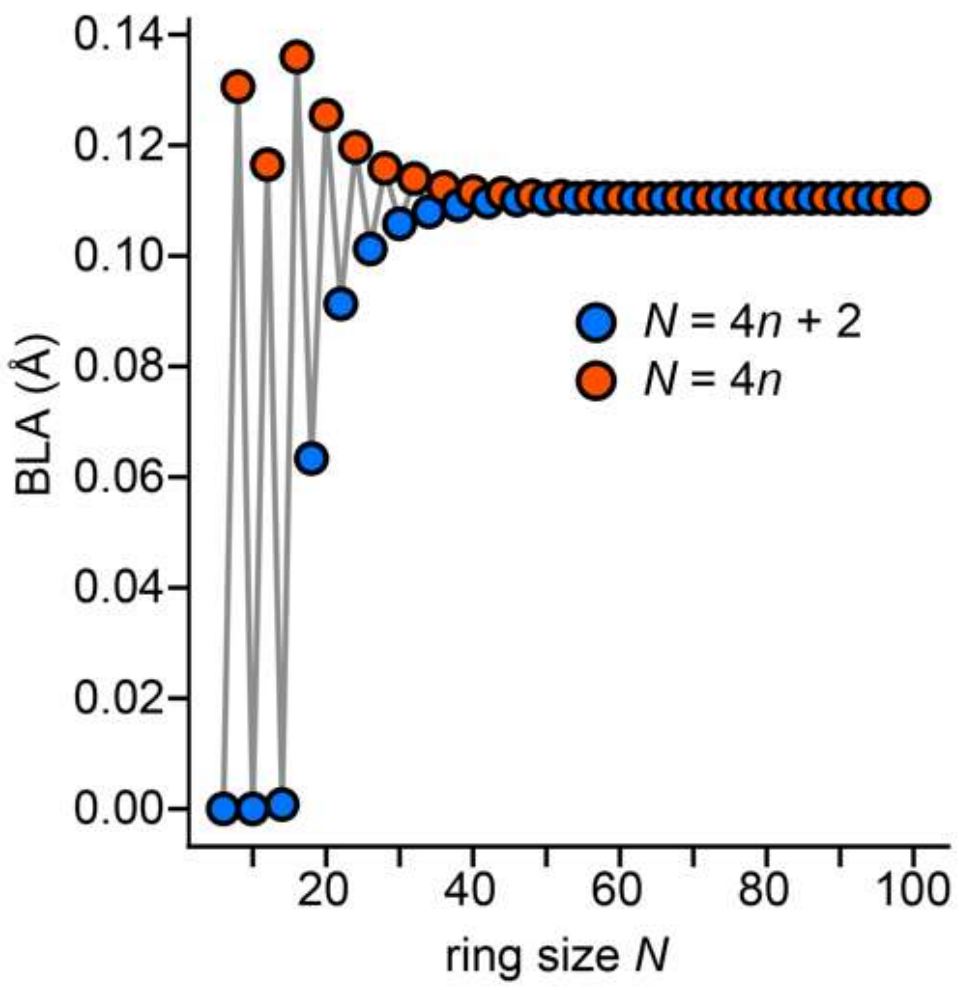

**Figure S38.** Bond-length alternation (BLA) for even-*N* cyclocarbons with *N* = 6–100 calculated at OX-BLYP30/def2-TZVP.

## 14. Aromatic stabilization energies:

**Cyclization.** The cyclization energy of cyclo[*N*]carbon ($C_N$) was evaluated relative to linear polyynic precursors using the isodesmic reaction in **Scheme S1**.

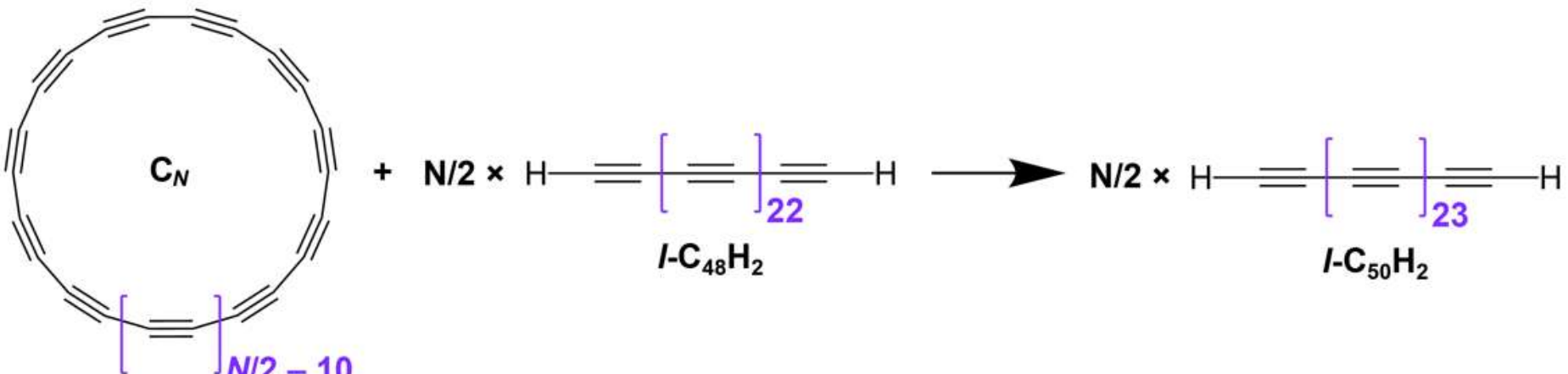

**Scheme S1.** Isodesmic reaction used to compute cyclocarbon cyclisation energies.

**Strain and Aromatic stabilization energy.** The cyclization energy $E_{\text{cyc}}$ can be written as a sum of the strain energy $E_{\text{strain}}$, which is always positive, and the aromatic stabilization energy ASE, which is positive for $N = 4n$, where $n$ is an integer, and negative for $N = 4n + 2$:

$$\Delta E_{\text{cyc}}(N) = E_{\text{strain}}(N) + \text{ASE}(N) \tag{4}$$

The average of two consecutive even-*N* cyclization energies is:

$$\frac{\Delta E_{\text{cyc}}(N)+\Delta E_{\text{cyc}}(N+2)}{2} = \frac{E_{\text{strain}}(N)+E_{\text{strain}}(N+2)}{2} + \frac{\text{ASE}(N)+\text{ASE}(N+2)}{2} \tag{5}$$

The first term on the right-hand side of eq. 5 is the approximate strain energy in the vicinity of $N + 1$. The second term will be close to zero, as the two ASE values will have comparable magnitude and opposite sign. Therefore:

$$\frac{\Delta E_{\text{cyc}}(N)+\Delta E_{\text{cyc}}(N+2)}{2} \approx E_{\text{strain}}(N+1) \tag{6}$$

The strain energy at *N* can then be evaluated as:

$$E_{\text{strain}}(N) \approx \frac{E_{\text{strain}}(N-1)+E_{\text{strain}}(N+1)}{2} \approx \frac{\Delta E_{\text{cyc}}(N)}{2} + \frac{\Delta E_{\text{cyc}}(N-2)+\Delta E_{\text{cyc}}(N+2)}{4} \tag{7}$$

Combining eqs. 7 and 4, we obtain:

$$\text{ASE}(N) \approx \frac{\Delta E_{\text{cyc}}(N)}{2} - \frac{\Delta E_{\text{cyc}}(N-2)+\Delta E_{\text{cyc}}(N+2)}{4} \tag{8}$$

which is equivalent to eq. (2) from ref. 23.

**15. Other supplementary figures:**

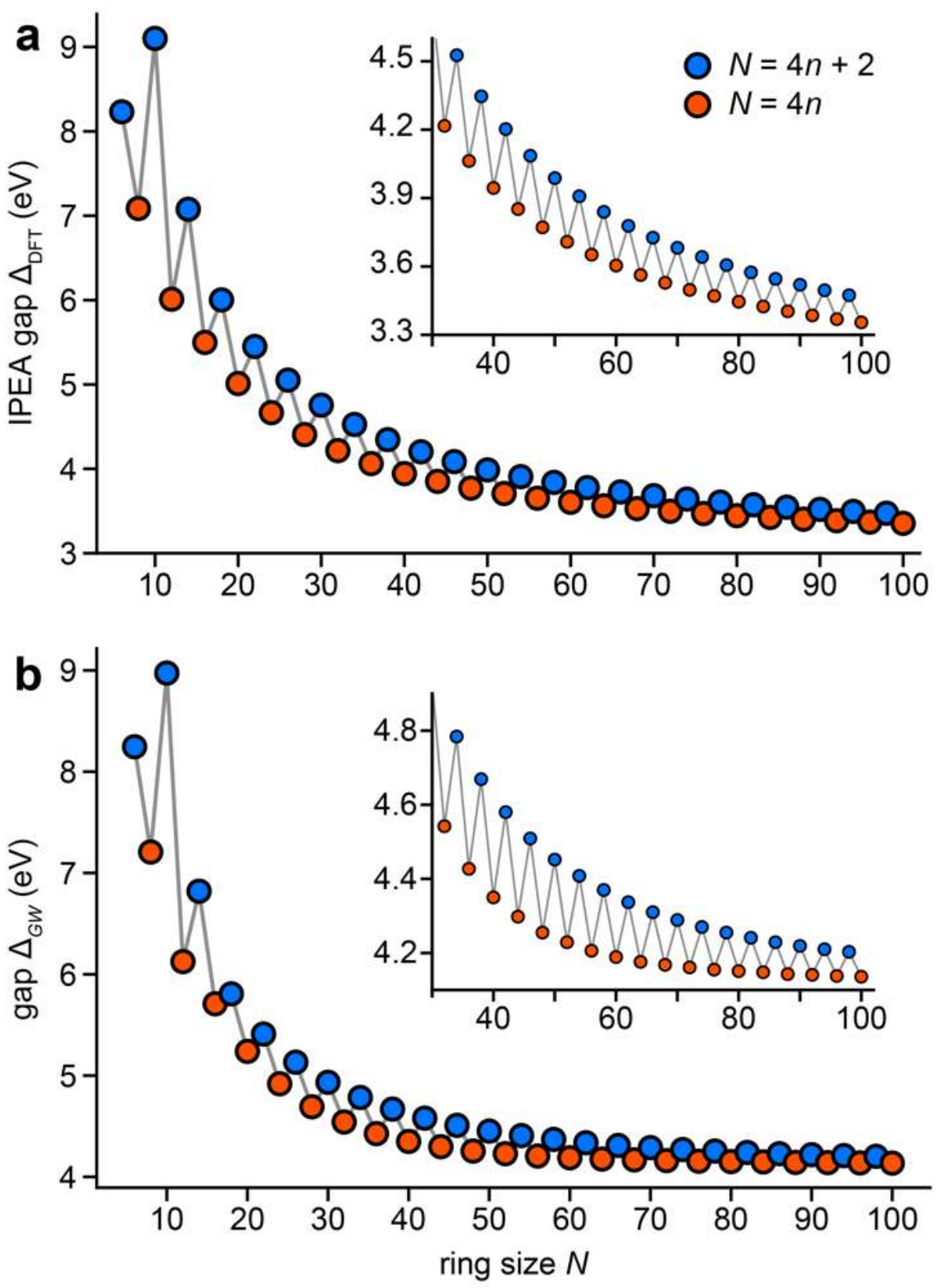

**Figure S39.** Transport gaps Δ calculated at OX-BLYP30/def2-TZVP (**a**) and $G_0W_0$@OX-BLYP30/def2-QZVP (**b**) level of theory for even-$N$ cyclocarbons with $N$ = 6–100.

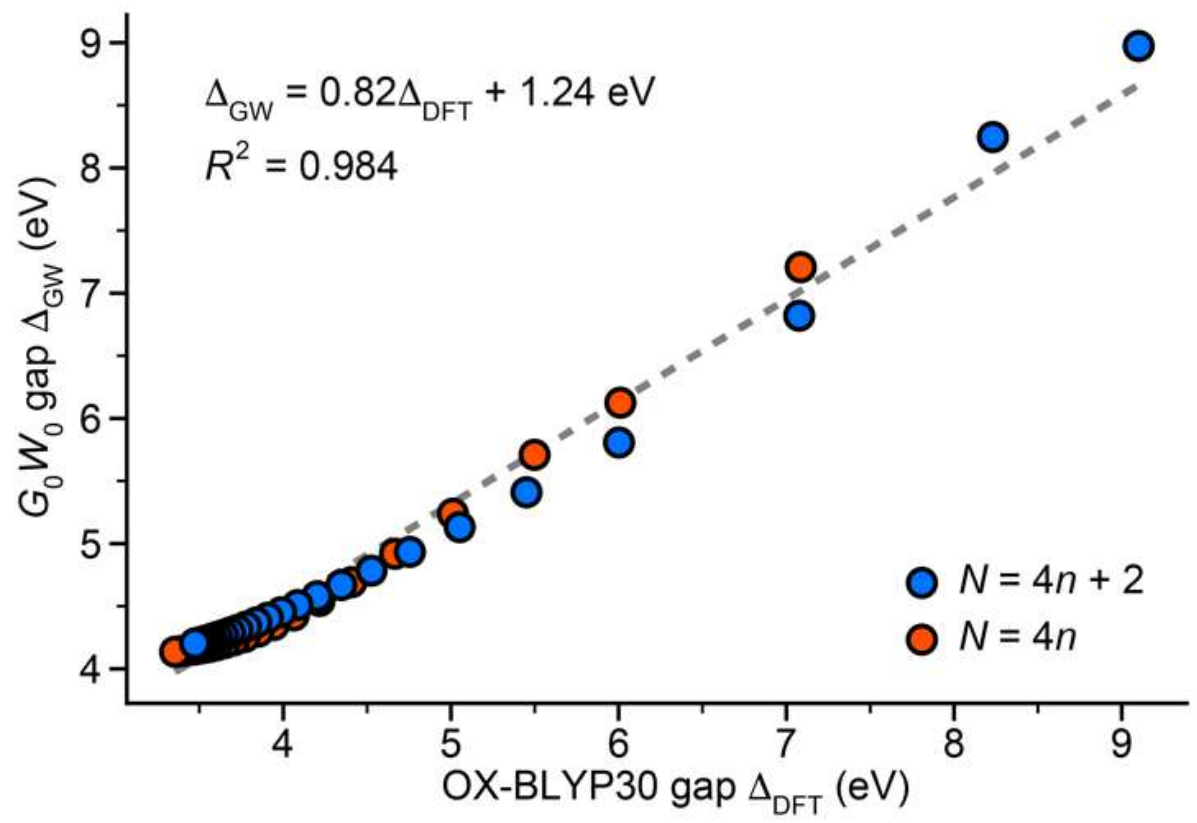

**Figure S40.** Linear regression between $G_0W_0$@OX-BLYP30/def2-QZVP gaps and OX-BLYP30/def2-TZVP IPEA gaps for even-$N$ cyclocarbons with $N$ = 6–100. The dashed line indicates the linear least-squares fit, with the slope and $R^2$ value shown. Good linear correlation shows that the use of the $G_0W_0$ method is appropriate.[24]

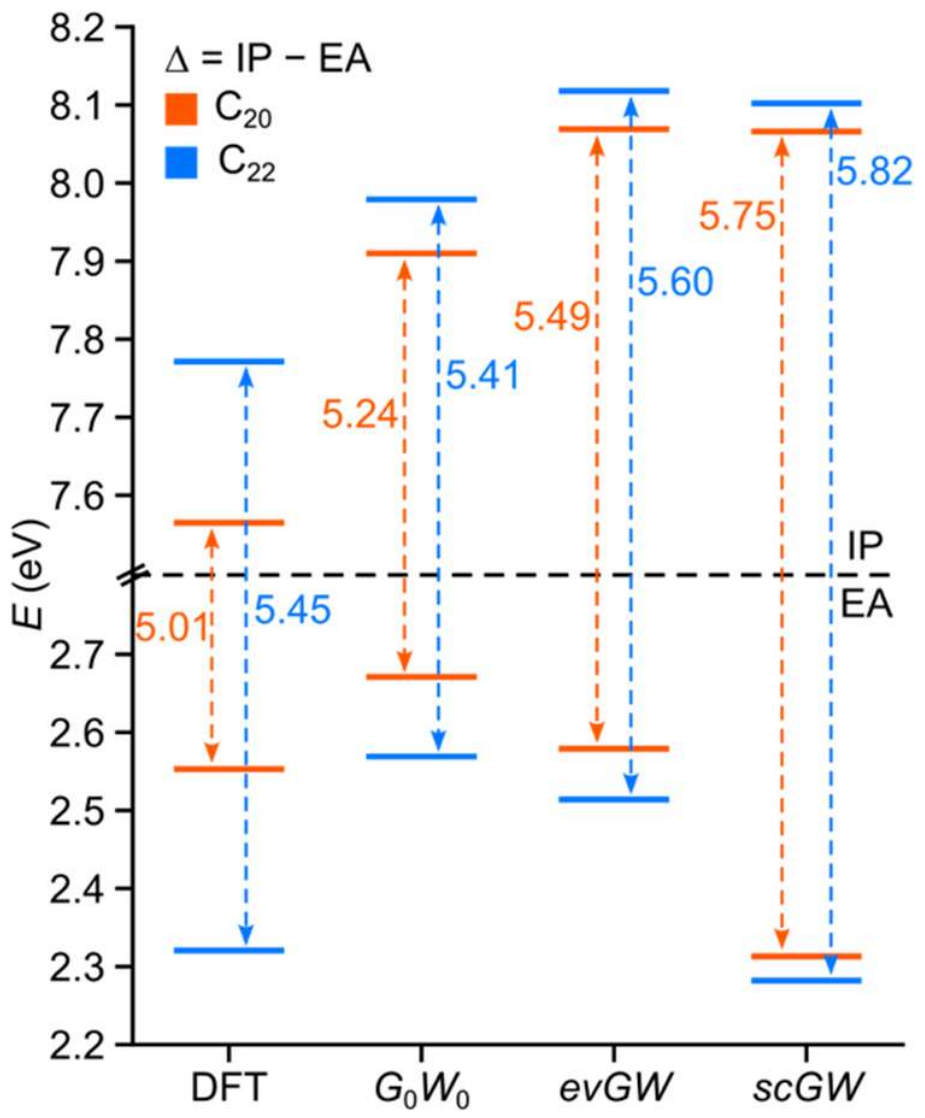

**Figure S41.** Ionization potentials (IP), electron affinities (EA), and transport gaps (Δ = IP − EA; arrows) for $C_{20}$ (orange) and $C_{22}$ (blue), calculated using OX-BLYP30/def2-TZVP (DFT), and several flavors of *GW* method: single-shot $G_0W_0$@OX-BLYP30/def2-QZVP ($G_0W_0$), eigenvalue-based *evGW*@OX-BLYP30/def2-QZVP (*evGW*), and fully self-consistent *scGW*@OX-BLYP30/def2-TZVP (*scGW*), where *evGW* and *scGW* reduce the dependence of *GW* on the DFT wavefunction used as the starting point.

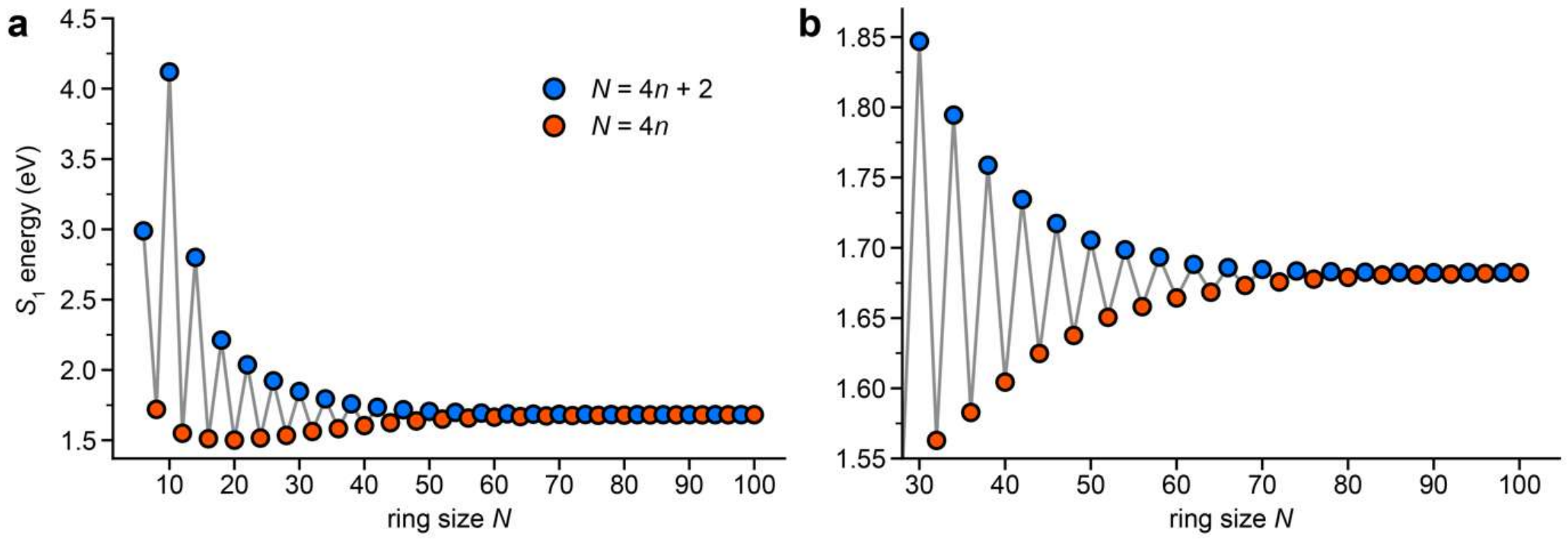

**Figure S41.** Lowest ($S_1$) vertical excitations energies for even-$N$ cyclocarbons ($N$ = 6–100) calculated at OX-BLYP30/def2-TZVP. Panel (**a**) shows the full range; panel (**b**) shows a zoomed-in view.

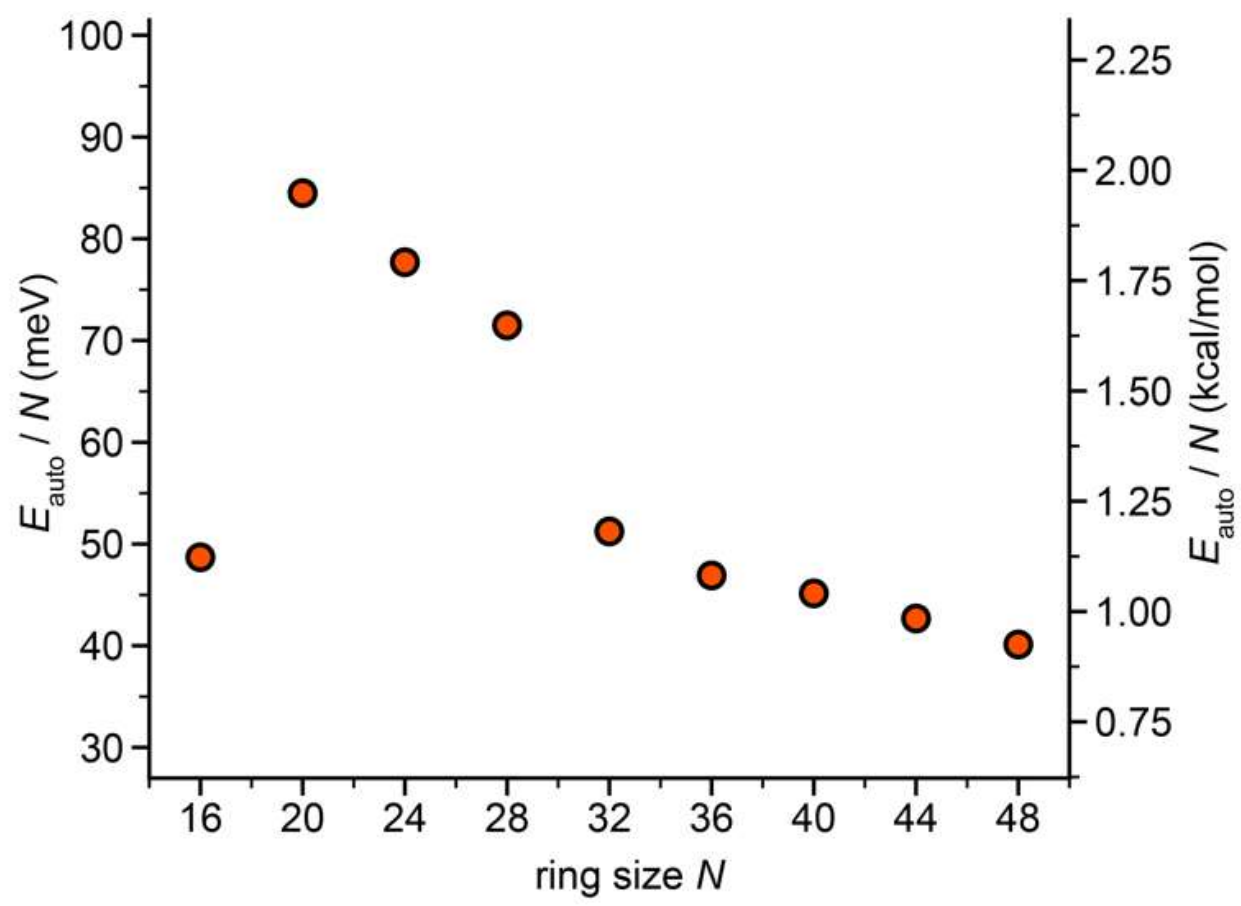

**Figure S42.** Automerization barriers of $N = 4n$ cyclocarbons, calculated at OX-BLYP30/def2-TZVP. The relatively low barrier for $C_{16}$ is likely due to the lowering of strain in the transition structure relative to the minima. The large decrease in the automerization barrier per bond observed from $C_{28}$ to $C_{32}$ can be rationalized by a qualitative change in the nature of the transition-state structures. For smaller rings ($16 < N \leq 28$), the automerization transition states retain cumulenic bonding pattern and remain planar. For larger rings ($N \geq 32$), the transition states exhibit significant bond-length alternation, accompanied by additional symmetry-breaking distortions. These include out-of-plane bending as well as deviations from circular symmetry toward slightly elliptical geometries. The combined emergence of bond-length alternation and geometric symmetry breaking lowers the automerization barrier by stabilizing the transition state relative to the high-symmetry, nearly planar structures observed for smaller rings, leading to the observed drop in energy from $N = 28$ to $N = 32$.

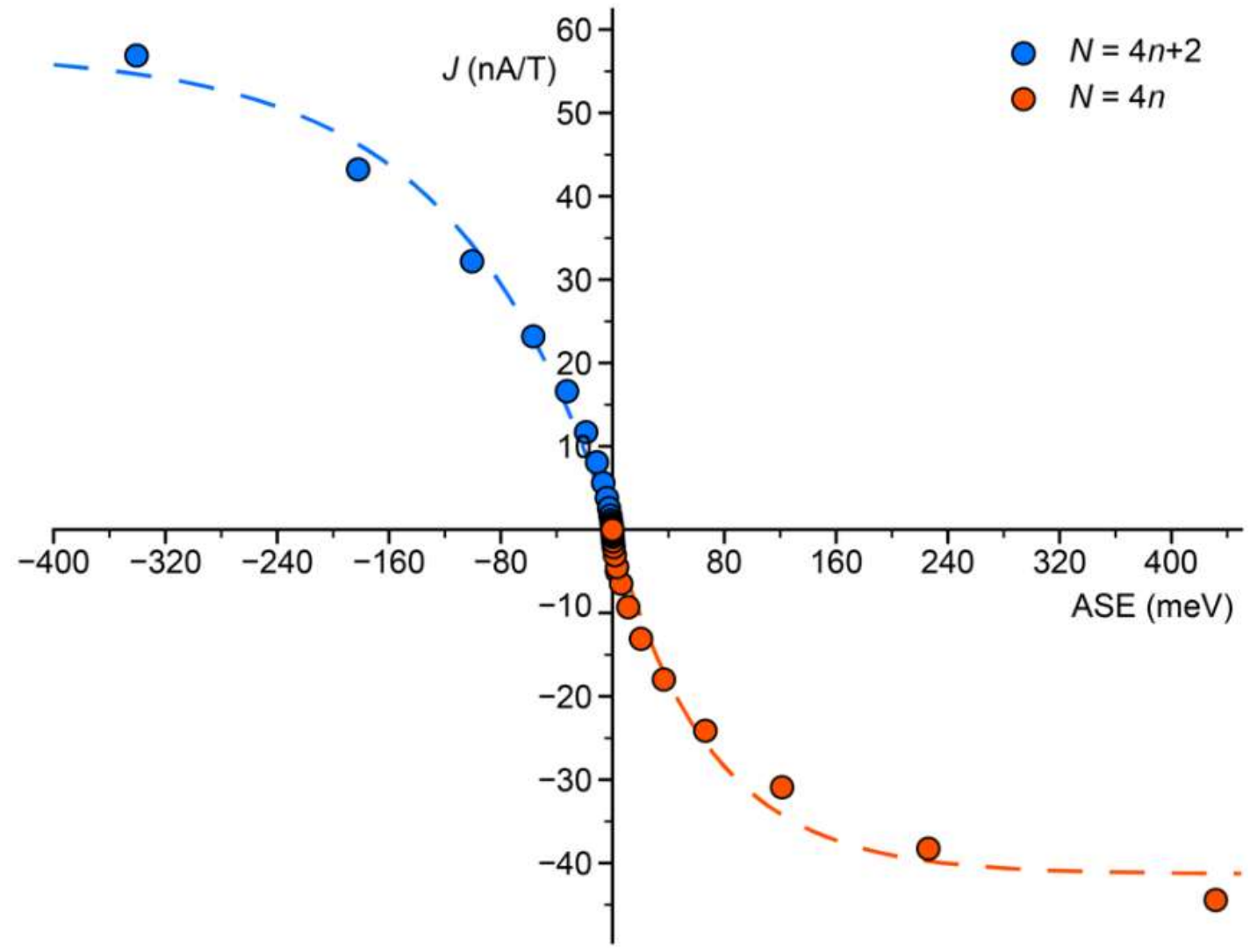

**Figure S43.** Aromatic stabilization energy ASE (in meV) and ring current $J$ (in nA/T) for even-$N$ cyclocarbons ($N$ = 16–98) computed at the OX-BLYP30/def2-TZVP level of theory. Data are shown for aromatic ($4n$+2, blue) and anti-aromatic ($4n$, orange) systems. The blue and orange lines are exponential fits to guide the eye. An approximately linear relationship is obtained for large values of $N$ (small ASE), as expected from a simple Hückel model for particles on a ring.[18]

## 16. Quantum computing

To investigate electron correlation, we performed single-point calculations on BLYP30/def2-TZVP-optimized geometries of $C_{20}$ and $C_{22}$ at the Hartree-Fock (HF) and the complete active space self-consistent field (CASSCF) levels of theory. These calculations were done using OpenMolcas[24] and the cc-pVDZ[25] basis set.

**Active space considerations.** As even-numbered cyclocarbons can be described with a double particle-on-a-ring model[26], for each angular momentum $k > 0$ they will have four orbitals (two in-plane and two out-of-plane) with $k$ nodal planes perpendicular to the plane of the ring. In $N = 4n + 2$ cyclo[$N$]carbons, all orbitals with $k \leq n$ will be occupied and those with $k > n$ will be unoccupied, so the smallest sensible half-filled active space is (8,8), that is 8 electrons in 8 orbitals, which captures all orbitals with angular momentum $k = n$ (occupied) and $k = n+1$ (unoccupied). In $N = 4n$ cyclo[$N$]carbons, orbitals with $k = n$ will be half-filled[27], making (4,4) the smallest sensible half-filled active space. Larger active spaces for both types of even-numbered cyclocarbons can be obtained by adding $4m$ occupied orbitals and $4m$ unoccupied orbitals, where $m$ is an integer smaller than $n$.

**Table S1. Sensible half-filled active space (AS) sizes for cyclocarbons.**

| ***N* / orbs in AS** | ***k* = *n*** | ***k* ∈ [*n*, *n*+1]** | ***k* ∈ [*n*–1, *n*, *n*+1]** | ***k* ∈ [*n*–1, *n*, *n*+1, *n*+2]** |
|---|---|---|---|---|
| 4*n* | 4,4 | – | 12,12 | – |
| 4*n* + 2 | – | 8,8 | – | 16,16 |

With this in mind, as a starting point we used CASSCF(12,12) and CASSCF(16,16) calculations for $C_{20}$ and $C_{22}$, respectively. Cations and anions were computed by only changing the number of electrons by one and otherwise keeping the active space the same. These results were used to prepare one- and two-electron integrals (FCIDump files) in expanded (28,28) and (32,32) active spaces for $C_{20}$ and $C_{22}$, respectively.

**Large active space configuration interaction.** Approximate configuration interaction (CI) calculations in the expanded (28,28) and (32,32) active spaces for $C_{20}$ and $C_{22}$, respectively, as well as for their cations and anions, were performed using the semi-stochastic heat bath CI (HCI) method, as implemented in Dice[28], and the SqDRIFT algorithm[29,30], executed on quantum hardware (**Fig. S45a**). For both methods we kept the subspace size, i.e. the number of determinants in projected CI matrix, at around $10^7$ (**Fig. S45b,c**).

**SqDRIFT computational details.** The quantum calculations were performed on an IBM Heron r3 quantum processor (*ibm_boston*), which features 156 qubits, a median two-qubit gate error of $1.16 \cdot 10^{-3}$, and a median readout error of $5 \cdot 10^{-3}$. The molecular Hamiltonian was mapped onto the quantum hardware using the Jordan–Wigner transformation, yielding quantum circuits with 56 and 64 qubits for $C_{20}$ and $C_{22}$, respectively. For each carbon ring, three molecular species were considered: neutral, anionic, and cationic. The corresponding circuits were prepared with 20 fermionic excitations and evaluated at three different evolution times, $t = 1.0,\ 2.0, 3.0$ (in atomic units), with 500 randomizations per configuration. Each circuit was measured 1024 times, resulting in a total of approximately 1.5M shots per molecular species.

**Discussion of the different theoretical methods used.** SqDRIFT surpasses best exact active space diagonalization-based conventional methods, that is FCI(20,20)@CASSCF(12,12) for $C_{20}$ and CASSCF(16,16) for $C_{22}$. HCI yields the lowest absolute energies; however, we note that SqDRIFT yields transport gap values within ~0.5 eV of the HCI estimate. These results are encouraging because the performance of SqDRIFT and similar algorithms is rapidly improving with quantum hardware advances[30], while conventional approaches do not benefit as much from improvements in classical hardware due to the exponential scaling of the CI problem.

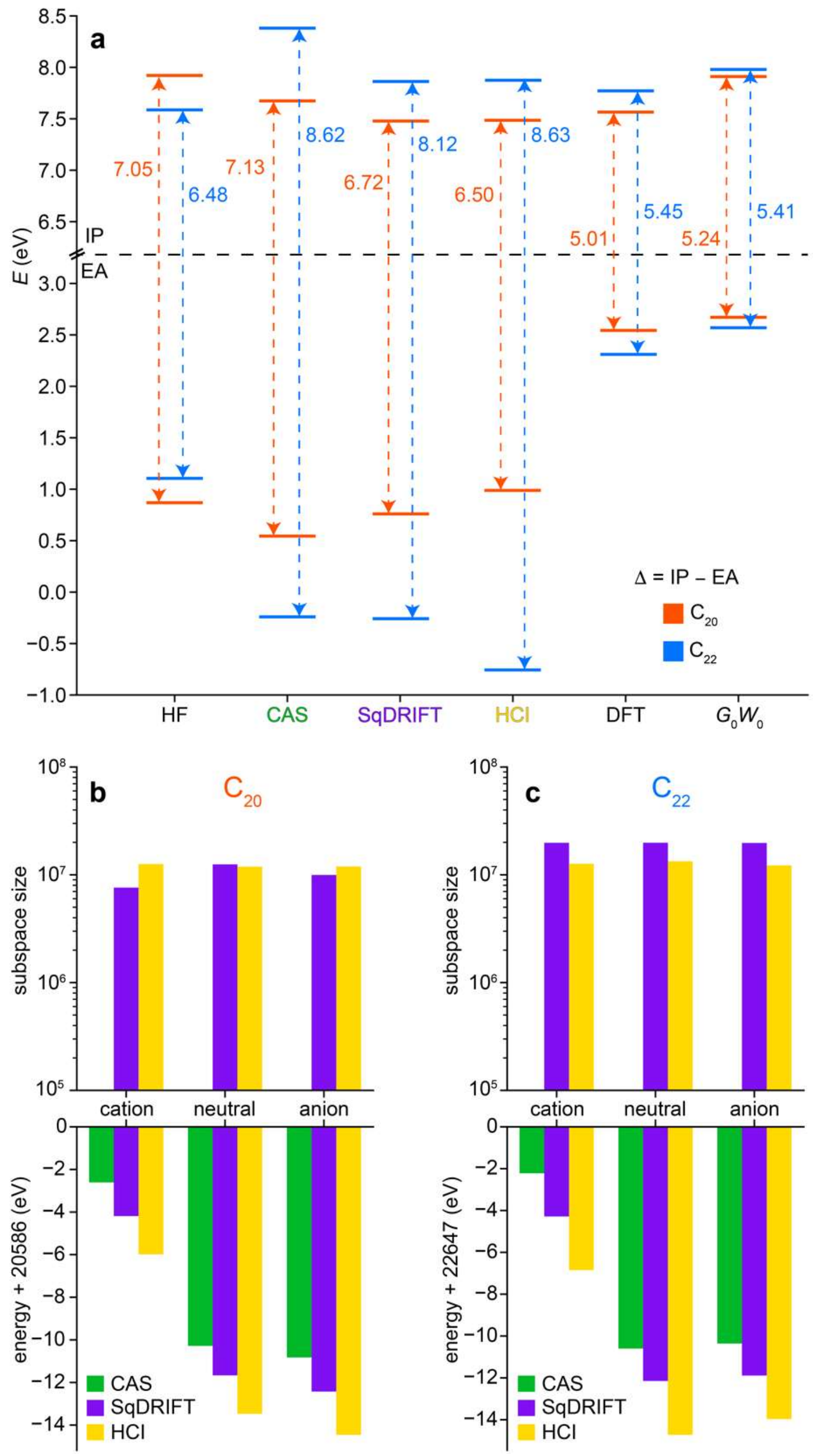

**Figure S45.** SqDRIFT results. **a**, ionization potentials (IP), electron affinities (EA), and transport gaps (arrows) obtained using Hartree-Fock (HF), CASSCF (CAS, green): CI(20,20)@CASSCF(12,12) for $C_{20}$ (orange) and CASSCF(16,16) for $C_{22}$ (blue), SqDRIFT (purple), and HCI (yellow). The active space for SqDRIFT and HCI was (28,28)@CASSCF(12,12) for $C_{20}$ and (32,32)@CASSCF(16,16) for $C_{22}$. **b** and **c**, absolute energies for $C_{20}$ (**b**) and $C_{22}$ (**c**) obtained using CAS (green), SqDRIFT (purple), and HCI (yellow) on the negative y-axis. The positive y-axis compares the subspace size for SqDRIFT (purple), and HCI (yellow).

## 17. References of Supplementary Information